\documentclass[aps,twocolumn,prl,floatfix,superscriptaddress]{revtex4-2}

\usepackage{graphicx}
\usepackage{amsmath,amsfonts,amssymb,bm}

\usepackage{lmodern}
\usepackage{xcolor}
\usepackage[normalem]{ulem}
\usepackage{float}
\usepackage{braket}
\usepackage{dsfont}

\makeatletter
\usepackage{pdfpages} 
\AtBeginDocument{\let\LS@rot\@undefined}
\makeatother

\usepackage[colorlinks,
            linkcolor=blue,    
            citecolor=blue,  
            urlcolor=blue,
 	        bookmarks=true,        
	        bookmarksopen=true,    
	        bookmarksnumbered=true,
]{hyperref}

\newcommand{\bs}[1]{\boldsymbol{#1}}
\newcommand{\beq}{\begin{equation}}
\newcommand{\eeq}{\end{equation}}

\definecolor{magenta}{rgb}{1.0, 0.0, 1.0}

\begin{document}

\title{Identifying, and constructing, complex magnon band topology}

\author{Alberto Corticelli}
\affiliation{Max Planck Institute for the Physics of Complex Systems, N\"{o}thnitzer Str. 38, 01187 Dresden, Germany}
\author{Roderich Moessner}
\affiliation{Max Planck Institute for the Physics of Complex Systems, N\"{o}thnitzer Str. 38, 01187 Dresden, Germany}
\author{Paul~A.~McClarty}
\affiliation{Max Planck Institute for the Physics of Complex Systems, N\"{o}thnitzer Str. 38, 01187 Dresden, Germany}

\begin{abstract}
Magnetically ordered materials tend to support bands of coherent propagating spin wave, or magnon, excitations. Topologically protected surface states of magnons offer a new path towards coherent spin transport for spintronics applications. In this work we explore the variety of topological magnon band structures and provide insight into how to efficiently identify topological magnon bands in materials. We do this by adapting the topological quantum chemistry approach that has used constraints imposed by time reversal and crystalline symmetries to enumerate a large class of topological electronic bands. We show how to identify physically relevant models of gapped magnon band topology by using so-called decomposable elementary band representations, and in turn discuss how to use symmetry data to infer the presence of exotic symmetry enforced nodal topology. 
\end{abstract}

\maketitle

{\it Introduction} $-$ There have been considerable efforts in the last few years to provide a taxonomy of nontrivial topological band structures enforced or allowed by time reversal and crystalline symmetries \cite{fang2012,parameswaran2013,watanabe2015,po2016,watanabe2016,bradlyn2017,kruthhof2017,watanabe2018,cano2018,song2018a,song2018b,elcoro2020magnetic,xu2020}. This work has brought powerful new concepts that tie crystal and magnetic structures to band topology. At the same time these ideas provide efficient methods to
efficiently search for topological materials resulting in a vast database of {\it ab initio} driven predictions of new electronic topological materials \cite{vergniory2019,topmat}. Such materials include gapless and gapped bulk topological matter with protected boundary states and anomalous transport properties.  The culmination of these efforts to classify band topology based on symmetry and to use symmetry data to diagnose topological bands is called  topological quantum chemistry (TQC) \cite{bradlyn2017,elcoro2020magnetic}. 

In a similar time frame, there has been increasing interest in exploring the role of band topology in {\it magnetic} excitations and how it affects the properties of magnetic materials \cite{mcclartyreview2022,bonbien2021topological,malki2020topological}. The pioneering work in this area has mainly been in devising  physically well-motivated models of magnon band topology such as Chern insulators  and Weyl magnons \cite{zhang2013,shindou2013edge,shindou2013topological,mook2016dynamics,mook2015waveguide,romhanyi2015hall,mook2019coplanar,mook2014magnon,mook2014edge,diaz2019skyrmion,Rold_n_Molina_2016,owerre2016first,mcclarty2018,Li2016,mook2016weyl}. This has inspired early experimental efforts to characterize magnon topology in materials \cite{mcclarty2017topological,boyuan2020,elliot2021,scheie2022,scheie2022b,weber2022}. All this work has gone hand-in-hand with the exploration of unusual transport properties in insulating magnets such as the thermal Hall effect \cite{onose2010observation,katsura2010theory,matsumoto2011prl,matsumoto2011rotational,matsumoto2014thermal,murakami2016thermal,kasahara2017thermal,kasahara2018,hentrich2019} and the exploration of signatures of bulk nodal magnon topology in neutron scattering experiments \cite{shivam2017,elliot2021,scheie2022}. On the horizon, there are exciting potential spintronics developments to be made detecting and manipulating topological magnon boundary states \cite{feldmeier2020local,malz2019topological,ruckriegel2018,mitra2021}.

In this paper, we show that the TQC approach can be adapted to magnon band topology, providing a classification of {\it symmetry-determined} topological bands in spin wave Hamiltonians. The ideas can be used to diagnose magnon topology on one hand, and on the other to build models and identify candidate topological magnon materials. The physical foundation for this work is that topological bands by definition cannot be built from a Wannier basis while preserving all underlying symmetries. Topological quantum chemistry rests on an enumeration of all possible Wannierizable band structures through so-called elementary band representations (EBRs), to be described in more detail below so that, essentially by elimination, one may establish whether some set of bands is topologically nontrivial. 

{\it Ab initio} methods are central to TQC. The closest analogue in widespread use to study magnetic excitations is linear spin wave theory which is based on an expansion, to quadratic order, of the spins in fluctuations around some magnetic structure. The goal of this paper is to show how to pass from elementary symmetry information $-$ the crystal structure and the magnetic order $-$ to linear spin wave models with nontrivial topology. 

Our starting point is to establish how crystal and time reversal symmetries are implemented within linear spin wave theory. In contrast to electronic systems, the band structures of interest emerge from an effective exchange Hamiltonian. We describe how this Hamiltonian, in conjunction with the minimal energy magnetic structure, fixes the symmetries of the problem. These are encoded in some magnetic space group. We then outline how to build band representations for magnons starting from the local moments on each magnetic site giving a complete table of all site symmetry groups compatible with magnetic order. Band representations minimally encode symmetry information in the magnon band structure. With these ingredients, we are in a position to identify constraints that magnons place on the possible symmetry data and hence on the possible topological bands. In particular, it turns out that magnons in systems with {\it significant spin-orbit coupling}  form a subset of all electronic topological bands. 

With these foundations, we then show, first in general and then through a series of examples, how to use symmetry information alone to build exchange models whose elementary excitations have nontrivial gapped and nodal magnon topology and to identify candidate materials. Examples include Chern bands, antiferromagnetic topological insulators, and three-fold and six-fold nodal points. Crucially, our workflow can be straightforwardly reversed, to diagnose nontrivial topology from spin wave fits to experimental data.

{\it EBRs and Topology} $-$ Before getting into the specifics for magnons, we give a lightning introductory review of TQC. We refer the reader to the supplementary section \cite{SM} for more technical details that will not, however, be necessary to appreciate the remainder of this paper. 

The essential symmetry ingredients of TQC are nothing more than the symmetry group $G_M$ of the magnetic structure and the Wyckoff positions of the magnetic ions that appear in any structural refinement of a magnetic material. The group $G_M$ is generally one of the magnetic space groups that encodes combinations of crystallographic point group symmetries, lattice translations, time reversal symmetry and perhaps non-symmorphic elements. 
To each Wyckoff position $\boldsymbol{q}$, we may assign a site symmetry group (SSG) $G_{\boldsymbol{q}}$ defined as the subgroup of $G_M$ that leaves the site invariant. This is generally isomorphic to a magnetic point group. 

We then need to include some information about the underlying lattice degrees of freedom $-$ the nature of the atomic orbitals. These necessarily transform under some representation of $G_{\boldsymbol{q}}$.  Following Zak, from these representations of the magnetic SSG we may arrive at a representation of the full $G_M$ group by the standard process of {\it induction} \cite{zak1981}. The result is a so-called band representation (BR). The BR is a momentum space representation of all elements of $G_M$ that contains information about the connectivity of the bands and the topology. To connect to topology we define elementary band representations (EBRs) to be BRs that are not unitarily equivalent to a direct sum of two or more BRs. These hold a distinguished place in relation to topology because they are the elementary units from which all Wannierizable band structures can be built for a given symmetry group. Any set of bands that cannot be built from EBRs is necessarily topological overall. All EBRs for all magnetic space groups have been tabulated $-$ each one characterized by eigenvalues of all symmetry operations at high symmetry momenta. For all $1,651$ magnetic space groups, there are roughly $20,000$ EBRs. In order to diagnose topological bands, one should in principle determine whether each energetically isolated set of bands can be written as a direct sum of EBRs with non-negative integer coefficients. If so, the bands are trivial. If not, they are symmetry-determined topological bands. A more fine-grained determination of the nature of the topology then requires further analysis. Symmetry enforced nodal topological bands can be read off directly from the dimension greater than one irreducible representations at high symmetry points, lines and planes.

{\it Magnons and Symmetry} $-$ Building on the principles behind TQC we now discuss the ideas in relation to magnons. In this work we are mainly interested in crystalline solids with localized magnetic moments and nonvanishing local dipolar order parameter $\langle S_i^\alpha\rangle$ for site $i$ and component $\alpha$. The magnon or spin wave excitations are the transverse fluctuations of the local ordered moments. We restrict our attention to the typical case where these form coherent propagating bands. This means we neglect the role of multi-magnon states and possible interesting questions of novel topology \cite{mcclartyrau2019} and fragility that arise from such states. We also neglect magnetic excitations beyond the ground state multiplet that could be handled within a multi-boson formalism (see e.g.~\cite{elliot2021}) to which TQC ideas may also be applied.

The symmetries of the magnon bands are descended from those of the magnetic Hamiltonian $H_{\rm M}$ considered to be composed of exchange couplings, dipolar couplings, single ion anisotropies and perhaps an external magnetic field. The magnetic order breaks the symmetries of the magnetic Hamiltonian down to a subgroup. It is important to note that the relevant symmetry groups for magnons are single-valued because the bands are spinless or bosonic. These are the groups that are relevant to weakly spin-orbit coupled electronic systems. However, in the context of magnons, these groups are relevant to the case where the moments and the spatial transformations are locked, which can only happen when spin-orbit coupling at the microscopic level is significant. The spin-orbit coupling is reflected in the appearance of anisotropies in the magnetic Hamiltonian. As is well-known, there are many cases where the magnetic Hamiltonian has discrete or continuous rotation symmetries. In such cases, magnetic order may lead to residual symmetries described by the spin-space groups \cite{BrinkmanElliott1966,BrinkmanElliott1966b,Brinkman1967,corticelli2021spinspace}. Topological quantum chemistry applied to such groups is beyond the scope of this work. We consider the case where these residual symmetries are those of a magnetic space group $G_M$ with $n_S$ sublattices in the magnetic primitive cell leading to $n_S$ bands considered to be computed from linear spin wave theory based on Hamiltonian $H_{\rm LSW}=\frac{S}{2} \sum_{\boldsymbol{k}}  \hat{\boldsymbol{\Upsilon}}^{\dagger}(\boldsymbol{k}) \boldsymbol{M}(\boldsymbol{k}) \hat{\boldsymbol{\Upsilon}}(\boldsymbol{k})  $ where the transformation properties of $2n_S$ component $\hat{\boldsymbol{\Upsilon}}(\boldsymbol{k})$ can be inferred from the transformations of the $S_i^{\pm}$ transverse spin components in a frame where $S^z$ is the direction of the ordered moment. For reference, explicit formulas are given in the Supplementary Section \cite{SM}. 

To build band representations, we must first identify the SSG from that of the Wyckoff position of the magnetic ions by requiring that the on-site $S^z$ transforms as the total symmetric irrep of the SSG. This constraint reduces the possible $122$ magnetic point groups to a set of $31$ groups isomorphic to SSGs. The relevant orbital content is given by the local frame transverse spin components $S_i^{\pm}$. We give a complete list of the magnetic SSGs in the Supplementary Section together with the irreducible representations of the SSG for which $S_i^{\pm}$ form a basis \cite{SM}.

Given this information, one may build a band representation for magnons and, again, explicit formulas are given in the Reference Material \cite{SM}. Given an energetically isolated set of magnon bands one may then ask whether this decomposes into EBRs. The EBRs relevant to magnons corresponding to all magnetic structures and significant spin-orbit exchange are tabulated. In the remainder of this paper we give concrete examples of how to use the tabulated EBRs to build models of topological magnons. We take two main routes. The first is to focus on cases where the symmetry information about band connectivity allows EBRs to split up into disconnected bands. By definition at least one of the resulting bands must be topological. Our second focus will be on nodal topology. Several models are known with Dirac and Weyl magnon touching points \cite{mcclartyreview2022}. But symmetry can enforce higher order degeneracies $3$, $4$ and $6$-fold degeneracies, and we show how to build models with such degeneracies.

{\it Magnon topology from decomposable EBRs} $-$ To build models of decomposable EBRs we focus on cases where the magnetic ions live on {\it maximal} Wyckoff positions, i.e.\ positions of maximal magnetic point group symmetry for a given $G_M$. These are distinguished by the fact that BRs induced from such sites are themselves EBRs and not composites of EBRs (apart from some well-understood exceptional cases). We give a complete table of decomposable EBRs that can be obtained from maximal Wyckoff positions and the allowed SSGs organized by magnetic space group and Wyckoff position \cite{SM}. The utility of this table is that one may couple moments living on such Wyckoff positions and be sure that there will be nontrivial topology in the resulting magnon bands provided free parameters are tuned to avoid accidental degeneracies and provided the number of free parameters is adequate to reduce the symmetries to the required $G_M$. This approach is a highly efficient means to build models of magnon topology and contrasts to generic cases of nontrivial topology where, in practice, one should compute so-called symmetry indicators as a function of free couplings to diagnose the topology.

We take an example to illustrate the main ideas $-$ the well-established case of Chern magnon bands in the Kitaev-Heisenberg honeycomb model with $[111]$ polarized moments \cite{mcclarty2018,joshi2018}. We reverse the usual logic to show how the model might have been inferred from the tabulated decomposable EBRs. Let us consider magnetic space group $F\bar{3}1m'$ ($\#162.77$ in the BNS convention) and Wyckoff position 2c corresponding to honeycomb layers. The magnetic site symmetry group is $32'$ and the moments are perpendicular to the honeycomb planes. The orbital basis on the $2c$ positions $(J_{\boldsymbol{q}}^+,J_{\boldsymbol{q}}^-)$ transforms under the $^1E+^2E$ irreps of the SSG. Consultation of tables in the Supplementary Section \cite{SM} or on the Bilbao crystallographic server \cite{bcs1,bcs2} reveals that induction to the full space group yields a single EBR that is decomposable into two bands. From symmetry alone we have therefore inferred the presence of nontrivial magnon band topology. A guide to using the Bilbao tables is given in the Supplementary Section \cite{SM}.

With this established, we may now build a model hosting the decomposable EBR and further characterize the nature of the topology. To do this, one should write down couplings between the magnetic moments that both stabilize the required magnetic structure and respect the resulting magnetic space group symmetries. Both conditions are important. For example, it is straightforward to stabilize the structure with ferromagnetic Heisenberg exchange but the resulting model has higher symmetry than $F\bar{3}1m'$ owing to a spin-space symmetry coming from the spin rotation symmetry of the underlying Hamiltonian. One may systematically compute all exchange couplings allowed by symmetry. To nearest neighbor these are the Heisenberg, Kitaev, $\Gamma$ and $\Gamma'$ terms \cite{rau2014,mcclarty2018}. Kitaev and Heisenberg are sufficient to respect $F\bar{3}1m'$ and a magnetic field may be applied along $[111]$ to stabilize the structure if necessary.  A linear spin wave calculation then reveals two propagating magnon bands with a gap between them. For decomposable EBRs the topology is not necessarily symmetry indicated but it turns out that the $C_3$ symmetry indicator formula \cite{fang2012} for the Chern number characterizes the topology in this case:
\beq
\exp\left( \frac{2\pi i}{3}C \right) = \prod_n \Theta_n(\Gamma) \Theta_n(K) \Theta_n(K')
\eeq
where the product is over $n$ bands and $\Theta_n(\boldsymbol{k})$ is the eigenvalue of $C_3$ at wavevector $\boldsymbol{k}$ in band $n$. This reveals that the model has two magnon bands with Chern numbers $\pm 1$, the order depending on the sign of the Kitaev exchange.  

\begin{figure}[tp]
  \centering
 \includegraphics[width=0.9\columnwidth]{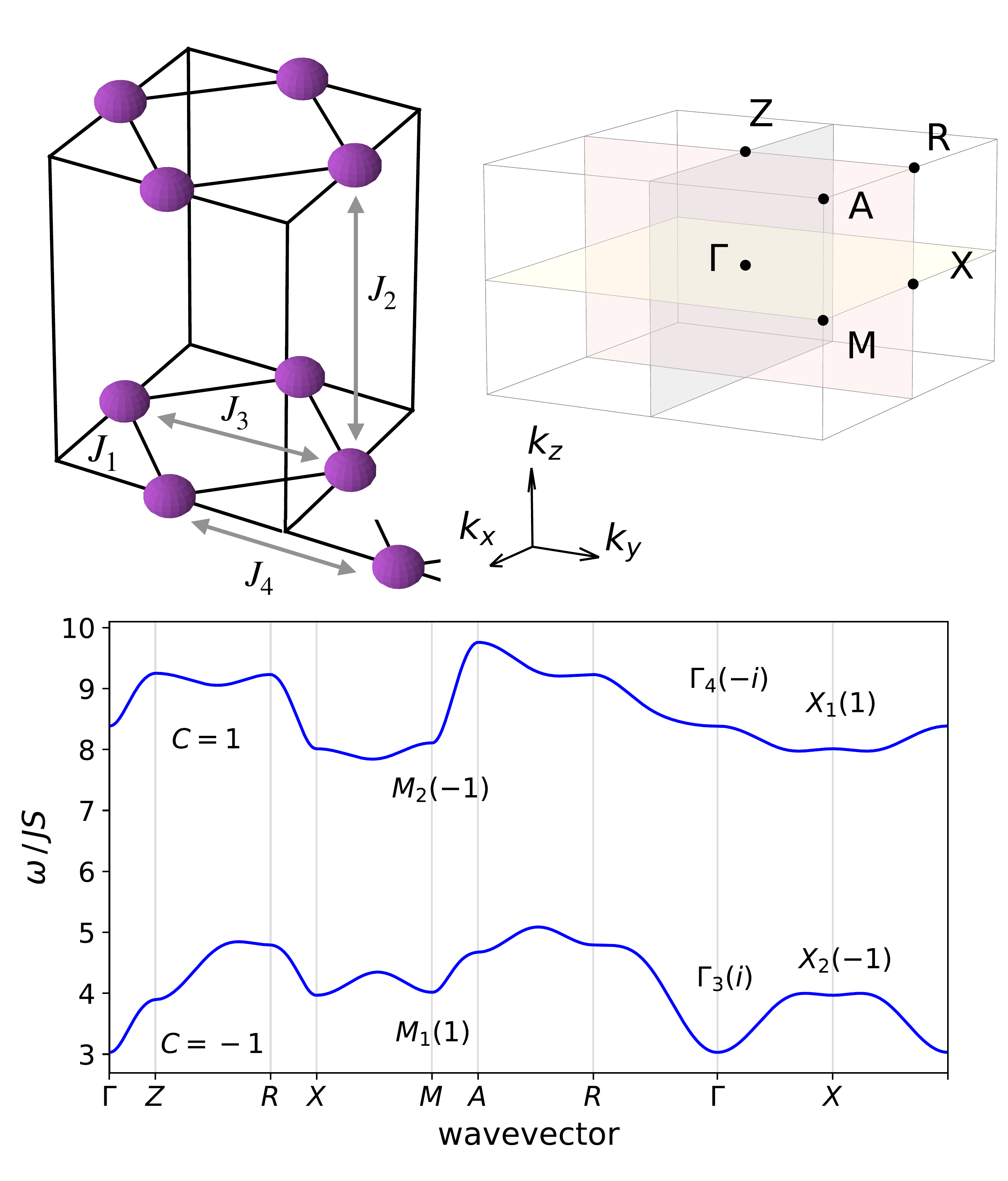}
\caption{Figure showing the magnetic sublattices of space group $75$ Wyckoff position $2c$ and the accompanying Brillouin zone. The lower panel shows magnon dispersion relations along high symmetry directions with the $C_2$ and $C_4$ eigenvalues given according to the symmetry indicator formula Eq.~\ref{eq:SIformC4}.
}
\label{fig:SG75} 
\end{figure}

We now sketch another example of gapped band topology working from the table of decomposable EBRs but this time without reference to an example already in the literature.  Consider space group $P4$ ($\# 75.1$, a type I MSG) with Wyckoff position $2c$ and irreps $2B$ for the transverse spin components. This again leads to a single decomposable EBR, now with SSG $C_2$ compatible with ferromagnetic $[001]$ magnetic order. The lattice is tetragonal with a basis $(0,1/2,0)$ and $(1/2,0,0)$. We compute all symmetry-allowed exchange couplings for first up to fourth nearest neighbors and choose some set of couplings that stabilizes the required magnetic structure. The linear spin wave spectrum has two dispersive gapped bands and the Chern number can, once again, be computed from a symmetry indicator formula
\beq
i^C = \prod_{n} \xi_n(\Gamma) \, \xi_n(M) \,\zeta_n(X)
\label{eq:SIformC4}
\eeq
where $C$ is the Chern number of the $n$ band(s), while $\xi(\bs{k})$ and $\zeta(\bs{k})$ are the eigenvalues respectively of $C_4$ and $C_2$. Fig.~\ref{fig:SG75} shows the lattice structure and the band structure with the eigenvalues indicated. The computed Chern numbers are $\pm 1$. 

The method is not restricted to diagnosing Chern bands as we show now with a third example. We take space group $P_c6/mcc$ ($\#192.252$) and Wyckoff position $4c$ which has SSG $-6m'2'$. This corresponds to an AA stacked honeycomb lattice with moments perpendicular to the plane that are ferromagnetically ordered in the plane and antiferromagnetically aligned between planes. Crucially this system is symmetric under time reversal times a translation that maps one layer to the next. The two magnon bands within each layer each carry a net Chern number which reverses between layers. One may show \cite{SM} that the coupled four magnon bands correspond to a single EBR that is decomposable. The result is an antiferromagnetic topological insulator that can be realized with an anisotropic exchange model for the in-plane moments with Heisenberg exchange between the layers. An explicit calculation of the band structure is provided for reference \cite{SM} (see also \cite{kondo2021}).

{\it Symmetry enforced nodal topology} $-$ In this part, we turn our attention to nodal topology focussing on exotic degeneracies that are enforced by symmetry: magnonic analogues of multifold fermion degeneracies \cite{bradlyn2016,canomultifold2019}. In the supplementary section we show how to use the Bilbao tables \cite{bcs1,bcs2} to establish symmetry-enforced degeneracies and give extensive tables of such degeneracies for magnons \cite{SM}. Here we show how to build models based on the symmetry information.

The first example is for magnetic space group $227.131$ $-$ a type III group $-$ and Wyckoff position $16d$ corresponding to all-in/all-out (AIAO) order on the A site of pyrochlore materials as realized in Nd$_2$M$_2$O$_7$ (M=Sn,Hf,Ir,Zr) \cite{bertin2015,anand2015,tomiyasu2012,lhotel2015,xu2015}, Sm$_2$Ir$_2$O$_7$ \cite{donnerer2016}, Eu$_2$Ir$_2$O$_7$ \cite{sagayama2013}, Cd$_2$Os$_2$O$_7$ \cite{yamaura2012} as well as FeF$_3$ \cite{reimers1992}. The magnetic structure has a magnetic $2-$fold screw and a magnetic S$_4$ symmetry. The single-valued symmetry group enforces a $3-$fold degenerate point at $\Gamma$ \cite{bcs1,bcs2}. We may establish this fact directly from a simple model for the magnons consisting of antiferromagnetic Heisenberg coupling with a weak $\langle 111\rangle$ Ising anisotropy in the exchange that lifts the considerable degeneracy of the Heisenberg model \cite{moessner1998} in favor of the AIAO structure. A linear spin wave calculation based on this model \cite{SM} reveals four dispersive modes with a spectral gap and the three-fold degenerate point at $\Gamma$. The existence of this quadratically dispersing three-fold point has previously been noted in Ref.~\cite{jian2018} as a parent state for Weyl fermions upon symmetry breaking with strain or an applied magnetic field.

\begin{figure}[tp]
  \centering
  \includegraphics[width=\columnwidth]{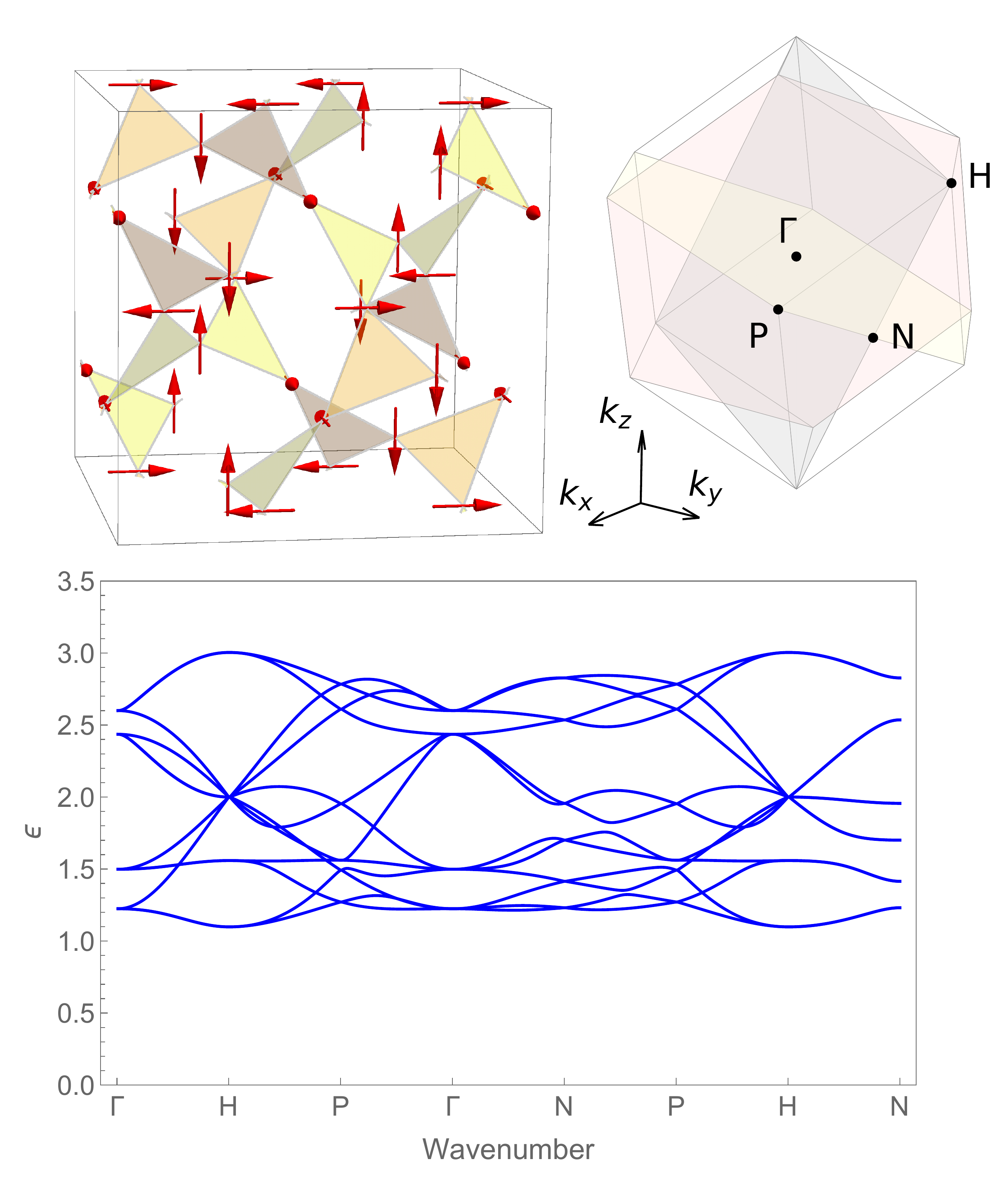}
  \caption{\label{fig:garnet}
Magnetic structure on the garnet hyperkagome lattice with $230.148$ magnetic space group symmetry (top left) and (right) the Brillouin zone with high symmetry points indicated. Bottom: spin wave spectrum with multi-fold magnons at $\Gamma$ and $H$. 
  }
\end{figure}

Our next example has both three-fold and six-fold degenerate magnons. Inspection of the table of degeneracies \cite{SM} reveals six-fold degeneracies for magnetic space group $230.148$ and Wyckoff position $24$c. The nearest neighbor exchange leads to two decoupled magnetic sublattices of corner-sharing triangles. This is the hyperkagome structure that arises on the R sites of garnets with chemical formula R$_3$M$_5$O$_{12}$. The magnetic structure compatible with $230.148$ is shown in Fig.~\ref{fig:garnet}. The moments are oriented along three cubic directions on each triangular face. This structure is observed in the material Dy$_3$M$_5$O$_{12}$ (M=Al,Ga) \cite{hastings1965,norvell1969,kibalin2020}. The $24$ Wyckoff sites are composed of $12$ magnetic sublattices plus a translation through $(1/2,/1/2,1/2)$ as the lattice is bcc. We therefore expect $12$ magnon modes. We compute the symmetry-allowed exchange couplings to nearest neighbor. There are six such couplings and one of these is an effective Ising exchange with easy axes along the cubic directions on different sublattices in the pattern required to stabilize the magnetic structure. With this as the dominant coupling, we consider a model with all six nearest neighbor couplings included and with antiferromagnet Heisenberg exchange coupling the two hyperkagome sublattices. A sample spin wave spectrum is shown in Fig.~\ref{fig:garnet}. This has several multi-fold bosonic points including four $3-$fold points at $\Gamma$ with quadratic dispersion and one $6-$fold point at $H$ on the zone boundary with linear dispersion that is a doubled spin-$1$ Weyl point. All the degeneracies in the spectrum are compatible with the group theory analysis.  

{\it Discussion} $-$ The classification of topological materials based on crystalline and time reversal symmetries is at a mature stage. In the foregoing we have connected the symmetry-based classification scheme based on elementary band representations to topological magnons. To do this, we showed how symmetries are inherited by magnons from those of the underlying exchange Hamiltonian and indicated how to build band representations for magnons. We have given conditions for the existing tables of EBRs to be applicable to topological magnons. We have shown through several examples that one can use the computed decomposable elementary band representations for single-valued magnetic space groups to build realistic, non-fine-tuned models of topological magnon band structures. We have also used tabulated symmetry-enforced degeneracies as a guide to building exchange models of exotic nodal topology such as six-fold degenerate touching points. Magnons provide an excellent platform to explore the interplay of magnetic symmetries and topology in conjunction with inelastic neutron scattering. In addition to model-building and experimental discovery within the framework laid out here, important open avenues are to explore magnon topology beyond the decomposable EBR paradigm within the TQC framework and to extend TQC to the spin-space groups that are applicable to Heisenberg models among other systems.

\begin{acknowledgments}
PM acknowledges useful discussions with Alexei Andreanov on magnetism in the garnets. This work was in part supported by the Deutsche Forschungsgemeinschaft  under grants SFB 1143 (project-id 247310070) and the cluster of excellence ct.qmat (EXC 2147, project-id 390858490).
\end{acknowledgments}

\bibliography{ebrmagnonrefs}
\bibliographystyle{apsrev4-2}

\end{document}


\title{Supplementary Material for Identifying, and Constructing, Complex Magnon Band Topology}

\author{A.~Corticelli}
\affiliation{Max Planck Institute for the Physics of Complex Systems, N\"{o}thnitzer Str. 38, 01187 Dresden, Germany}
\author{R. Moessner}
\affiliation{Max Planck Institute for the Physics of Complex Systems, N\"{o}thnitzer Str. 38, 01187 Dresden, Germany}
\author{P.~A.~McClarty}
\affiliation{Max Planck Institute for the Physics of Complex Systems, N\"{o}thnitzer Str. 38, 01187 Dresden, Germany}

\begin{abstract}
This section contains supporting information for the paper ``Identifying, and constructing, complex magnon band topology". Section~\ref{sec:EBRTopology} briefly introduces magnetic space groups and their band representations. EBRs are introduced and their role in accounting for topological bands. Section~\ref{sec:Magnons} discusses band representations in relation to magnons, enumerates all relevant site symmetry groups, reviews linear spin wave theory, the Berry phase for bosons and the implementation of magnetic symmetries within this formalism. Subsection D also describes how to use the Bilbao tables to extract information about EBRs and nodal topology. Section~\ref{sec:EBRExamples} describes, in detail, several examples of decomposable EBRs for magnons. Finally, Section~\ref{sec:nodaltopology} discusses nodal topology originating from EBRs. 
\end{abstract}

\maketitle

\section{Overview of Elementary Band Representations and Topology}
\label{sec:EBRTopology}

We have described briefly the essential ideas behind the EBR approach to topological band structures. We now make these ideas more precise by first reviewing aspects of the theory of space groups and band representations of these groups. 

\subsection{Basic definitions and properties of space groups}

A space group $G$ is a group of crystal lattice symmetries. There are $230$ such groups in three dimensions that each have a coset decomposition
\beq
\mathbf{G}=\bigcup_{\alpha}\element{R_\alpha}{\bs{t}_{\alpha}}\mathbf{T}
\eeq
where $\mathbf{T}$ are the primitive lattice translations (forming a normal subgroup) and those of the form $\element{g}{\bs{t}}$ combining point group element $g$ and non-Bravais translation $\bs{t}$. The combination rules are
\begin{align}
& \element{R_1}{\bs{t}_1}\element{R_2}{\bs{t}_2} = \element{R_1 R_2}{R_1 \bs{t}_2 + \bs{t}_1 }  \\
& \element{R}{\bs{t}}^{-1} = \element{R^{-1}}{- R^{-1}\bs{t}}.
\label{eq:sgcomb}
\end{align}

A site symmetry group of real space point $\bs{q}$, $G_{\bs{q}}$ is the finite subgroup of $G$ that leaves the point invariant. $G_{\bs{q}}$ is isomorphic to a point group. A Wyckoff position is the set of points inside the primitive cell whose site symmetry groups are the same (or, more precisely, in the same conjugacy class). A random point will tend to have only the identity as its site symmetry group and it is then labelled as a general position. 

Each point $\bs{q}$ has an orbit which is the set of points reached from $\bs{q}$ through elements $g$ of the space group. Each Wyckoff position has a multiplicity that counts the number of points in the orbit of the position that live in the same cell. 

The above definitions refer to the crystal in real space. But for constructing representations of the space group it is better to move to reciprocal space where the translations are diagonalized. The analogue of a site symmetry group in reciprocal space is the little group which is the set of space group elements that leave momentum $\bs{k}$ invariant up to reciprocal lattice vectors. This group contains all translations and it is often useful to mod them out to obtain the so-called little co-group of $\bs{k}$. The analogue of a Wyckoff position in reciprocal space is the star of $\bs{k}$, denoted $\bs{k}^*$ which is the set of points in the Brillouin zone reachable by acting with space group elements on $\bs{k}$.  

Including magnetic degrees of freedom and time reversal symmetry expands the number of groups from $230$ to $1651$. The time reversal operator is denoted $\mathcal{T}$ or by apostrophe as $\element{E}{\bs{0}}'$ . Of the $1651$ magnetic space groups, there are four types:
\begin{enumerate}
    \item Type I: the ordinary nonmagnetic space groups $G$,
    \item Type II: grey groups of the form $G+\mathcal{T}G$ which are relevant to paramagnets,
    \item Type III: groups of the form $H+\mathcal{T}(G-H)$ where $G$ and $H$ are non-magnetic,
    \item Type IV: black and white groups $G+\mathcal{T}\element{E}{\bs{t}} H$ composed of time-reversed sublattices.
\end{enumerate}
Magnetically ordered crystals have symmetry groups belonging to classes $I$, $III$ and $IV$.  

\subsection{Space group band representations}


Following Zak, we construct band representations of a space group $\rho_G(g)$ where $g$ is an arbitrary space group element. These are induced from a representation of a subgroup $-$ the site symmetry group for a given position. Physically, we are interested in the case where a lattice is populated with ions carrying some localized orbital degrees of freedom and tight-binding models derived from these.
To induce a representation we need a coset decomposition of the space group 
\beq
G = \bigcup_\alpha g_{\alpha}H
\eeq
where $H$ is a subgroup of $G$ $-$ in this case the site symmetry group at $\bs{q}$. The elements $g_\alpha$ have the effect of mapping $\bs{q}$ to $\bs{q}_\alpha$ which are all sites in the Wyckoff position of $\bs{q}$. 

To see what the band representation looks like, suppose we are given a complete basis of Wannier functions $W_{i\alpha}(\bs{r})$ where $i=1,\ldots,d$ runs over onsite degrees of freedom $-$ essentially a set of orbitals, and $\alpha$ runs over different positions in the primitive for a given Wyckoff position ($\alpha$ therefore runs over $1$ to $n$, the multiplicity of that position). Translations map the Wannier functions to different primitive cells ($\mathcal{N}$ in number) giving $d\times n \times \mathcal{N}$ distinct Wannier functions. 

Going to momentum space
\beq
v_{i\alpha}(\bs{k},\bs{r}) = \sum_{\bs{t}} e^{i\bs{k}\cdot\bs{t}} W_{i\alpha}(\bs{r}-\bs{t})
\eeq
provides a basis of states on which the band representation can act. 
\beq
\left[ \rho_G(g) \right] v_{i\alpha}(\bs{k},\bs{r}) = e^{-i (R\bs{k})\cdot \bs{t}_{\beta\alpha}} \sum_{i=1}^d \left[\rho(h) \right]_{ji} v_{j\beta}(R\bs{k},\bs{r}).
\eeq
In this expression space group element $g$ and site symmetry group element $h$ are related by
\beq
g = g_{\alpha}^{-1} \element{E}{\bs{t}_{\beta\alpha}} g_{\beta} h
\eeq
where $\bs{t}_{\beta\alpha}=g \bs{q}_\alpha - \bs{q}_\beta$ and $R$ is the point group element in $g$. So the $\alpha$ and $\beta$ coset representatives are fixed given $g$ and $h$. 

Evidently the band representation links $\bs{k}$ and $R\bs{k}$. In the case where $R\bs{k}$ is the same as $\bs{k}$ up to a reciprocal lattice vector, the corresponding block in the band representation is a representation of the little group at $\bs{k}$. However, the band representation has off-diagonal blocks that contain information about how different points in the zone are connected. 

\subsection{Elementary Band Representations}

A band representation constructed via the method detailed in the previous section may be decomposable into the direct sum of two or more band representations. If this is the case it is called {\it composite} and otherwise {\it elementary}. More precisely, we first define an equivalence between two band representations $\rho^{(1)}_G(h,\bs{k})$, $\rho^{(2)}_G$ if it is possible to find unitary $S(\lambda; h,\bs{k})$ such that $\lambda:[0,1]$ tunes smoothly from one of the two band representations to the other: $S(0; h,\bs{k})=\rho^{(1)}_G(h,\bs{k})$ and  $S(2; h,\bs{k})=\rho^{(2)}_G(h,\bs{k})$. Such a function preserves the quantization of any Wilson loops in momentum space. This notion of equivalence is explicitly realized by inducing a BR from two distinct sites $\bs{q}_1$, $\bs{q}_2$ with respective site symmetry groups $G_{\bs{q}_1}$ and  $G_{\bs{q}_2}$. A line between the two points is associated with SSG $G_{\bs{q}_1}\cap G_{\bs{q}_2}$ and by moving along this line the induced band representation defines $S(\lambda; h,\bs{k})$. It follows that equivalence of band representations amounts to being able to find a site that interpolates between the SSGs of the endpoint BRs. With this notion of equivalence, we now define composite BRs to be those that are equivalent to direct sums of BRs. An EBR can be characterized by the multiplicity of irreps at all high symmetry momenta.

Elementary band representations (EBRs), thus defined, are the fundamental symmetry-derived bands built from localized orbitals. In contrast, as we noted in the main text, the key distinguishing feature of topological bands is that they are not Wannier localizable. 

The foundation of TQC is a complete enumeration of the EBRs for all $1651$ magnetic space groups together with the compatibility relations that constrain how little groups at particular momenta are connected. This task, while considerable, is possible at all because the number of EBRs is finite, bounded by the number of irreps of SSGs at all Wyckoff positions of all magnetic space groups. In addition, many BRs induced in this way are actually composite. It turns out that to capture all EBRs it suffices (modulo some carefully characterized exceptions) to consider only the irreps of so-called {\it maximal} SSGs. Maximal SSGs are defined as SSGs $G_{\rm q}$ such that there is no finite group $H$ for space group $G$ for which $G_{\rm q}\subset H \subset G$. Given each EBR, one may further ask whether it is decomposable or not by computing the compatibility relations for the constituent bands. 

The result is that there are $20206$ magnetic EBRs belonging to the $1651$ (single-valued or spinless) magnetic space groups of which $1907$ are decomposable. For our purposes, these are the relevant magnetic space groups. A similar enumeration has been carried out also for the doubled (or spinful) magnetic space groups. A complete tabulation of these EBRs organized by magnetic space group may be found on the Bilbao Crystallographic Server \cite{bcs1, bcs2}. 

Given a set of bands that are energetically isolated, one may then assess whether they can be decomposed, on their own, into a combination of EBRs (with non-negative integer coefficients). If so, the bands are topological trivial (or fragile). If not they are topologically non-trivial. In this situation, should the coefficients be integer-valued including negative integers the topology is fragile and otherwise it is stable. 

In addition, single EBRs may be composed of multiple bands that are not forced to be connected by compatibility relations. In the main text we assign particular importance to such cases. These {\it decomposable} EBRs have the property that at least one of the disconnected component set of bands must be topological. In cases where one component is trivial the decomposable EBR is a self-contained case of fragile topology. In the examples we have explored, the two disconnected components are both topological. 

EBR are also useful for assessing the existence of topological semimetals. These arise from connected EBRs where, in electronic systems, the bands are filled up to touching points or lines within the EBR.

All these insights have been put to use diagnosing band topology in the electronic band structures of materials. Given the symmetry group of a crystalline material and the Wyckoff positions and orbitals of the constituent ions, symmetry places strong constraints on the EBRs that may occur in the band structure. From the computed band structure (usually performed along high symmetry directions in momentum space), one may compute the multiplicities of the irreps at these momenta. From the identities of the tabulated EBRs one may then make the assessment of whether some given set of bands is reducible into EBRs. 

This approach massively generalizes the Fu-Kane criterion for two-dimensional topological insulators that, in the original formulation, allows one to compute the Chern number purely from discrete data at high symmetry momenta. By now, analogous formulas called symmetry indicator formulas are known for all space groups and all (double-valued) magnetic space groups each of which allows one to diagnose directly from irrep multiplicities whether the band or group of bands is trivial or not.


\section{Magnetic Symmetry and Magnons}
\label{sec:Magnons}

In this section, we briefly review the essential facts about magnons and their symmetries. Magnons are to be understood as coherent magnetic excitations about some spontaneous or field-induced magnetic structure. We focus our attention on commensurate magnetic order characterized by some periodic arrangement of moments with nonzero vacuum expectation value $\langle J_i^\alpha \rangle$ for sites $i$ and components $\alpha$.

To understand magnon symmetries it is helpful to begin with the magnetic Hamiltonian describing coupled magnetic moments on a lattice. The lattice itself has symmetries specified by one of the $230$ space groups. The Hamiltonian may have higher symmetry however: spin rotation symmetry for Heisenberg couplings or time reversal symmetry when the couplings are of even degree in the moments. In the most general case, the magnetic Hamiltonian has a spin-space symmetry composed of elements with somewhat decoupled spin rotation symmetries. However, in this paper, we restrict our attention to the case of strong spin-orbit coupled moments so that the moments are locked to spatial transformations. Under this assumption, the magnetic space groups are adequate to describe all the relevant symmetries. An important implication of this assumption is that we are explicitly or implicitly considering the case where the magnetic interactions of all types allowed by spin-space-locked symmetry are present and significant. In other words, the exchange is considered to be maximally anisotropic. 

From the point of view of materials, the restriction to magnetic space groups is strictly speaking correct as spin-orbit coupling is always present even when the orbital moment is quenched at the single ion level. Anisotropies in the exchange will be present even in such instances. However, for practical purposes, this assumption is too severe as there are many materials where the interactions are experimentally indistinguishable from the Heisenberg limit or where the spin-orbit is weak enough that a residual spin-space symmetry remains. Such cases are discussed in greater detail in Ref.~\onlinecite{corticelli2021spinspace}.

For such cases, we stress that the techniques we employ can be used straightforwardly to study spin-space groups also. But since they have not yet been tabulated we leave a systematic study of their topological properties as a task for the future. 

As discussed above, magnetic space groups can be classified into four different types. The magnetic space group symmetries of the magnetic Hamiltonian fall into classes I or II. Respectively these are the ordinary space groups (I) and groups of the form $G+\mathcal{T} G$ where $\mathcal{T}$ is the time reversal operation (II). 

The interplay between spatial and time reversal symmetries can become more elaborate in the presence of magnetic order. Magnetic order breaks the symmetry group of the Hamiltonian $G_H$ down to $G_M$ which is the set of symmetry elements $-$ combinations of point group elements (e.g. rotations, mirrors), translations and time reversal $-$ that leave the magnetic structure, $\langle J_i^z \rangle$, invariant where the $z$ component refers to a local frame aligned with the ordered moment. Since the magnetic order breaks physical time reversal symmetry spontaneously if it is not explicitly broken at the Hamiltonian level, $G_M$ cannot be a type II magnetic space group. This leaves the $1421$ type I, III and IV magnetic space groups as possible magnon symmetry groups.

To build a band representation we require the magnetic site symmetry groups. These can be viewed as the set of elements of $G_M$ that both leave the site invariant up to a primitive translation and leave the magnetic order invariant. Thus, given $\bs{q}_1$ an orbit of the Wyckoff position, applying the elements of magnetic site symmetry group $G_{\bs{q}_1}$ the condition that the magnetic order by left invariant is:
\beq
g_{\alpha}J^{z}_{\bs{q}_1} = J^{z}_{\bs{q}_1} ~~~~~ \forall g_{\alpha} \in G_{\bs{q}_1}.
\eeq
Expressed another way, $J^{z}$ must transform like the total symmetric irrep of $G_{\bs{q}_1}$. Those Wyckoff positions that do not satisfy this constraint are not compatible with order and must reduce to Wyckoff positions of a less symmetric magnetic space group.
Using this constraint we recover the magnetic structures compatible with the magnetic space group by listing the Wyckoff positions compatible with order and applying:
\beq
g_{i}J^{z}_{\bs{q}_1}: R_i^{z\alpha} J^{\alpha}_{\bs{q}_i} ~~~~~\forall g_{i} ~:~ G = \bigcup_i g_{i} (G_{\bs{q}_i} \ltimes T)
\eeq
where $\bs{q}_i$ is the orbit of the Wyckoff position relative to $g_{i}$ and $R_i^{z\alpha}$ is the rotation matrix associated to $g_{i}$.

The site symmetry group of the magnetic ions must be isomorphic to one of the $122$ magnetic point groups. These are divided into the $32$ crystallographic point groups, the $32$ grey point groups with pure time-reversal operators and $58$ black and white point groups with mixed time-reversal elements. The constraint that the $J_z$ component transform as the totally symmetric representation of the site symmetry group restricts the possibilities to a subset of the magnetic point groups.
Of the $32$ crystallographic point groups, $13$ preserve the moment, of the $58$ black and white point groups there are $18$ moment preserving groups. None of the grey groups are possible since time-reversal does not preserve the magnetic order. In total, $31$ out of the $122$ magnetic point groups are possible magnetic site symmetry groups. These are listed in Table~\ref{tab:axialgroup}.

\begin{table*}[!htb]
\centering
\begin{tabular}{ |c|c| }
 \hline
 \multicolumn{1}{|c|}{Magnetic Site Symmetry Group}  & $(J^+, J^-)$ irreps   \\ 
 \hline
$1$,~ $2'$,~ $m'$      &  $2A$   \\
$-1$,~ $2'/m'$     &  $2A_g$ \\
$2$,~ $2'2'2$,~ $m'm'2$      & $2B$ \\
$m$,~ $m'm2'$       &  $2A''$ \\
$2/m$,~ $m'm'm$      &  $2B_g$ \\
$4$,~ $42'2'$,~ $4m'm'$  &  $^1E + {^2E}$ \\
$-4$, ~ $-42'm'$     &  $^1E + {^2E}$ \\
$4/m$,~ $4/mm'm'$     &  $^1E_g + {^2E_g}$ \\
$3$,~ $32'$,~ $3m'$     &  $^1E + {^2E}$ \\
$-3$,~ $-3m'$     &  $^1E_g + {^2E_g}$ \\
$6$,~ $62'2'$,~ $6m'm'$ &  $^1E_2 + {^2E_2}$ \\
$-6$,~ $-6m'2'$     &  $^1E'' + {^2E''}$ \\
$6/m$,~ $6/mm'm'$     &  $^1E_{2g} + {^2E_{2g}}$ \\
\hline
\end{tabular}
\caption{
The $31$ site symmetry groups compatible with magnetic ordered systems. The groups with same unitary subgroup are list in the same line and the relative unitary irreps of the transverse spin components for spin wave are listed in the right column.
The notation of the irreps follows Ref.~\cite{bradley2009mathematical}.
}
\label{tab:axialgroup}
\end{table*}

We are now in a position to discuss magnons. These are transverse modes built from the $J^{\pm}_i$ components. In order to build up band representations for magnons, the starting point is the set of site symmetry groups for which $J_z$ transforms as the totally symmetric representation listed in Table~\ref{tab:axialgroup}. Given these groups, we may establish how the transverse spin components transform and this information completely fixes the allowed representations of the SSG for the purposes of building the band representation. These SSG representations are also given in Table~\ref{tab:axialgroup}.

From the table we see that $J^{\pm}_i$ will, in general, induce a pair of EBRs, which are the same if real or complex conjugates of one other if complex. This fact reflects distinctive paraunitarity of the bosonic Bogoliubov diagonalization that produces two set of bands at positive and negative energies with complex conjugated eigenvector thus redundantly encoding information about the band structure. 
It is important to note that the Herring criterion forbids these complex conjugated EBRs from pairing within a single EBR when the anti-unitary elements are considered. This is a result of the magnetic order constraint preserving $J^z$, which translate in the prohibition to mix of $J^+$ and $J^-$.

What distinguishes band representations induced from these SSG representations? 
We have seen that the interesting band representations are the EBRs obtained from the maximal Wyckoff positions since all the other BRs can be seen as composite of these.
Therefore the EBRs can be divided into two groups for magnons. One group is induced from maximal Wyckoff positions compatible with magnetic order $-$ they can be induced directly starting from the orbitals $J^{\pm}_{\bs{q}}$ on the maximal Wyckoff positions ${\bs{q}}$ themselves. 
The second group has maximal Wyckoff positions that are not compatible with magnetic order or comes from representations describing different orbitals than $J^{\pm}_{\bs{q}}$. These can only be induced as a part of a composite representation from orbitals $J^{\pm}_{\bs{q}_L}$ from a less symmetric Wyckoff position ${\bs{q}_L}$.
In practice, the first group allows one to construct magnon spectra composed of a single EBR, with straightforward topological identification once a gap is present. All the {\it decomposable} EBRs which can appear among this kind of single EBRs in magnon band structures are listed in Tab.~\ref{tab:EBRsplitted}.

\subsection{Linear Spin Wave Theory}

So far we have discussed the transformation properties of magnons at a relatively abstract level. To connect to the magnon band structures of materials we use the standard Holstein-Primakoff identity 
\begin{align}
\hat{J}^{z} & = S - \hat{a}^\dagger \hat{a} \\
\hat{J}^{+} & = \sqrt{2S} \sqrt{1 - \frac{\hat{a}^\dagger \hat{a}}{2S}} \hat{a} = \sqrt{2S} \left( 1- \frac{\hat{a}^\dagger \hat{a}}{4S} \right) \hat{a} + \ldots  \\
\hat{J}^{-} & = \sqrt{2S} \hat{a}^\dagger \sqrt{1 - \frac{\hat{a}^\dagger \hat{a}}{2S}} = \sqrt{2S} \hat{a}^\dagger \left( 1- \frac{\hat{a}^\dagger \hat{a}}{4S} \right) + \ldots
\end{align}
where the spins are of length $S$ and the bosons $a$, $a^\dagger$ satisfy the usual commutation relations $[\hat{a},\hat{a}^\dagger]=1$. Linearizing these gives
\begin{align}
& \hat{J}^{+}_{\bs{k}a} \rightarrow \sqrt{2S} \hat{a}_{\bs{k}a} \\
& \hat{J}^{-}_{\bs{k}a} \rightarrow \sqrt{2S} \hat{a}^\dagger_{-\bs{k}a}.
\label{eq:LSWtranverse}
\end{align}

Using this bosonic representation for the spins and expanding the magnetic Hamiltonian around the mean field ground state leads to the quadratic Hamiltonian
\begin{widetext}
\beq
\mathsf{H}_{\rm SW} = \frac{S}{2} \sum_{\boldsymbol{k}} \hat{\boldsymbol{\Upsilon}}^{\dagger}(\boldsymbol{k}) \left( \begin{array}{cc} \boldsymbol{\mathsf{A}}(\boldsymbol{k}) &  \boldsymbol{\mathsf{B}}(\boldsymbol{k}) \\  \boldsymbol{\mathsf{B}}^{\star}(-\boldsymbol{k}) &  \boldsymbol{\mathsf{A}}^{\star}(-\boldsymbol{k})  \end{array} \right) \hat{\boldsymbol{\Upsilon}}(\boldsymbol{k}) \equiv   \frac{S}{2} \sum_{\boldsymbol{k}}  \hat{\boldsymbol{\Upsilon}}^{\dagger}(\boldsymbol{k}) \boldsymbol{M}(\boldsymbol{k}) \hat{\boldsymbol{\Upsilon}}(\boldsymbol{k}) 
\label{eq:SWH}
\eeq
\end{widetext}
where 
\beq
\arraycolsep=0.8pt\def\arraystretch{1.0}
\hat{\boldsymbol{\Upsilon}}^{\dagger}(\boldsymbol{k}) = \left( \begin{array}{cccccc} \hat{a}^{\dagger}_{\boldsymbol{k}1} & \ldots & \hat{a}^{\dagger}_{\boldsymbol{k}m} & \hat{a}_{-\boldsymbol{k}1} & \ldots & \hat{a}_{-\boldsymbol{k}m} \end{array}   \right) \hspace{0.1cm}
\eeq
and the $\mathsf{A}_{ab}(\boldsymbol{k})$ and $\mathsf{B}_{ab}(\boldsymbol{k})$ depend on the exchange couplings in the local quantization frame as follows:
\begin{align}
\mathsf{A}_{ab}(\boldsymbol{k}) & =\tilde{\mathsf{J}}_{ab}^{+-}(\boldsymbol{k}) - \delta_{ab} \sum_{c} \tilde{\mathsf{J}}_{ac}^{zz}(\boldsymbol{0}) \label{eq:Ablock} \\
\mathsf{B}_{ab}(\boldsymbol{k}) & = \frac{1}{2}\left( \tilde{\mathsf{J}}_{ab}^{xx}(\boldsymbol{k}) - \tilde{\mathsf{J}}_{ab}^{yy}(\boldsymbol{k}) -i \tilde{\mathsf{J}}_{ab}^{xy}(\boldsymbol{k}) - i\tilde{\mathsf{J}}_{ab}^{yx}(\boldsymbol{k})   \right) \nonumber \\
& =  \tilde{\mathsf{J}}_{ab}^{--}(\boldsymbol{k}).
\label{eq:Bblock}
\end{align}
Note that these expressions with the factor one-half define $\tilde{\mathsf{J}}_{ab}^{\alpha\beta}$ for $\alpha, \beta = \pm$.

The diagonalizing transformation 
\beq
\boldsymbol{V}^\dagger (\boldsymbol{k})  \boldsymbol{M}(\boldsymbol{k}) \boldsymbol{V}(\boldsymbol{k}) = \boldsymbol{\Lambda}(\boldsymbol{k})
\eeq
on Eq.~\ref{eq:SWH} to find the spin wave spectrum must preserve the commutation relations 
\beq
\left[ \Upsilon_a , \Upsilon_b^\dagger \right] = \eta_{ab}
\eeq
where $\eta_{ab}=1$ if $a=b\leq m$ and $\eta_{ab}=-1$ if $a=b\geq m+1$ and zero otherwise. Here 
\beq
\boldsymbol{\Lambda}(\boldsymbol{k}) = \left( \begin{array}{cccccc} \epsilon_1(\boldsymbol{k}) & \epsilon_2(\boldsymbol{k}) & \ldots \epsilon_1(-\boldsymbol{k}) & \epsilon_2(-\boldsymbol{k}) & \ldots \end{array} \right)
\eeq
In order for both conditions to be satisfied $\boldsymbol{V}$ is not unitary in general, as would be the case for fermions, but {\it paraunitary} meaning that
\begin{align*}
\boldsymbol{V}(\boldsymbol{k}) \boldsymbol{\eta}\boldsymbol{V}^\dagger (\boldsymbol{k}) = \boldsymbol{\eta}.
\end{align*}
The transformation is unitary only in the case where the number non-conserving terms in the Hamiltonian vanish. 

\subsection{Berry Phase and Berry Curvature}

The Berry phase is of central importance to band topology. For bosonic systems the Berry phase for band $n$ is 
\beq
A_\mu^{(n)}(\boldsymbol{k}) = i {\rm Tr}\left[ \bs{\Gamma}^{(n)} \bs{\eta} \boldsymbol{V}^\dagger(\boldsymbol{k}) \bs{\eta} \partial_{k_\mu} \boldsymbol{V}(\boldsymbol{k}) \right]. 
\eeq
where 
\beq
\bs{P}_\mu = \boldsymbol{V}(\boldsymbol{k}) \bs{\Gamma}^{(n)} \bs{\eta}\boldsymbol{V}^\dagger(\boldsymbol{k}) \bs{\eta}
\eeq
is the projector onto the $n$th band. 

From this one may compute the Berry curvature for the $n$th band
\beq
F_\mu^{(n)}(\boldsymbol{k}) = \epsilon_{\mu\rho\sigma}  \partial_{k_\rho}A_\sigma^{(n)}(\boldsymbol{k}).
\eeq
The integral of this curvature over a 2D slice through the Brillouin zone is quantized and deformable only by closing a band gap.
\beq
C^{(n)} = \frac{1}{2\pi} \int_{2D BZ} d^2\bs{k} F^{(n)}(\boldsymbol{k})
\eeq

\subsection{Symmetries}

In this section, we show how to build a representation of the group elements for magnons. The relevant group is composed of those elements of the full magnetic Hamiltonian that leave the magnetic structure invariant, $\mathbf{G}_M$. The spin wave Hamiltonian is expressed in the basis of transverse spin components and is invariant under this group. Thus in the local quantization frame,
\begin{align}
& \hat{J}^{+}_{\bs{k}a} \rightarrow \sqrt{2S} \hat{a}_{\bs{k}a} \\
& \hat{J}^{-}_{\bs{k}a} \rightarrow \sqrt{2S} \hat{a}^\dagger_{-\bs{k}a}.
\label{eq:transversetoHP}
\end{align}
Leaving time reversal aside for now, a group element takes the form $S=\element{g}{\bs{t}}$ acting on a lattice site $\bs{R}_i+\bs{r}_a$ as
\beq
\bs{R}_i+\bs{r}_a \rightarrow g(\bs{R}_i+\bs{r}_a) + \bs{t} = \bs{R}_j + \bs{r}_b.
\eeq
Under this transformation, local moments are mapped from one lattice position to another preserving the moment orientation and, in general, rotating the transverse components. Under a $C_n$ rotation of the transverse components about the moment orientation
\beq
C_n \hat{J}^{\pm}_{ia}C_n^{-1} = e^{-2\pi i/n} \hat{J}^{\pm}_{ia}.
\eeq
and we further note that inversion leaves moments invariant as they are pseudo-vectors and that reflections are equivalent to an inversion times a $C_2$ with axis perpendicular to the mirror plane. 

The space group element $S$ acting on $\hat{J}^{\pm}_{ia}$ gives
\beq
U_S \hat{J}^{\pm}_{ia} U_S^{-1} = \left[ U^{\pm}_g \right]_{ab} \hat{J}^{\pm}_{jb} 
\eeq
where $\left[ U^{\pm}_g \right]_{ab}$ permutes sublattices and carries out local rotations. We then Fourier transform as
\beq
J_{ia}^\alpha = \frac{1}{\sqrt{N}} \sum_{\bs{k}} J_{\bs{k}a}^\alpha e^{i\bs{k}\cdot \left(\bs{R}_i+ \nu \bs{r}_a \right)}
\eeq
where $\nu=0,1$ keeps track of both standard conventions. A short calculation reveals that
\beq
U_S \hat{J}^{\pm}_{\bs{k}a} U_S^{-1} = \left[ U^{\pm}_g \right]_{ab} \hat{J}^{\pm}_{g\bs{k}b}  \exp\left[ -i (g\bs{k})\cdot \left( (1-\nu)(\bs{r}_b - g\bs{r}_a) - \bs{t} \right) \right].
\eeq
On the basis of this transformation law and the invariance of $H_{\rm SW}$, one may show that
\beq
M_{ab}(\bs{k}) = \left[ U^{\pm}_g \right]^\dagger_{am}M_{mn}(g^{-1}\bs{k})\left[ U^{\pm}_g \right]_{nb}. 
\eeq

Including time reversal symmetries as follows 
\begin{align}
& \hat{\mathcal{T}}i\hat{\mathcal{T}}^{-1} = -i \\
& \hat{\mathcal{T}}\hat{J}^{\pm}_{ia}\hat{\mathcal{T}}^{-1} = -\hat{J}^{\mp}_{ia} \\
& \hat{\mathcal{T}}\hat{J}^{\pm}_{\bs{k}a} \hat{\mathcal{T}}^{-1} = -\hat{J}^{\mp}_{-\bs{k}a}
\label{eq:TR}
\end{align}
completes the set of transformations on transverse components of the magnetic moment allowing one to construct a representation on the basis of $a_{\bs{k}a}$ and $a^\dagger_{-\bs{k}a}$.

\subsection{Use of Bilbao Crystallographic Server MBANDREP tool for magnons}

\begin{figure}[!htb]
\hbox to \linewidth{ \hss
\includegraphics[width=0.99\linewidth, trim=0.2cm 2.2cm 0.2cm 0.2cm, clip]{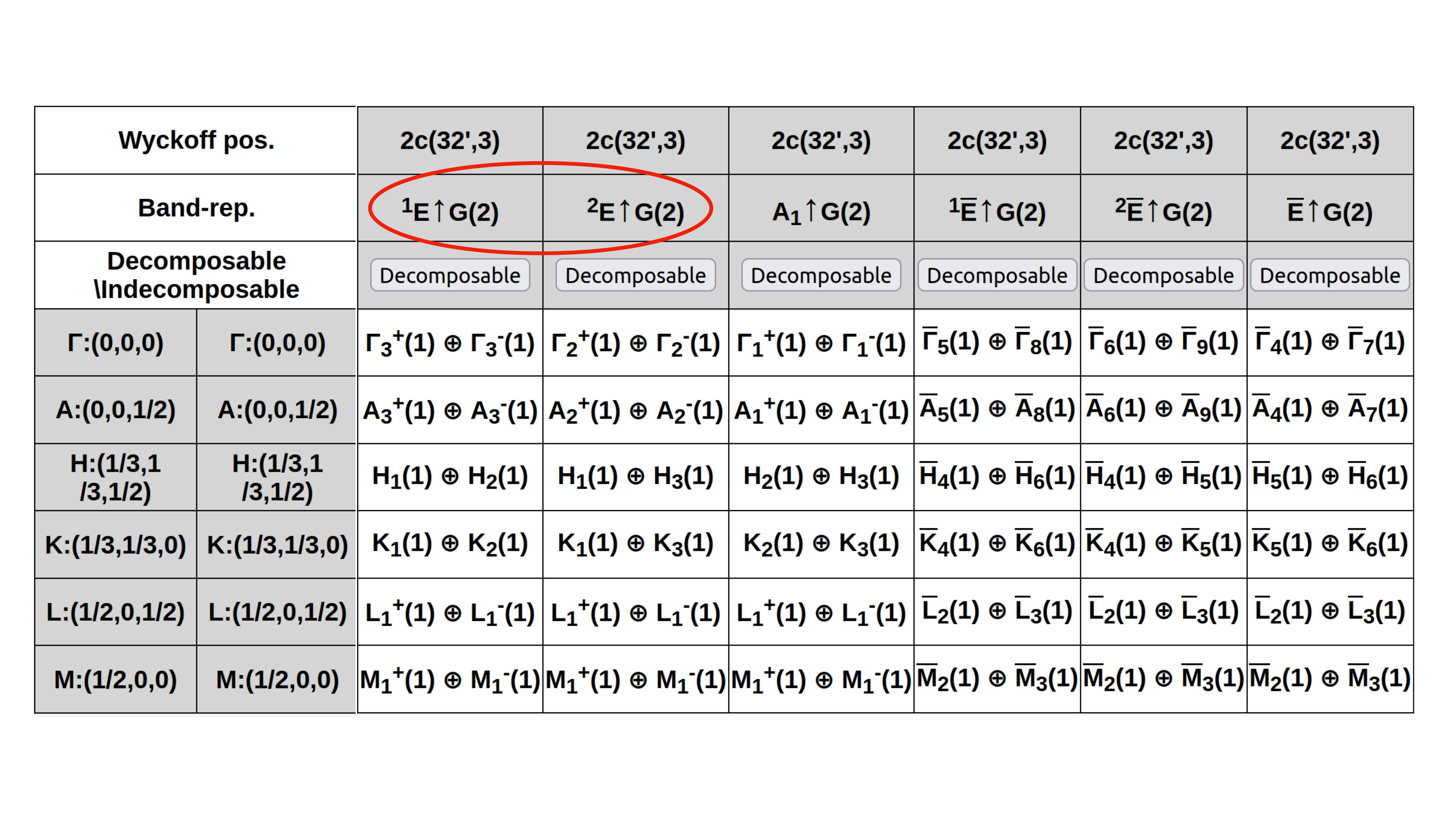}
\hss}
\caption{
Screenshot from MBANDREP tool on Bilbao Crystallographic Server for the magnetic space group $P\bar{3}1m'$ ($\# 162.77$) and maximal Wyckoff position $2c$. In the red circle the relevant band representations for magnons. The use of the table is discussed in the main text.
}
\label{Fig:BANDREP_1}
\end{figure}

\begin{figure}[!htb]
\hbox to \linewidth{ \hss
\includegraphics[width=0.6\linewidth, trim=0.2cm 1.2cm 0.2cm 0.2cm, clip]{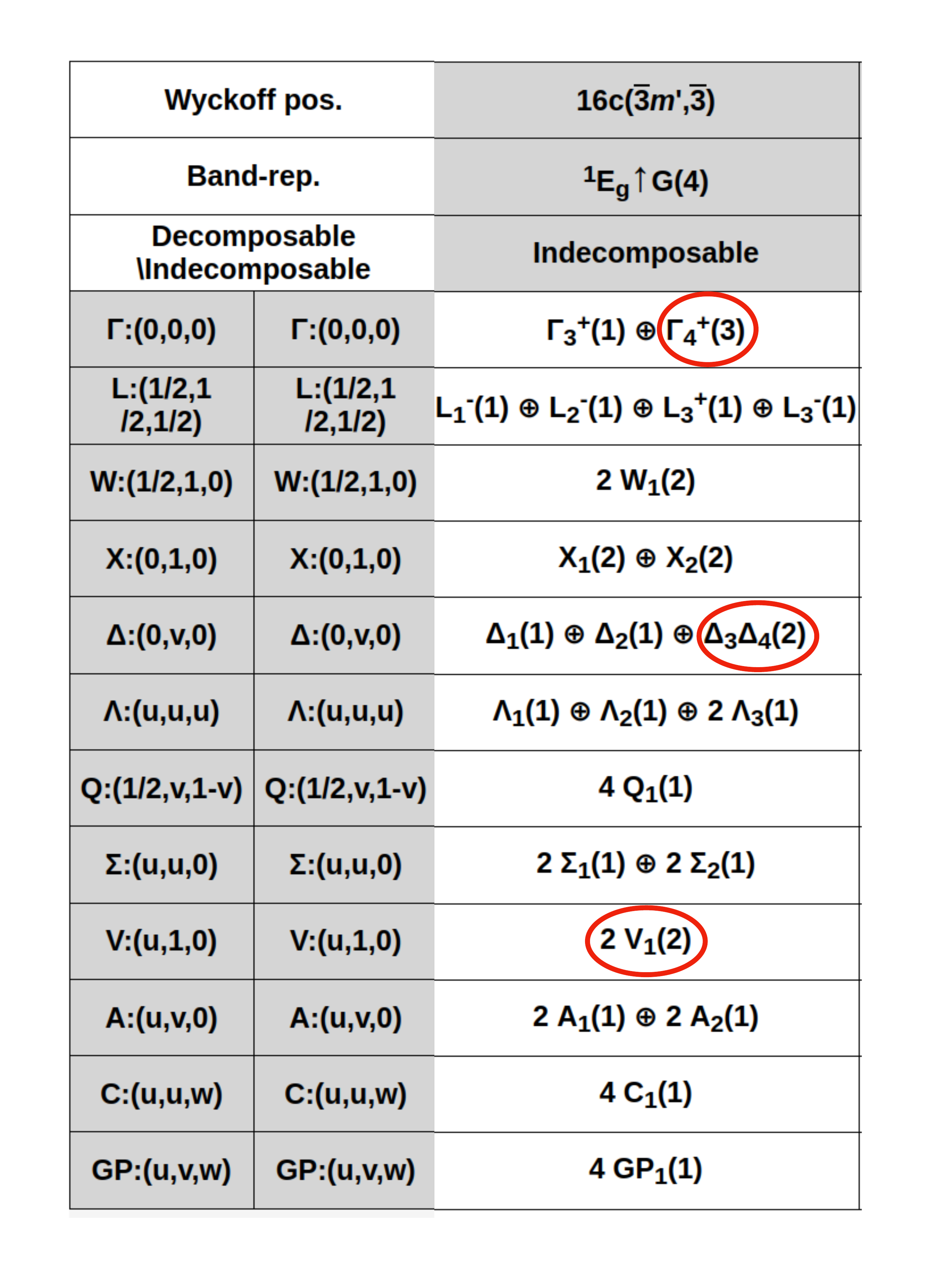}
\hss}
\caption{
Screenshot from MBANDREP tool on Bilbao Crystallographic Server for the magnetic space group $227.131$ Wyckoff position $16c$ and EBR induced from ${^1E_g}$. In the red circles the representations showing the 3-fold nodal point at $\Gamma$ and the two 2-fold nodal lines at $\Delta$ and $V$.
}
\label{Fig:BANDREP_2}
\end{figure}

Here we explain briefly how to make practical use of the magnetic band representation tool \href{https://www.cryst.ehu.es/cgi-bin/cryst/programs/mbandrep.pl}{MBANDREP} on the Bilbao Crystallographic Server for magnon systems \cite{elcoro2020magnetic,xu2020}.
This tool is very useful for topological magnons for two main purposes: identifying decomposable (topological gapped) EBRs and finding compatibility relations constraining possible topological nodal features. Once interesting cases have been established on symmetry grounds one may use the symmetry data to build models with nontrivial magnon band topology as we show.

Given a magnetic space group in the BNS convention, this tool shows which band representations can be induced from a particular Wyckoff position. Each Wyckoff position may induce either an elementary or a composite band representation.

As an example, suppose we want to know if a topological gap is possible for the honeycomb Heisenberg-Kitaev FM $[111]$ without an explicit calculation (see also Sec. \ref{subsec:HoneycombHK}). The essential ingredients to make this determination are the symmetries of the underlying crystal lattice, which are conserved by the exchange coupling, and the magnetic order.
In this case, the honeycomb lattice can be seen as a 3D space group $P6/mmm$, with associated point group $D_{6h}$ ($6/mmm$), and Wyckoff position $2c$. The Kitaev coupling does not preserve the $C_6$ rotation, breaking the point group down to $D_{3d}$ ($-3m$). In addition, the FM $[111]$ ordering out of plane breaks the two-fold rotations in the plane and relative mirrors, which can only be restored in combination with time-reversal symmetry thus forming the magnetic point group $-3m'$.
Therefore, the relevant magnetic space group is $P\bar{3}1m'$ ($\# 162.77$).
In listing the symmetries of the exchange Hamiltonian we should also consider possible  spin-rotation symmetries. For the case of the Heisenberg-Kitaev model there are extra spin rotations symmetries that form a $D_2$ group but these are killed off by the $[111]$ magnetic order as discussed in Ref.~\onlinecite{corticelli2021spinspace}.
Having established the magnetic space group and the Wyckoff position, we can use the MBANDREP tool and analyze the relevant band representations.

In Fig.~\ref{Fig:BANDREP_1} we give an example of a MBANDREP table.
In the first row the Wyckoff position ($2c$) is indicated together with its magnetic site-symmetry group and unitary subgroup ($32',3$).
In the second row, the different band representations from unitary irreps of the site-symmetry group are shown.
In the third row, the band representation is labelled as either decomposable (possible topological gap) or indecomposable (no gap possible) as determined from the compatibility relations.
In case it is decomposable, the tool further shows all the possible split branches.
Finally, the subsequent rows show the band representation subduced to irreps of the maximal high symmetry points in reciprocal space.

Of the six possible band representations here we can immediately neglect the double valued ones (last 3 columns with barred irreps) and focus on the single valued ones as is appropriate for spin waves.
In particular, the orbitals $J^{\pm}_{\bs{q}}$ transform under the unitary subgroup of the site-symmetry group $3$ as irreps ${^1E}, {^2E}$ as we can see in Tab.~\ref{tab:axialgroup}. Therefore we are interested in the EBR in the red circle ${^1E} \uparrow G(2)$ and ${^2E} \uparrow G(2)$, where the number in parentheses indicates the number of bands given by this Wyckoff position (here Wyckoff $2c$ is a two sublattices basis producing two bands). One of the irreps ${^1E}$ or ${^2E}$ is associated to the positive energies bands while the complex conjugated one to the negative bands produced by Bogoliuobov diagonalization.
We therefore see immediately that the orbital $J^{\pm}_{\bs{q}}$ induces a decomposable EBR, meaning that a gap is topological. A list of all possible single decomposable EBRs for magnons can be found in Tab.~\ref{tab:EBRsplitted}.
This table can be used to straightforwardly identify a topological gap, without accessing the full server details or computing explicitly the representations of the band structure.
Indeed if the magnetic material one is looking at (or the model intended to be constructed) is present in the table, non-trivial magnon topology is guaranteed once a gap is present.

One may, with this information, build a model exhibiting this topological gap. To do this, it is necessary to enforce the symmetries of the lattice at the level of the magnetic Hamiltonian taking care not to allow spin-space symmetries. Physically this means that the model should be spin-orbit coupled and include, in principle, all symmetry-allowed exchange couplings. The Kitaev-Heisenberg model includes only a subset of the allowed couplings to nearest neighbor but (i) the magnetic order is sufficient to break down the residual symmetries and (ii) including further couplings such as $\Gamma$, $\Gamma'$ \cite{mcclarty2018} preserves the band topology.

Finally we detail how to identify topological nodal models using the magnetic band representation. As an example we take the magnetic group $227.131$ Wyckoff position $16c$ relevant for pyrochlore AIAO order discussed in Sec.~\ref{subsec:PyrochloreNodal}.
From the Bilbao server, we extract the information in Fig.~\ref{Fig:BANDREP_2}. Here we are interested in the EBR ${^1E_g} \uparrow G(4)$ having 4 bands (or alternatively the conjugated one, showing the same enforced degeneracies).
To identify possible symmetry enforced degeneracies we need to inspect the subduced irreps at the high symmetry points/lines/surface in the first Brillouin zone. At the $\Gamma$ point the EBR will be composed of two irreps, $\Gamma_3^1(1) + \Gamma_4^+(3)$, where the number in parentheses is the dimension of the irrep, indicating the presence of a $3$-fold degenerate nodal point.
In addition we can identify one $2$-fold nodal line at $\Delta = (0,v,0) = [\Gamma-X]$ and 2 others at $V = (u,1,0) = [X-W]$ , marked with red circles (the coordinate are given in reciprocal lattice basis).
These symmetry enforced degeneracies can then be checked by carrying out a direct LSW calculation, as in Fig.~\ref{sec:nodaltopology}.

\section{Decomposable EBR for Magnons: Examples}
\label{sec:EBRExamples}

\subsection{Honeycomb Heisenberg-Kitaev FM $[111]$ model}
\label{subsec:HoneycombHK}

For simplicity the first case considered is the two dimensional Honeycomb Heisenberg-Kitaev ferromagnet. It has already been shown in Ref.~\cite{mcclarty2018,joshi2018} that the model hosts a gap with a non-trivial Chern invariant for the out of plane polarization, $[111]$ direction in Kitaev spin coordinate system. In fact the gap closes only for polarizations along $[xy0]$, $[x0z]$, $[0yz]$ (where an emergent spin-space symmetry enforces the gap closure \cite{corticelli2021spinspace}). In this section, we focus on the case where the moment is along $[111]$ and show that the nontrivial topology can be understood from the perspective of a decomposable EBR.

\subsubsection{Crystal structure}

The crystal structure we consider is an honeycomb lattice with an edge-shared octahedral environment around the sites so that Kitaev couplings are allowed by symmetry. The magnetic moments are polarized perpendicular to the honeycomb plane. 
This magnetic structure is described by the magnetic group $F\bar{3}1m'$ ($\#162.77$ in BNS setting) with Wyckoff position $2c$ that has site symmetry group $32'$.
The lattice primitive vectors are:
\beq
\bs{a}_1 = (\sqrt{3}/2,3/2),~~ \bs{a}_2 = (\sqrt{3}/2,-3/2)
\eeq
and basis coming from Wyckoff position $2c$ is (origin at the center of the hexagon):
\beq
\boldsymbol{\delta}_1 = (\sqrt{3}/2,-1/2),~ \boldsymbol{\delta}_2 = (\sqrt{3}/2,1/2).
\eeq
%

\begin{figure}[!htb]
\hbox to \linewidth{ \hss
\includegraphics[width=0.90\linewidth, trim=0.2cm 2.2cm 0.2cm 0.2cm, clip]{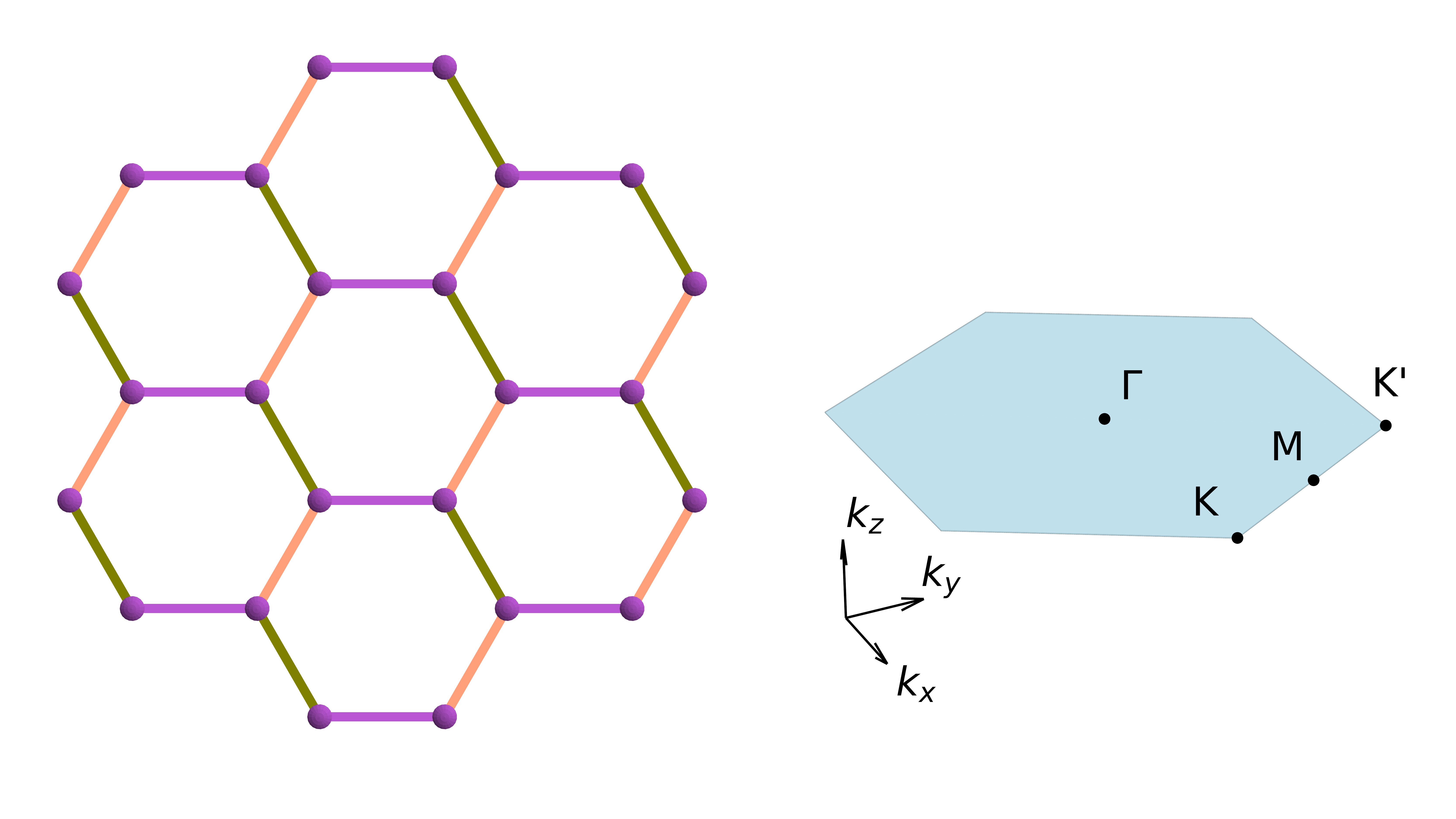}
\hss}
\caption{
The honeycomb lattice and its first Brillouin zone. In the lattice the bond colors (purple, orange, green) indicate the $x$, $y$, $z$ pattern of the Kitaev couplings. In the Brillouin zone the high symmetry points are: $\Gamma =$ $(0,0,0)$, $M =$ $(\frac{\pi}{\sqrt{3}},\frac{\pi}{3})$,
$K =$ $(\frac{4\pi}{3\sqrt{3}}, 0)$, $K'=$ $(\frac{2\pi}{3\sqrt{3}},\frac{2\pi}{3})$.
}
\label{Fig:LatticeHoneycombHK}
\end{figure}

Taking into account the octahedral environment the relevant point
group under which the exchange is left invariant is $D_{3d}$
with nontrivial symmetries including three-fold rotations
about the hexagonal centers, three two-fold rotational
symmetries about axes through opposite hexagonal vertices and inversion symmetry again about the hexagonal
centers. Once the out of plane $[111]$ magnetization is taken into consideration, the two-fold rotations are symmetries only in combination with time-reversal to restore the spin direction.
The generators of the group are:
\beq
G/T =  \element{E}{{\bf 0}}, ~ \element{3^+_{001}}{{\bf 0}}, ~ \element{2_{1\text{-}10}}{{\bf 0}}', ~ \element{-1}{{\bf 0}}
\label{eq:symmhc}
\eeq
where $z$ corresponds to $[111]$.

Also we define for later use in the exchange coupling matrix the bonds joining nearest neighbors:
\begin{align}
\boldsymbol{\delta}_x & = \left( 0, 1 \right), \\
\boldsymbol{\delta}_y & = \left(-\sqrt{3}/2, -1/2 \right), \\
\boldsymbol{\delta}_z & = \left(\sqrt{3}/2, -1/2  \right).
\label{eq:deltaxyz}
\end{align}

\subsubsection{Exchange Hamiltonian}
The nearest neighbor model on this lattice has, as symmetry allowed exchange terms, the following on the $x$, $y$, $z$ bonds in Eq.~\ref{eq:deltaxyz}:

\begin{align}
\hspace{0px}
\mathsf{J}_x = & 
\begin{pmatrix} 
J + 2K    & \Gamma' &  \Gamma'    \\
\Gamma'   & J       &  \Gamma    \\
\Gamma'   & \Gamma  & J
\end{pmatrix} 
&
\mathsf{J}_y = & 
\begin{pmatrix} 
J         & \Gamma' &  \Gamma    \\
\Gamma'   & J + 2K  &  \Gamma'    \\
\Gamma   & \Gamma'  & J
\end{pmatrix} 
&
\mathsf{J}_z = & 
\begin{pmatrix} 
J         & \Gamma &  \Gamma'    \\
\Gamma   & J       &  \Gamma'    \\
\Gamma'   & \Gamma'  & J + 2K
\end{pmatrix}. 
&
\label{Eq:Jcoupling_HoneycombHK}
\end{align}

In addition we allow for a magnetic field of magnitude $h$ in the $[111]$ direction. The linear spin wave Hamiltonian approximation is:
\begin{align}
\mathsf{A}(\boldsymbol{k}) & = \left( \begin{array}{cc}   
-3J - 2K + h &  \left( J + \frac{2K}{3} \right) \left( e^{i\boldsymbol{k}\cdot\boldsymbol{\delta}_x} + e^{i\boldsymbol{k}\cdot\boldsymbol{\delta}_y} + e^{i\boldsymbol{k}\cdot\boldsymbol{\delta}_z}  \right)  \\ 
 \left( J + \frac{2K}{3} \right) \left( e^{-i\boldsymbol{k}\cdot\boldsymbol{\delta}_x} + e^{-i\boldsymbol{k}\cdot\boldsymbol{\delta}_y} + e^{-i\boldsymbol{k}\cdot\boldsymbol{\delta}_z}  \right)& -3J - 2K + h
\end{array} \right) \\
\mathsf{B}(\boldsymbol{k}) & = \left( \begin{array}{cc}   
0 & \frac{2K}{3} \left( e^{i\boldsymbol{k}\cdot\boldsymbol{\delta}_x  + \frac{2\pi i}{3}} + e^{i\boldsymbol{k}\cdot\boldsymbol{\delta}_y - \frac{2\pi i}{3}} + e^{i\boldsymbol{k}\cdot\boldsymbol{\delta}_z}  \right) \\
\frac{2K}{3}  \left( e^{-i\boldsymbol{k}\cdot\boldsymbol{\delta}_x  + \frac{2\pi i}{3}} + e^{-i\boldsymbol{k}\cdot\boldsymbol{\delta}_y - \frac{2\pi i}{3}} + e^{-i\boldsymbol{k}\cdot\boldsymbol{\delta}_z}  \right) & 0
\end{array} \right).
\end{align}
This Hamiltonian contains only the $J$, $K$ and $h$ couplings. The $\Gamma$ and $\Gamma'$ terms merely renormalize the $J$, $K$, $h$ model so we have omitted them for simplicity.

\subsubsection{Band topology}

\begin{figure}[tp]
  \centering
  \includegraphics[width=0.75\columnwidth]{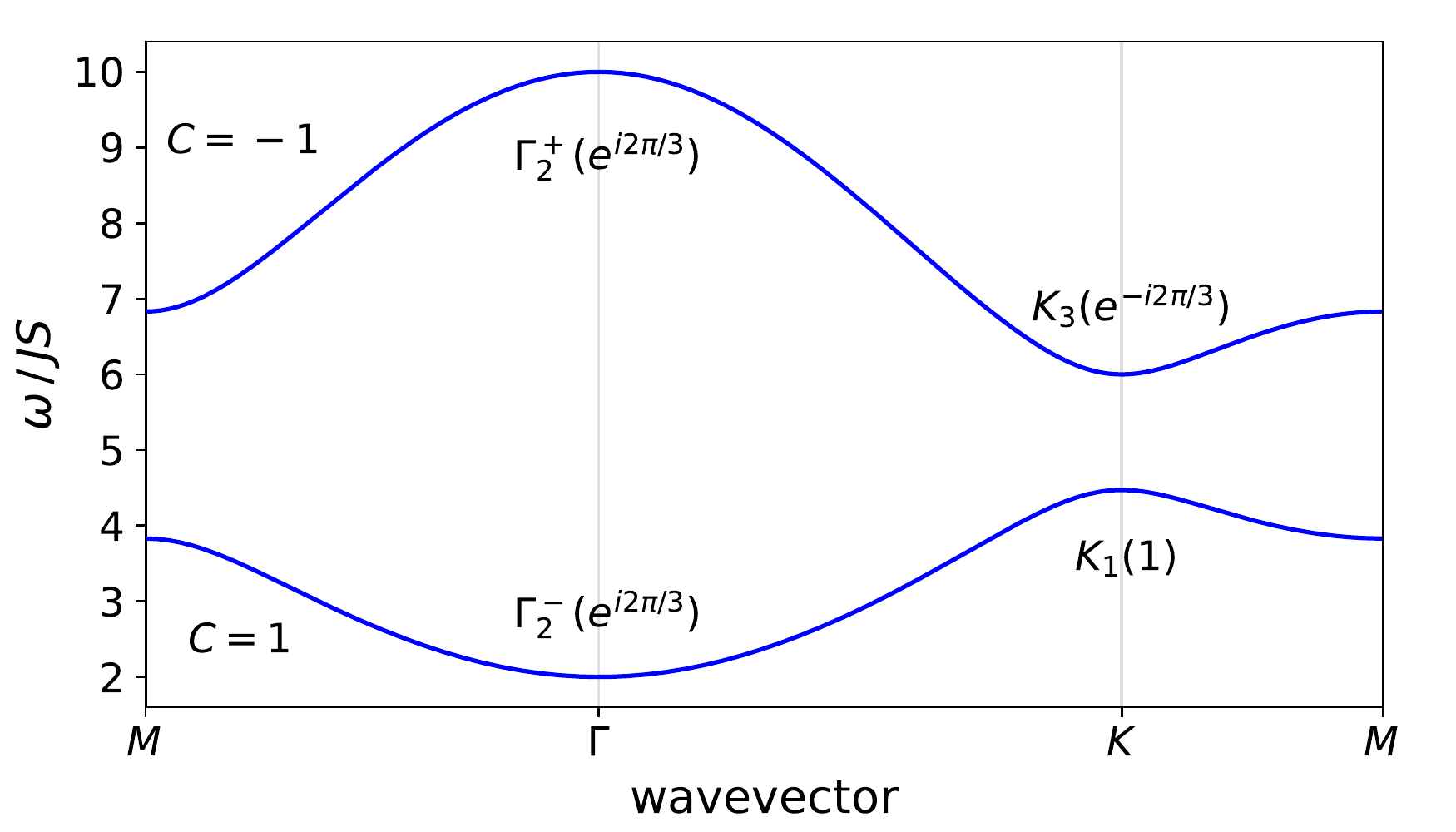}
  \caption{\label{fig:Dispersion_HoneycombHK}
Dispersion relations of magnons in the honeycomb lattice FM $[111]$ for $J=0$, $K=1$ and $h=5$. At the $\Gamma$ and $K$ point band representations are indicated with their respective $C_3$ eigenvalue. The band Chern number is indicated on the left, highlighting the topological gap.
}
\end{figure}

The orbital basis $(J_{\bs{q}}^+, J_{\bs{q}}^-)$ lives on the Wyckoff position $2c$ with associated site symmetry group generators (here we use the primitive lattice basis):
\begin{align}
&\bs{q}_1^{2c} = g_1 \, \bs{q}_1^{2c} = (1/3, 2/3) ~~~~ G_{\bs{q}_1^{2c}} = \element{3^+_{001}}{{0, 1, 0}} ,\, \element{2_{1-10}}{{1, 1, 0}}' \nonumber \\
&\bs{q}_2^{2c} = g_2 \, \bs{q}_1^{2c} = (2/3, 1/3) ~~~~ G_{\bs{q}_2^{2c}} = \element{3^+_{001}}{{1, 1, 0}} ,\, \element{2_{1-10}}{{1, 1, 0}}' 
\end{align}
where we have indicated also the orbit transformations $g_1 = \element{E}{{\bf 0}}$ and $g_2 = \element{-1}{1, 1, 0}$ forming the coset decomposition $G= \bigcup_\alpha g_{\alpha} (G_{\bs{q}_1^{2c}} \ltimes T)$.
The group $G_{\bs{q}_1^{2c}}$ is therefore isomorphic to $32'$ and the orbitals transform under the representation:
\beq
\rho_{32'} [(J_{\bs{q}}^+, J_{\bs{q}}^-)] = {^1E} + {^2E}
\eeq
which induces a two-band decomposable elementary band representation (see Table \ref{tab:EBRsplitted}).
The two bands, once split, produce a topological gap with chiral surface states. 
The system is a Chern insulator with a bulk invariant associated with a non-trivial Wilson loop that can be computed through the symmetry indicated formulas of the point group $C_3$:
\beq
e^{i\frac{2 \pi}{3}C} = \prod_{n} \Theta_n(\Gamma) \, \Theta_n(K) \,\Theta_n(K')
\eeq
where $C$ is the Chern number of the $n$ band(s) and $\Theta(\bs{k})$ are the eigenvalues of $C_3$.
The eigenvalues and irreps of the two bands are:
\begin{align}
& \text{band 1}: ~~~\Gamma_2^-(C_3^+)=e^{i\frac{2 \pi}{3}} \,, ~K_1(C_3^+)=1 \,, ~K'_1(C_3^+)=1 ~~~~~~~~~~~~\, \rightarrow C=1 \nonumber \\
& \text{band 2}: ~~~\Gamma_2^+(C_3^+)=e^{i\frac{2 \pi}{3}} \,, ~K_3(C_3^+)=e^{-i\frac{2 \pi}{3}} \,, ~K'_3(C_3^+)=e^{-i\frac{2 \pi}{3}} ~~\rightarrow C=-1
\end{align}
which corresponds to the induced $^2E \uparrow G$ band representation (while $^1E \uparrow G$ can be found for the negative eigenvalues).

\subsection{Honeycomb XYZ-DM FM $[001]$ model}
\label{subsec:HoneycombHDM}

The honeycomb lattice offers another famous topological gapped model, the Haldane model, which we study here in the context of EBR.
The honeycomb isotropic Heisenberg has Dirac cones due to $P\mathcal{T}$ symmetry pinned at $K$ by $C_3$.
To lift this degeneracy, the spin-space time-reversal symmetry present in Heisenberg need to be broken. 
This can be achieved either by having anisotropic Heisenberg (XYZ model) or by introducing a next nearest neighbour DM interaction with out of plane magnetic order (spin wave analog of Haldane model).
In both case a gap with non-trivial Chern number will arise. For completeness here we analyze the full model XYZ-DM from an EBR perspective.

\subsubsection{Crystal structure}

The crystal structure we consider is an honeycomb lattice with magnetic moments polarized perpendicular to the honeycomb plane $[001]$.
Both the anisotropic $XYZ$ model and the next NN out of plane DM interaction preserve all the honeycomb symmetry, space group $P6/mmm$.
Nevertheless the $[001]$ magnetic order will reduce the group to the type III $P6/mm'm'$ ($\#191.240$). The Wyckoff position is $2c$ with site-symmetry group $-6m'2'$.
The generators of the group are:
\beq
G/T =  \element{E}{{\bf 0}}, ~ \element{3^+_{001}}{{\bf 0}}, ~
\element{2_{001}}{{\bf 0}}, ~\element{2_{1\text{-}10}}{{\bf 0}}', ~ \element{-1}{{\bf 0}}
\label{eq:symmhc}
\eeq

The primitive lattice and basis are the same as \ref{subsec:HoneycombHK}. Here we define additionally for later use the bonds joining the next nearest neighbors:
\begin{align}
\boldsymbol{\delta}_{2x} & = \left( \sqrt{3}, 0 \right), \\
\boldsymbol{\delta}_{2y} & = \left(-\sqrt{3}/2, 3/2 \right), \\
\boldsymbol{\delta}_{2z} & = \left(-\sqrt{3}/2, -3/2  \right).
\label{eq:delta2xyz}
\end{align}

\subsubsection{Exchange Hamiltonian}
The nearest neighbor anisotropic Heisenberg interaction respect the symmetry of the honeybomb lattice and reads for the $x$, $y$, $z$ bonds in Eq.~\ref{eq:deltaxyz}:

\begin{align}
\hspace{0px}
\mathsf{J}_x = & 
\begin{pmatrix} 
J_x & 0   &  0    \\
0   & J_y &  0   \\
0   & 0   & J_z
\end{pmatrix} 
&
\mathsf{J}_y = & 
\begin{pmatrix} 
\frac{J_x}{4} + \frac{3 J_y}{4}  & \frac{\sqrt{3} J_x}{4} - \frac{\sqrt{3} J_y}{4} &  0    \\
\frac{\sqrt{3} J_x}{4} - \frac{\sqrt{3} J_y}{4}   & \frac{3 J_x}{4} + \frac{J_y}{4}  &  0    \\
0   & 0  & J_z
\end{pmatrix} 
&
\mathsf{J}_z = & 
\begin{pmatrix} 
\frac{J_x}{4} + \frac{3 J_y}{4}  & -\frac{\sqrt{3} J_x}{4} + \frac{\sqrt{3} J_y}{4} &  0    \\
-\frac{\sqrt{3} J_x}{4} + \frac{\sqrt{3} J_y}{4}   & \frac{3 J_x}{4} + \frac{J_y}{4}  &  0    \\
0   & 0  & J_z
\end{pmatrix}. 
&
\label{Eq:Jcoupling_HoneycombHDM}
\end{align}

The next nearest neighbor DM interaction has $\bs{D} = D \hat{z}$ and exchange hamiltonian for bonds $2x$, $2y$, $2z$ bonds in Eq.~\ref{eq:delta2xyz}:
\beq
\hspace{0px}
\mathsf{J}_{2x} = \mathsf{J}_{2y} = \mathsf{J}_{2z} =
\begin{pmatrix} 
0 & D   &  0    \\
-D   & 0 &  0   \\
0   & 0   & 0
\end{pmatrix} 
\label{Eq:Jcoupling_HoneycombHDM_NNN}
\eeq

The linear spin wave Hamiltonian approximation is:
\begin{align}
\mathsf{A}(\boldsymbol{k}) & = \left( \begin{array}{cc}   
-3J_z + D \, \gamma(\bs{k})  &  \left( \frac{J_{x} + J_{y}}{2} \right) \left( e^{i\boldsymbol{k}\cdot\boldsymbol{\delta}_x} + e^{i\boldsymbol{k}\cdot\boldsymbol{\delta}_y} + e^{i\boldsymbol{k}\cdot\boldsymbol{\delta}_z}  \right)  \\ 
 \left( \frac{J_{x} + J_{y}}{2} \right) \left( e^{-i\boldsymbol{k}\cdot\boldsymbol{\delta}_x} + e^{-i\boldsymbol{k}\cdot\boldsymbol{\delta}_y} + e^{-i\boldsymbol{k}\cdot\boldsymbol{\delta}_z}  \right) & -3J_z - D \, \gamma(\bs{k})
\end{array} \right) \\
\mathsf{B}(\boldsymbol{k}) & = \left( \begin{array}{cc}   
0 & \left( \frac{J_{x} - J_{y}}{2} \right) \left( e^{i\boldsymbol{k}\cdot\boldsymbol{\delta}_x} + e^{i\boldsymbol{k}\cdot\boldsymbol{\delta}_y + \frac{2\pi i}{3}} + e^{i\boldsymbol{k}\cdot\boldsymbol{\delta}_z - \frac{2\pi i}{3}}  \right) \\
\left( \frac{J_{x} - J_{y}}{2} \right)  \left( e^{-i\boldsymbol{k}\cdot\boldsymbol{\delta}_x} + e^{-i\boldsymbol{k}\cdot\boldsymbol{\delta}_y + \frac{2\pi i}{3}} + e^{-i\boldsymbol{k}\cdot\boldsymbol{\delta}_z - \frac{2\pi i}{3}}  \right) & 0
\end{array} \right).
\end{align}
where:
\beq
\gamma(\bs{k}) = \sin(\boldsymbol{k}\cdot\boldsymbol{\delta}_{2x}) + \sin(\boldsymbol{k}\cdot\boldsymbol{\delta}_{2y}) + \sin(\boldsymbol{k}\cdot\boldsymbol{\delta}_{2z})
\eeq

\subsubsection{Band topology}

\begin{figure}[tp]
  \centering
  \includegraphics[width=0.75\columnwidth]{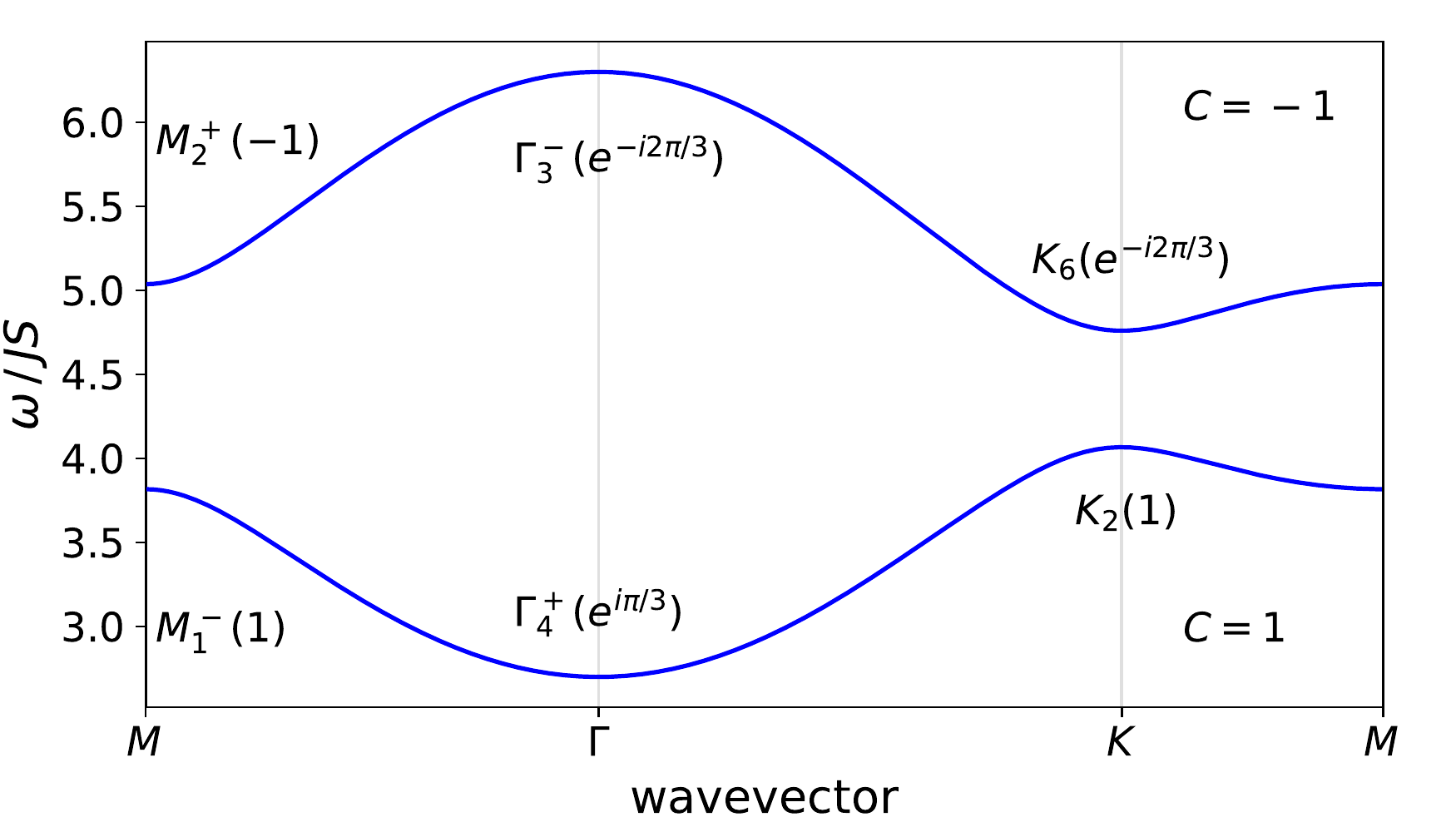}
  \caption{\label{fig:Dispersion_HoneycombHDM}
Dispersion relations of magnons in the honeycomb lattice FM $[001]$ for $J_x=-1$, $J_y=-0.2$, $J_z=-1.5$, $D=0.1$. At the $\Gamma$, $K$, $M$ points band representations are indicated with their respective $C_6$, $C_3$, $C_2$ eigenvalue respectively. The band Chern number is indicated on the right, highlighting the topological gap.
}
\end{figure}

The orbital basis $(J_{\bs{q}}^+, J_{\bs{q}}^-)$ lives on the Wyckoff position $2c$ with associated site symmetry group generators (here we use the primitive lattice basis):
\begin{align}
&\bs{q}_1^{2c} = g_1 \, \bs{q}_1^{2c} = (1/3, 2/3) ~~~~ G_{\bs{q}_1^{2c}} = \element{3^+_{001}}{{0, 1, 0}} ,\, \element{m_{001}}{\bs{0}} ,\, \element{2_{1\text{-}10}}{{1, 1, 0}}' \nonumber \\
&\bs{q}_2^{2c} = g_2 \, \bs{q}_1^{2c} = (2/3, 1/3) ~~~~ G_{\bs{q}_2^{2c}} = \element{3^+_{001}}{{1, 1, 0}},\, \element{m_{001}}{\bs{0}} ,\, \element{2_{1\text{-}10}}{{1, 1, 0}}' 
\end{align}
where we have indicated also the orbit transformations $g_1 = \element{E}{{\bf 0}}$ and $g_2 = \element{-1}{1, 1, 0}$ forming the coset decomposition $G= \bigcup_\alpha g_{\alpha} (G_{\bs{q}_1^{2c}} \ltimes T)$.
The group $G_{\bs{q}_1^{2c}}$ is therefore isomorphic to $-6m'2'$ and the orbitals transform under the representation:
\beq
\rho_{-6m'2'} [(J_{\bs{q}}^+, J_{\bs{q}}^-)] = {^1E''} + {^2E''}
\eeq
which induces a two-band decomposable elementary band representation (see Table \ref{tab:EBRsplitted}).
The two bands, once split, produce a topological gap with chiral surface states.
The system is a Chern insulator with a bulk invariant associated with a non-trivial Wilson loop that can be computed through the symmetry indicated formulas of the point group $C_6$:
\beq
e^{i\frac{2 \pi}{6}C} = \prod_{n} \eta_n(\Gamma) \, \Theta_n(K) \,\zeta_n(K')
\label{eq:C6invariant}
\eeq
where $C$ is the Chern number of the $n$ band(s) and $\eta(\bs{k})$, $\Theta(\bs{k})$, $\zeta(\bs{k})$ are the eigenvalues of $C_6$, $C_3$, $C_2$.
The eigenvalues and irreps of the two bands are:
\begin{align}
& \text{band 1}: ~~~\Gamma_4^+(C_6^+)=e^{i\frac{\pi}{3}} \,, ~K_2(C_3^+)=1 \,, ~M_1^-(C_2)=1 ~~~~~~~~~~~~ \rightarrow C=1 \nonumber \\
& \text{band 2}: ~~~\Gamma_3^-(C_6^+)=e^{-i\frac{2 \pi}{3}} \,, ~K_6(C_3^+)=e^{-i\frac{2 \pi}{3}} \,, ~M_2^+(C_2)=-1 ~~\rightarrow C=-1
\label{eq:rep_honeycombHDM}
\end{align}
which corresponds to the induced $^2E'' \uparrow G$ band representation (while $^1E'' \uparrow G$ can be found for the negative eigenvalues).

\subsection{Stacked honeycomb AFM topological insulator}
\label{subsec:HoneycombAFTI}

We now consider a system with $AA$ stacked honeycomb planes with anisotropic couplings within each layer and AFM Heisenberg exchange between layers. This results in a magnonic topological crystalline insulator as noted in \cite{kondo2021}.
The in-plane model (with decoupled layers) is, from a symmetry perspective, identical to that studied in Sec.~\ref{subsec:HoneycombHDM}, which has two split bands with opposite Chern number. Adding a stacked layer with AFM order produces two further bands, but with reversed Chern number. If the two layers are not coupled, the surface state of opposite chiralities survive independently. 
If the layer are coupled, for example through AFM Heisenberg, then the respective bulk bands can mix, rendering the Chern invariant ill-defined. Nevertheless the effective anti-unitary symmetry $\mathcal{T}\,T_{\bs{a_3}/2}$ imposes a Kramers degeneracy in the plane $k_z = \pi$, which protects the hybridized surface states from gapping, leading to a topological insulator with a $\mathbb{Z}_2$ invariant \cite{kondo2021}.

\subsubsection{Crystal structure}

The crystal structure is a stacked honeycomb lattice with magnetic moments along $[001]$ anti-aligned between layers.
In the previous section we have shown how the single layer correspond to group $P6/mm'm'$, but here we have an additional black and white translation between the two AFM layers with magnetic space group type IV $P_c6/mcc$ ($\#192.252$).
The Wyckoff position is $4c$ with site-symmetry group $-6m'2'$.
The generators of the group are:
\beq
G/T =  \element{E}{{\bf 0}}, ~ \element{3^+_{001}}{{\bf 0}}, ~
\element{2_{001}}{{\bf 0}}, ~\element{2_{1\text{-}10}}{0, 0, 1/2}, ~ \element{-1}{{\bf 0}}, ~ \element{1}{0, 0, 1/2}'
\label{eq:symmhc}
\eeq
The lattice primitive vectors are:
\beq
\bs{a}_1 = (\sqrt{3}/2,3/2,0),~~ \bs{a}_2 = (\sqrt{3}/2,-3/2,0),~~ \bs{a}_3 = (0,0,1)
\eeq
and basis coming from Wyckoff position $4c$ is (origin at the center of the hexagon):
\beq
\boldsymbol{\delta}_1 = (\sqrt{3}/2,-1/2, 0),~ \boldsymbol{\delta}_2 = (\sqrt{3}/2,1/2, 0),~ \boldsymbol{\delta}_3 = (\sqrt{3}/2,-1/2, 1/2),~ \boldsymbol{\delta}_4 = (\sqrt{3}/2,1/2, 1/2)
\eeq

\subsubsection{Exchange Hamiltonian}

The model we consider has exchange 
\beq
J_x S_{i}^x S_{i+\hat{y}}^x + J_y S_{i}^y S_{i+\hat{y}}^y + J_z S_{i}^z S_{i+\hat{y}}^z
\eeq
on the honeycomb bond aligned with $y$ and the components refer to the crystallographic frame with $\hat{z}$ perpendicular to the honeycomb layers. We then tile all bonds using $C_3$ and translations exactly as in Eq. \ref{Eq:Jcoupling_HoneycombHDM}. We coupled the AA layers with a simple AFM Heisenberg $J_c$ coupling. We also include a second-neighbor in-plane Dzyaloshinskii-Moriya coupling considered in Sec.~\ref{subsec:HoneycombHDM}, even if not strictly necessary for the non-trivial topology here.

The linear spin wave Hamiltonian is:
\begin{align}
\mathsf{A}(\boldsymbol{k}) & = \left( \begin{array}{cccc}   
J_H + D \, \gamma_1(\bs{k})  &  \left( \frac{J_{x} + J_{y}}{2} \right) \gamma_2(\bs{k})  & 0 & 0  \\ 
 \left( \frac{J_{x} + J_{y}}{2} \right) \gamma_2^*(\bs{k}) & J_H - D \, \gamma_1(\bs{k}) & 0 & 0\\
0 & 0 & J_H - D \, \gamma_1(\bs{k})  &  \left( \frac{J_{x} + J_{y}}{2} \right) \gamma_2(\bs{k})    \\ 
0 & 0 & \left( \frac{J_{x} + J_{y}}{2} \right) \gamma_2^*(\bs{k}) & J_H + D \, \gamma_1(\bs{k})
\end{array} \right) \\
\mathsf{B}(\boldsymbol{k}) & = \left( \begin{array}{cccc}   
0 & \left( \frac{J_{x} - J_{y}}{2} \right) \beta_1(\bs{k}) & -2 J_c \cos(\bs{k}\cdot \bs{a_3}/2) & 0 \\
\left( \frac{J_{x} - J_{y}}{2} \right)  \beta_1(-\bs{k}) & 0 & 0 & -2 J_c \cos(\bs{k}\cdot \bs{a_3}/2) \\
-2 J_c \cos(\bs{k}\cdot \bs{a_3}/2) & 0 & 0 &
\left( \frac{J_{x} - J_{y}}{2} \right)  \beta_1(-\bs{k}) \\
0 & -2 J_c \cos(\bs{k}\cdot \bs{a_3}/2) &
\left( \frac{J_{x} - J_{y}}{2} \right)  \beta_1(\bs{k}) & 0 \\
\end{array} \right).
\end{align}
where:
\begin{align}
& J_H = -3J_z + 2J_c \\
&\gamma_1(\bs{k}) = \sin(\boldsymbol{k}\cdot\boldsymbol{\delta}_{2x}) + \sin(\boldsymbol{k}\cdot\boldsymbol{\delta}_{2y}) + \sin(\boldsymbol{k}\cdot\boldsymbol{\delta}_{2z}) \\
&\gamma_2(\bs{k}) = e^{i\boldsymbol{k}\cdot\boldsymbol{\delta}_x} + e^{i\boldsymbol{k}\cdot\boldsymbol{\delta}_y} + e^{i\boldsymbol{k}\cdot\boldsymbol{\delta}_z}   \\
&\beta_1(\bs{k}) = e^{i\boldsymbol{k}\cdot\boldsymbol{\delta}_x} + e^{i\boldsymbol{k}\cdot\boldsymbol{\delta}_y + \frac{2\pi i}{3}} + e^{i\boldsymbol{k}\cdot\boldsymbol{\delta}_z - \frac{2\pi i}{3}} 
\end{align}
where the $x$, $y$, $z$ and $2x$, $2y$, $2z$ bonds are the same as Eq.~\ref{eq:deltaxyz} and Eq.~\ref{eq:delta2xyz} and on all honeycomb layers.

\subsubsection{Band topology}

\begin{figure}[tp]
  \centering
  \includegraphics[width=0.75\columnwidth]{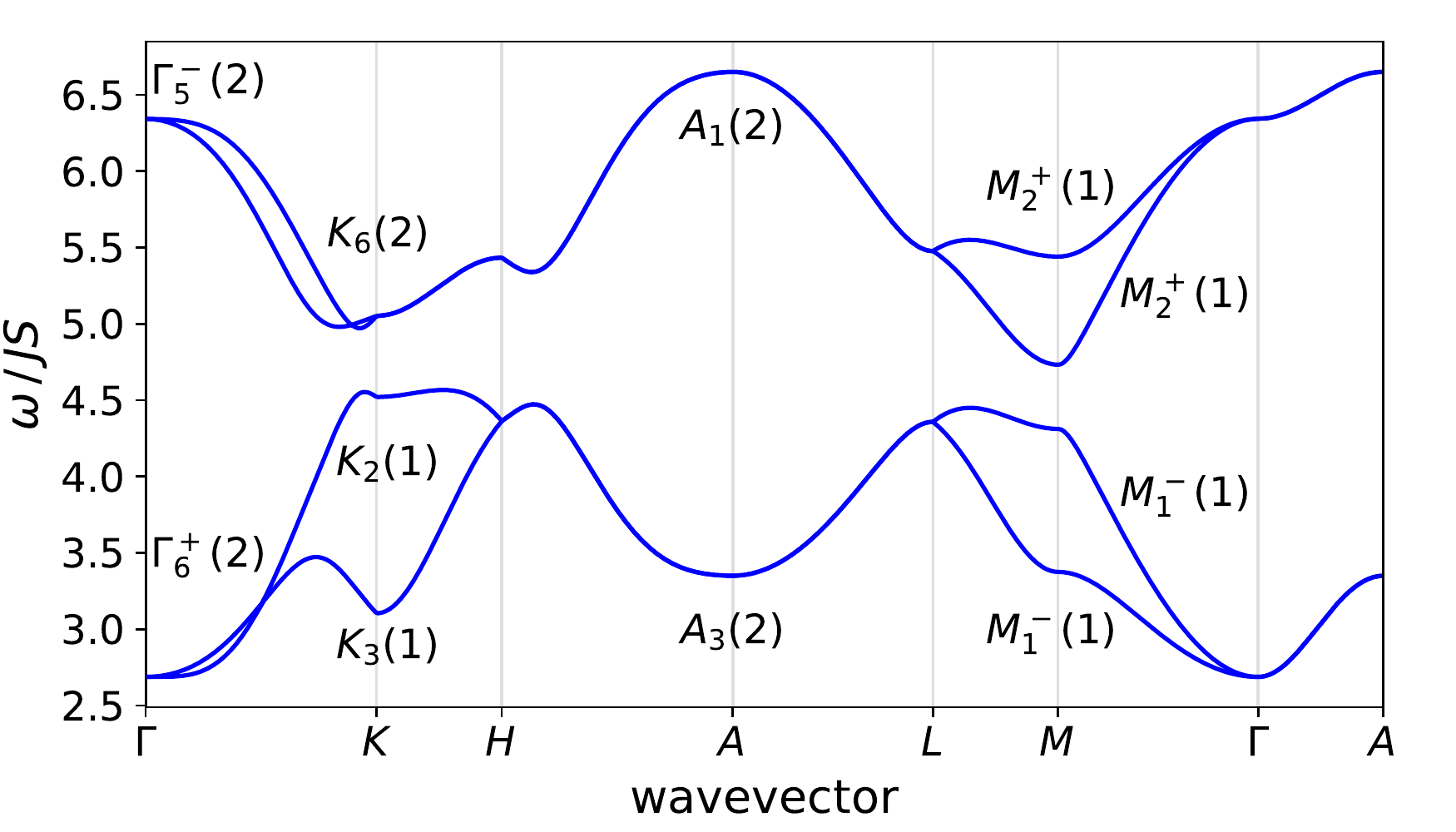}
  \caption{\label{fig:Dispersion_HoneycombAFTI}
Dispersion relations of magnons for the AA stacked honeycomb lattice ferromagnet with $[001]$ moments for $J_x=-1$, $J_y=-0.1$, $J_z=-1$, $D=0.5$, $J_c=-1$. At the $\Gamma$, $K$, $M$, $A$ points band representations are indicated with their respective dimension. The topological gap comes from the splitting of a 4-band EBR.
A nodal surface is present at $[H-A-L]$ and a nodal line along $[\Gamma-A]$.
}
\end{figure}

The orbital basis $(J_{\bs{q}}^+, J_{\bs{q}}^-)$ lives on the Wyckoff position $4c$ with associated site symmetry group generators (here we use the primitive lattice basis):
\begin{align}
&\bs{q}_1^{4c} = g_1 \, \bs{q}_1^{2c} = (1/3, 2/3, 0) ~~~~ G_{\bs{q}_1^{4c}} = \element{3^+_{001}}{{0, 1, 0}} ,\, \element{m_{001}}{\bs{0}} ,\, \element{2_{1\text{-}10}}{{1, 1, 0}}' \nonumber \\
&\bs{q}_2^{4c} = g_2 \, \bs{q}_1^{2c} = (2/3, 1/3, 0) ~~~~ G_{\bs{q}_2^{4c}} = \element{3^+_{001}}{{1, 1, 0}},\, \element{m_{001}}{\bs{0}} ,\, \element{2_{1\text{-}10}}{{1, 1, 0}}'
\\
&\bs{q}_3^{4c} = g_3 \, \bs{q}_1^{2c} = (1/3, 2/3, 1/2) ~~~~ G_{\bs{q}_3^{4c}} = \element{3^+_{001}}{{0, 1, 0}} ,\, \element{m_{001}}{{0, 0, 1}} ,\, \element{2_{1\text{-}10}}{{1, 1, 1}}' \nonumber \\
&\bs{q}_4^{4c} = g_4 \, \bs{q}_1^{2c} = (2/3, 1/3, 1/2) ~~~~ G_{\bs{q}_4^{4c}} = \element{3^+_{001}}{{1, 1, 0}},\, \element{m_{001}}{{0, 0, 1}} ,\, \element{2_{1\text{-}10}}{{1, 1, 1}}' 
\end{align}
where we have indicated also the orbit transformations $g_1 = \element{E}{{\bf 0}}$, $g_2 = \element{-1}{1, 1, 0}$, $g_3 = \element{1}{0, 0, 1/2}'$, $g_4 = \element{-1}{1, 1, 1/2}'$ forming the coset decomposition $G= \bigcup_\alpha g_{\alpha} (G_{\bs{q}_1^{4c}} \ltimes T)$.
The group $G_{\bs{q}_1^{4c}}$ is therefore isomorphic to $-6m'2'$ and the orbitals transform under the representation:
\beq
\rho_{-6m'2'} [(J_{\bs{q}}^+, J_{\bs{q}}^-)] = {^1E''} + {^2E''}
\eeq
which induces a four-band decomposable elementary band representation (see Table \ref{tab:EBR_honeycombAFTI}).
The subduced irreps in reciprocal space of the two branches (2 bands each) are:
\begin{align}
& \text{branch 1}: ~~\Gamma_6^+(2) \,, ~A_3(2)\,, ~H_3(2) \,, ~K_3(1)+K_2(1) \,, ~L_1(2) \,, ~M_2^-(1) + M_1^-(1)
 \nonumber \\
& \text{branch 2}: ~~\Gamma_5^-(2) \,, ~A_1(2)\,, ~H_2(2) \,, ~K_6(2) \,, ~L_2(2) \,, ~M_4^+(1) + M_3^+(1)
 \nonumber \\
\end{align}
which corresponds to the induced $^2E'' \uparrow G(4)$ band representation ($^1E'' \uparrow G(4)$ for negative eigenvalues).

The model has a nontrivial $\mathbb{Z}_2$ invariant linked to the black and white translation $\element{1}{0, 0, 1/2}'$ \cite{kondo2021}.
This is indeed reflected by the EBR picture.
When the two layers are decoupled, there are two decomposable EBRs $^2E'' \uparrow G(2)$ (layer spin up) and $^1E'' \uparrow G(2)$ (layer spin down) of the kind in Eq.~\ref{eq:rep_honeycombHDM}, which produce bands with opposite Chern number for the opposite layers.
When the two layers couple, the black and white translation $\element{1}{0, 0, 1/2}'$ pairs the two EBRs into a new single EBR which is nevertheless decomposable and therefore topological.

Finally the band representation, beside the topological gap, also predicts the nodal plane $E_1E_2(2)$ between each pair of bands.

We have established the topological character of the magnon bands based on symmetry. A more refined analysis reveals that this model has a nontrivial $\mathbb{Z}_2$ invariant that can be computed from the Berry phase $A_\mu^{(n)}(\boldsymbol{k})$ in the pairs of bands joined by Kramers degeneracies. Thus for bands $n=1,2,3,4$ where $n=1,2$ form the lower energy pair Ref.~\cite{kondo2021} show that the relevant invariant is built from $A_\mu^{(-)}(\boldsymbol{k})=A_\mu^{(1)}(\boldsymbol{k})+A_\mu^{(2)}(\boldsymbol{k})$ and $A_\mu^{(+)}(\boldsymbol{k})=A_\mu^{(3)}(\boldsymbol{k})+A_\mu^{(4)}(\boldsymbol{k})$
and z component of the Berry curvature $F_z^{(\pm)}(\boldsymbol{k})$
\beq
\nu^{\pm} = \frac{1}{2\pi}\left( \oint_{\partial HBZ} dk^\mu A_\mu^{(\pm)}(\boldsymbol{k}) - \int_{HBZ} d^2k F_z^{(\pm)}(\boldsymbol{k}) \right) {\rm mod}\hspace{2pt} 2
\eeq
where $HBZ$ refers to half of the zone such that the remainder is covered by $\boldsymbol{k}\rightarrow -\boldsymbol{k}$.

\begin{table*}[!htb]
\centering
\hbox to \linewidth{ \hss
\begin{tabular}{|c|c|c|}
 \hline
  \multicolumn{1}{|c|}{~EBR~}  & ${^1E''} \uparrow G(4)$ & ${^2E''} \uparrow G(4)$ \\ 
 \hline
  & Decomposable & Decomposable \\
 \hline
  $\Gamma: (0,0,0)$
 &$\Gamma_5^-(2) + \Gamma_6^+(2)$ 
 & $\Gamma_5^-(2) + \Gamma_6^+(2)$ \\
 $A: (0,0,1/2)$
 & $A_2(2) + A_4(2)$ & $A_1(2) + A_3(2)$ \\
 $H: (1/3,1/3,1/2)$
 & $H_1(2) + H_3(2)$ & $H_2(2) + H_3(2)$ \\
 $K: (1/3,1/3,0)$
 & $K_2(2) + K_3(1) + K_6(2)$  & 
 $K_2(2) + K_3(1) + K_6(2)$ \\
 $L: (1/2,0,1/2)$
 & $L_1(2) + L_2(2)$ & $L_1(2) + L_2(2)$ \\
 $M: (1/2,0,0)$
 & $M_1^-(2) + M_2^-(1) + M_3^+(1) + M_4^+(1)$ & 
 $M_1^-(2) + M_2^-(1) + M_3^+(1) + M_4^+(1)$ \\
 $E: (u,v,1/2)$
 & $2E_1E_2(2)$ & $2E_1E_2(2)$ \\
 \hline
\end{tabular}
\hss}
\caption{
Relevant magnon EBR for magnetic space group  type IV $P_c6/mcc$ ($\#192.252$) Wyckoff position $4c$ and site symmetry group $-6m'2'$. The table shows the representations subduced at the reciprocal space high symmetry points, with the dimension in the parenthesis.
Compatibility relation allow decomposability and therefore a topological gap. In the last line the 2-fold degenerate nodal plane representation.
}
\label{tab:EBR_honeycombAFTI}
\end{table*}

\subsection{Space group $P4$; FM $[001]$ model}

\begin{figure}[!htb]
\hbox to \linewidth{ \hss
\includegraphics[width=0.90\linewidth, trim=0.2cm 0.2cm 0.2cm 0.2cm, clip]{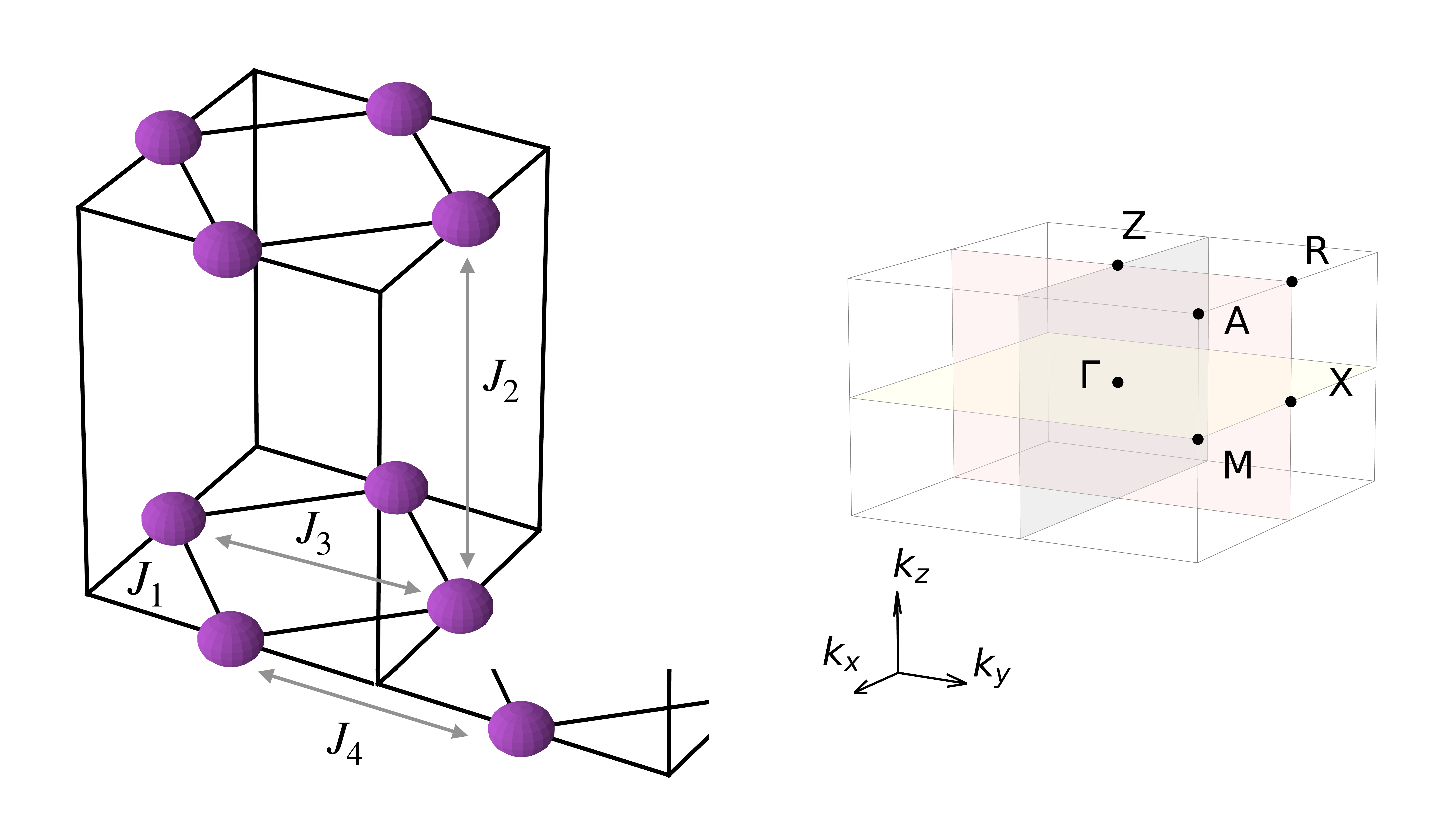}
\hss}
\caption{
The lattice of space group $P4$ ($\# 75.1$) and its simple tetragonal first Brillouin zone.
In the lattice the $J_i$ couplings mentioned in the text are showed.
In the Brillouin zone the high symmetry points are: $\Gamma =$ $(0,0,0)$, $X =$ $(0,\pi,0)$,
$Z =$ $(0,0,\pi)$, $M=$ $(\pi,\pi,0)$, 
$R =$ $(0,\pi,\pi)$, $A =$ $(\pi.\pi,\pi)$.
}
\label{Fig:Lattice75}
\end{figure}

\subsubsection{Crystal structure}
Here we consider the space group $P4~ (\#75.1)$ with Wyckoff position $2c$. The system is described by a single decomposable EBR with a site-symmetry group $C_2$ compatible with FM $[001]$ magnetic order and must host a topological gap once the EBR is split.

The lattice is a simple tetragonal with primitive vectors:
\beq
\bs{a}_1 = (a,0,0),~~ \bs{a}_2 = (0,a,0), ~~\bs{a}_3 = (0,0,c)
\eeq
The basis coming from Wyckoff position $2c$ is:
\beq
\bs{\delta}_1 = (0,1/2,z),~ \bs{\delta}_2 = (1/2,0,z)
\eeq
The lattice and the first Brillouin zone are shown in Fig.~ \ref{Fig:Lattice75}.

There are 4 symmetries in the group, all around the axial $z$ direction:
\beq
G/T =  \element{E}{{\bf 0}}, ~ \element{2_{001}}{{\bf 0}}, ~ \element{4^+_{001}}{{\bf 0}}, ~ \element{4^-_{001}}{{\bf 0}}
\label{eq:symm75}
\eeq

Also we define for later use in the exchange coupling matrix the additional lattice points:
\begin{align}
\bs{\delta}_{1x} = \bs{\delta}_1 +  \bs{a}_1 ,~ \bs{\delta}_{1y} = \bs{\delta}_1 +  \bs{a}_2, ~\bs{\delta}_{1z} = \bs{\delta}_1 +  \bs{a}_3  \nonumber \\
\bs{\delta}_{2x} = \bs{\delta}_2 +  \bs{a}_1 ,~ \bs{\delta}_{2y} = \bs{\delta}_2 +  \bs{a}_2, ~\bs{\delta}_{2z} =\bs{\delta}_2 +  \bs{a}_3
\end{align}

\subsubsection{Exchange Hamiltonian}
The model $J_1 +J_2 +J_3 +J_4$ on this lattice consists of $10$ different bond types:
\begin{itemize}
	\item Set $J_1$ : ${(1,2)_a,(1,2)_b,(1,2)_c,(1,2)_d}$
         \item Set $J_2$ : ${(1,1)_z,(2,2)_z}$
         \item Set $J_3$ : ${(1,1)_x,(2,2)_y}$
         \item Set $J_4$ : ${(1,1)_y,(2,2)_x}$.
\end{itemize}

Applying the symmetries in Eq.~\ref{eq:symm75} the exchange terms are constrained to 26 possible coupling parameters:
\begin{align}
\hspace{0px}
\mathsf{J}_{(1, 2)_a} = & 
\begin{pmatrix} 
\hphantom{-}J_1^{xx}     & \hphantom{-}J_1^{xy}       &  \hphantom{-}J_1^{xz}     \\
\hphantom{-}J_1^{yx}     & \hphantom{-}J_1^{yy}       &  \hphantom{-}J_1^{yz}     \\
\hphantom{-}J_1^{zx}     & \hphantom{-}J_1^{zy}        & \hphantom{-}J_1^{zz} 
\end{pmatrix} 
\hspace{40px}&
\mathsf{J}_{(1, 2)_b}  = &
\begin{pmatrix} 
\hphantom{-}J_1^{yy}     & -J_1^{xy}       &  \hphantom{-}J_1^{zy}     \\
-J_1^{yx}                              & \hphantom{-}J_1^{xx}       &  -J_1^{zx}     \\
\hphantom{-}J_1^{yz}     & -J_1^{xz}        & \hphantom{-}J_1^{zz} 
\end{pmatrix}
\label{Eq:Jcoupling12a}
\\[6px]
\mathsf{J}_{(1, 2)_c}  = &
\begin{pmatrix} 
\hphantom{-}J_1^{xx}     & \hphantom{-}J_1^{xy}       &  -J_1^{xz}     \\
\hphantom{-}J_1^{yx}     & \hphantom{-}J_1^{yy}       &  -J_1^{yz}     \\
-J_1^{zx}     & -J_1^{zy}        & \hphantom{-}J_1^{zz} 
\end{pmatrix}
&
\mathsf{J}_{(1, 2)_d}  = &
\begin{pmatrix} 
\hphantom{-}J_1^{yy}     & -J_1^{xy}       &  -J_1^{zy}     \\
-J_1^{yx}                              & \hphantom{-}J_1^{xx}       &  \hphantom{-}J_1^{zx}     \\
-J_1^{yz}     & \hphantom{-}J_1^{xz}        & \hphantom{-}J_1^{zz} 
\end{pmatrix}
 \\[6px]
\mathsf{J}_{(1, 1)_z} = & 
\begin{pmatrix} 
\hphantom{-}J_2^{xx}     & \hphantom{-}J_2^{xy}       &  \hphantom{-}0     \\
\hphantom{-}J_2^{yx}     & \hphantom{-}J_2^{yy}       &  \hphantom{-}0    \\
\hphantom{-}0     &\hphantom{-}0        & \hphantom{-}J_2^{zz} 
\end{pmatrix} 
&
\mathsf{J}_{(2, 2)_z} = & 
\begin{pmatrix} 
\hphantom{-}J_2^{yy}     & -J_2^{yx}       &  \hphantom{-}0     \\
-J_2^{xy}     & \hphantom{-}J_2^{xx}       &  \hphantom{-}0    \\
\hphantom{-}0     &\hphantom{-}0        & \hphantom{-}J_2^{zz} 
\end{pmatrix} 
 \\[6px]
\mathsf{J}_{(1, 1)_x} = & 
\begin{pmatrix} 
\hphantom{-}J_3^{xx}     & \hphantom{-}J_3^{xy}       &  \hphantom{-}J_3^{xz}     \\
\hphantom{-}J_3^{xy}     & \hphantom{-}J_3^{yy}       &  \hphantom{-}J_3^{yz}     \\
-J_3^{xz}     & -J_3^{yz}        & \hphantom{-}J_3^{zz} 
\end{pmatrix} 
\hspace{40px}&
\mathsf{J}_{(2, 2)_y}  = &
\begin{pmatrix} 
\hphantom{-}J_3^{yy}     & -J_3^{xy}       &  -J_3^{yz}     \\
-J_3^{xy}     & \hphantom{-}J_3^{xx}       &  \hphantom{-}J_3^{xz}     \\
\hphantom{-}J_3^{yz}     & -J_3^{xz}        & \hphantom{-}J_3^{zz} 
\end{pmatrix}
\\[6px]
\mathsf{J}_{(1, 1)_y} = & 
\begin{pmatrix} 
\hphantom{-}J_4^{xx}     & \hphantom{-}J_4^{xy}       &  \hphantom{-}J_4^{xz}     \\
\hphantom{-}J_4^{xy}     & \hphantom{-}J_4^{yy}       &  \hphantom{-}J_4^{yz}     \\
-J_4^{xz}     & -J_4^{yz}        & \hphantom{-}J_4^{zz} 
\end{pmatrix} 
\hspace{40px}&
\mathsf{J}_{(2, 2)_x}  = &
\begin{pmatrix} 
\hphantom{-}J_4^{yy}     & -J_4^{xy}       &  \hphantom{-}J_4^{yz}     \\
-J_4^{xy}     & \hphantom{-}J_4^{xx}       &  -J_4^{xz}     \\
-J_4^{yz}     & \hphantom{-}J_4^{xz}        & \hphantom{-}J_4^{zz} 
\end{pmatrix}
\label{Eq:Jcoupling_75}
\end{align}

We consider now a field $h$ polarized state in the $[001]$ direction and apply the LSW approximation obtaining $\bs{M}(\bs{k})$ with:
\begin{align}
\tilde{\mathsf{J}}_{11} = ~ & ~~~\mathsf{J}_{(1,1)_z} e^{i\bs{k}\cdot(\bs{\delta}_1-\bs{\delta}_{1z})}  +  \mathsf{J}_{(1,1)_x} e^{i\bs{k}\cdot(\bs{\delta}_1-\bs{\delta}_{1x})}  +  \mathsf{J}_{(1,1)_y} e^{i\bs{k}\cdot(\bs{\delta}_1-\bs{\delta}_{1y})}  \nonumber \\
&+ \mathsf{J}_{(1,1)_z}^T e^{i\bs{k}\cdot(\bs{\delta}_{1z} -\bs{\delta}_{1})}  +  \mathsf{J}_{(1,1)_x}^T e^{i\bs{k}\cdot(\bs{\delta}_{1x}-\bs{\delta}_{1})}  +  \mathsf{J}_{(1,1)_y}^T e^{i\bs{k}\cdot(\bs{\delta}_{1y}-\bs{\delta}_{1})}
 \\
\tilde{\mathsf{J}}_{22} = ~ & ~~~\mathsf{J}_{(2,2)_z} e^{i\bs{k}\cdot(\bs{\delta}_2-\bs{\delta}_{2z})}  +  \mathsf{J}_{(2,2)_x} e^{i\bs{k}\cdot(\bs{\delta}_2-\bs{\delta}_{2x})}  +  \mathsf{J}_{(2,2)_y} e^{i\bs{k}\cdot(\bs{\delta}_2-\bs{\delta}_{2y})}  \nonumber \\
&+ \mathsf{J}_{(2,2)_z}^T e^{i\bs{k}\cdot(\bs{\delta}_{2z} -\bs{\delta}_{2})}  +  \mathsf{J}_{(2,2)_x}^T e^{i\bs{k}\cdot(\bs{\delta}_{2x}-\bs{\delta}_{2})}  +  \mathsf{J}_{(2,2)_y}^T e^{i\bs{k}\cdot(\bs{\delta}_{2y}-\bs{\delta}_{2})}
 \\
\tilde{\mathsf{J}}_{12} = ~ & ~~~\mathsf{J}_{(1,2)_a} e^{i\bs{k}\cdot(\bs{\delta}_1-\bs{\delta}_{2})}  +  \mathsf{J}_{(1,2)_b} e^{i\bs{k}\cdot(\bs{\delta}_1-\bs{\delta}_{2y})} + \mathsf{J}_{(1,2)_c} e^{i\bs{k}\cdot(\bs{\delta}_{1x}-\bs{\delta}_{2y})} +  \mathsf{J}_{(1,2)_d} e^{i\bs{k}\cdot(\bs{\delta}_{1x}-\bs{\delta}_{2})}
 \\
\tilde{\mathsf{J}}_{21} = ~ & ~~~\mathsf{J}_{(1,2)_a}^T e^{i\bs{k}\cdot(\bs{\delta}_2-\bs{\delta}_{1})}  +  \mathsf{J}_{(1,2)_b}^T e^{i\bs{k}\cdot(\bs{\delta}_{2y}-\bs{\delta}_{1})} + \mathsf{J}_{(1,2)_c}^T e^{i\bs{k}\cdot(\bs{\delta}_{2y}-\bs{\delta}_{1x})} + \mathsf{J}_{(1,2)_d}^T e^{i\bs{k}\cdot(\bs{\delta}_{2}-\bs{\delta}_{1x})}
\end{align}
in Eqs.~\ref{eq:Ablock} and \ref{eq:Bblock}.

\subsubsection{Band topology}

The orbital basis $(J_{\bs{q}}^+, J_{\bs{q}}^-)$ sits here on the Wyckoff position $2c$ with associated site symmetry group:
\begin{align}
&\bs{q}_1^{2c} = g_1 \, \bs{q}_1^{2c} = (0,1/2,z) ~~~~ G_{\bs{q}_1^{2c}} = \element{2_{001}}{{0, 1, 0}} \nonumber \\
&\bs{q}_2^{2c} = g_2 \, \bs{q}_1^{2c} = (1/2,0,z) ~~~~ G_{\bs{q}_2^{2c}} = \element{2_{001}}{{1, 0, 0}} 
\end{align}
where we have indicated also the orbit transformations $g_1 = \element{E}{{\bf 0}}$ and $g_2 = \element{4^-_{001}}{{\bf 0}}$ forming the coset decomposition $G= \bigcup_\alpha g_{\alpha} (G_{\bs{q}_1^{2c}} \ltimes T)$.
The group $G_{\bs{q}_1^{2c}}$ is therefore isomorphic to $C_2$ and the orbital have representation:
\beq
\rho_{C_2} [(J_{\bs{q}}^+, J_{\bs{q}}^-)] = 2 B
\eeq
which induce a two-bands decomposable elementary band representation (see Table \ref{tab:EBR_75}).
The two bands once splitted produced a topological gap with chiral surface states.
Indeed the system is a Chern insulator with a bulk invariant associated with a non-trivial Wilson loop that can be computed through the symmetry indicated formulas of the point group $C_4$:
\beq
i^C = \prod_{n} \xi_n(\Gamma) \, \xi_n(M) \,\zeta_n(X)
\eeq
where $C$ is the Chern number of the $n$ band(s), while $\xi(\bs{k})$ and $\zeta(\bs{k})$ are the eigenvalues respectively of $C_4$ and $C_2$.
The eigenvalues and irreps of the two bands are:
\begin{align}
& \text{band 1}: ~~~\Gamma_3(C_4^+)=i \,, ~M_1(C_4^+)=1 \,, ~X_2(C_2)=-1 ~~~~\, \rightarrow C=-1 \nonumber \\
& \text{band 2}: ~~~\Gamma_4(C_4^+)=-i \,, ~M_2(C_4^+)=-1 \,, ~X_1(C_2)=1 ~~\rightarrow C=1
\end{align}
which corresponds to the induced $B \uparrow G$ band representation.

\begin{figure}[tp]
  \centering
  \includegraphics[width=0.75\columnwidth]{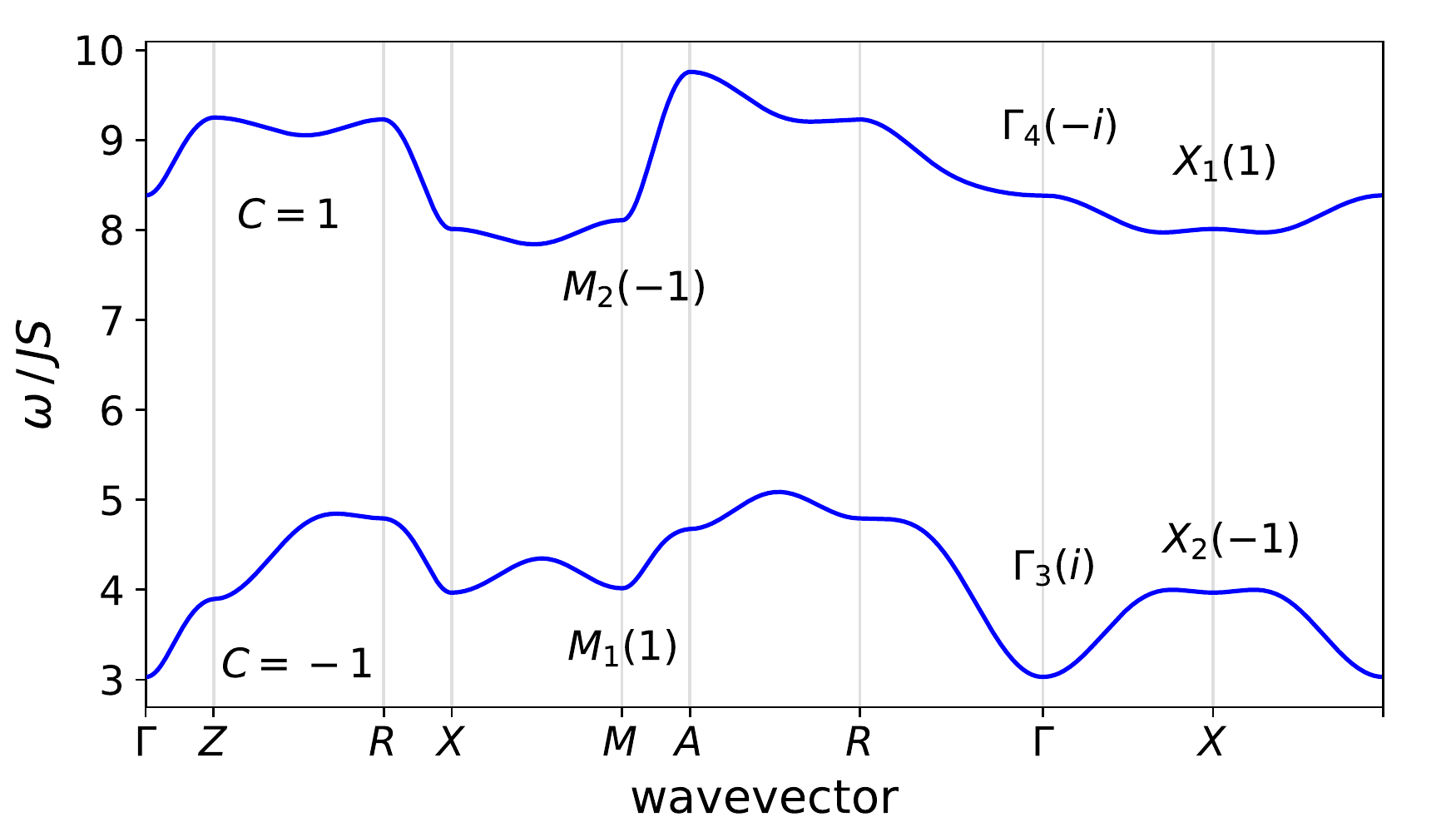}
  \caption{\label{fig:Dispersion75}
Dispersion relations of magnons in the space group $P4$ ($\#75.1$) FM $[001]$ lattice for random strong SO couplings $J_1 + J_2 + J_3 + J_4$ and $h=5$. At the $\Gamma$, $M$ and $X$ point band representations are indicated with their respective $C_4$ or $C_2$ eigenvalue. The band Chern number is indicated on the left, highlighting the topological gap.
}
\end{figure}

\begin{table*}[!htb]
\centering
\hbox to \linewidth{ \hss
\begin{tabular}{|c|c|}
 \hline
  \multicolumn{1}{|c|}{~EBR~}  
  & $B \uparrow G(2)$ \\ 
 \hline
  & Decomposable \\
 \hline
  $\Gamma: (0,0,0)$
 & $\Gamma_3(1) + \Gamma_4(1)$ \\
 $A: (1/2,1/2,1/2)$
 & $A_1(1) + A_2(1)$ \\
 $M: (1/2,1/2,0)$
 & $M_1(1) + M_2(1)$ \\
 $R: (0,1/2,1/2)$
 & $R_1(1) + R_2(1)$ \\
 $X: (0,1/2,0)$
 & $X_1(1) + X_2(1)$ \\
 $Z: (0,0,1/2)$
 & $Z_3(1) + Z_4(1)$ \\
 \hline
\end{tabular}
\hss}
\caption{
Relevant magnon EBR for space group $P4$ ($\# 75.1$) Wyckoff position $2c$ and site symmetry group $2$.  The table shows the representations subduced at the reciprocal space high symmetry points, with the dimension in the parenthesis.
Compatibility relation allow decomposability and therefore a topological gap.
}
\label{tab:EBR_75}
\end{table*}

\section{Nodal topology in magnons}
\label{sec:nodaltopology}

EBRs are also useful in determining the nodal topology of a given magnetic space group. While different EBRs can accidentally cross each other, degeneracies cannot be enforced between them. The only symmetry enforced degeneracies are inside EBRs themselves.
For magnons, all the single-valued EBRs are relevant, although only a subset can be directly induced as single EBR (due to compatibility with magnetic order). Those that cannot be induced in this way must be induced from lower symmetric Wyckoff positions as components of a composite of EBRs.

In Tab.~\ref{tab:nodalEBRstat} the total number of single-valued EBRs with enforced degeneracy is given with rows indicating the nature of the degeneracy (point, line or plane) and columns giving the order of the degeneracy.
Evidently, symmetry-enforced $2$-fold degenerate points are rather common among all EBRs. In addition, $3, 4$ and $6$-fold points can arise even among the singled-valued groups. A greater diversity of nodal topology is possible in the double magnetic groups that are important in spin-orbit coupled electronic systems but not generically for magnons. In addition to point-like degeneracies, we have listed the number of nodal lines and surfaces appearing among the single-valued EBRs.

The complete set of symmetry data required to obtain the various types of nodal topology can be found on the Bilbao Crystallographic Server \cite{bcs1,bcs2}. For convenience we provide tabulation of the the magnetic space group and Wyckoff positions relevant to magnons that enforce the more exotic nodal features:
\begin{itemize}
  \item Nodal point $6$-fold in Tab.~\ref{tab:NodalPoint6fold}.
  \item Nodal point $4$-fold in Tab.~\ref{tab:NodalPoint4fold}.
  \item Nodal point $3$-fold in Tab.~\ref{tab:NodalPoint3fold}.
  \item Nodal lines $4$-fold in  Tab.~\ref{tab:NodalLine4fold}.
  \item Nodal plane $2$-fold in Tab.~\ref{tab:NodalPlane2fold}.
\end{itemize}

Here we mention that if we rel
ax the locking between spin and space and we deal with spin-space groups, generally the degeneracies present in the system are much higher, producing more exotic nodal features which are not possible in magnetic space groups.

\begin{table*}[!htb]
\centering
\begin{tabular}{|c|c|c|c|c|c|}
 \hline
 \multicolumn{1}{|c|}{~~~Nodal~~~}  & ~~~~2~~~~ & ~~~~3~~~~ & ~~~~4~~~~ & ~~~~5~~~~ & ~~~~6~~~~  \\ 
 \hline
Point    & 16414  & 1621  & 3444   & /   & 289\\
Line     & 14151  & /   & 462    & /   & /   \\
Plane  & 5442  & /   & /     & /   & /   \\
 \hline
\end{tabular}
\caption{
Statistics of nodal features for single-valued EBRs. In total the single-valued EBRs are $20,206$. The rows indicated the reciprocal space dimension: high symmetry points, lines and planes. The columns are the order of degeneracy of the nodal feature and the valued listed are the number of different single-valued EBRs. The high symmetric nodal points (lines) embedded in a nodal line (surface) are included in the table.
}
\label{tab:nodalEBRstat}
\end{table*}

\subsection{Nodal Topology: Pyrochlore AIAO Order}
\label{subsec:PyrochloreNodal}

\begin{figure}[!htb]
\hbox to \linewidth{ \hss
\includegraphics[width=0.90\linewidth, trim=0.2cm 3.5cm 0.2cm 0.2cm, clip]{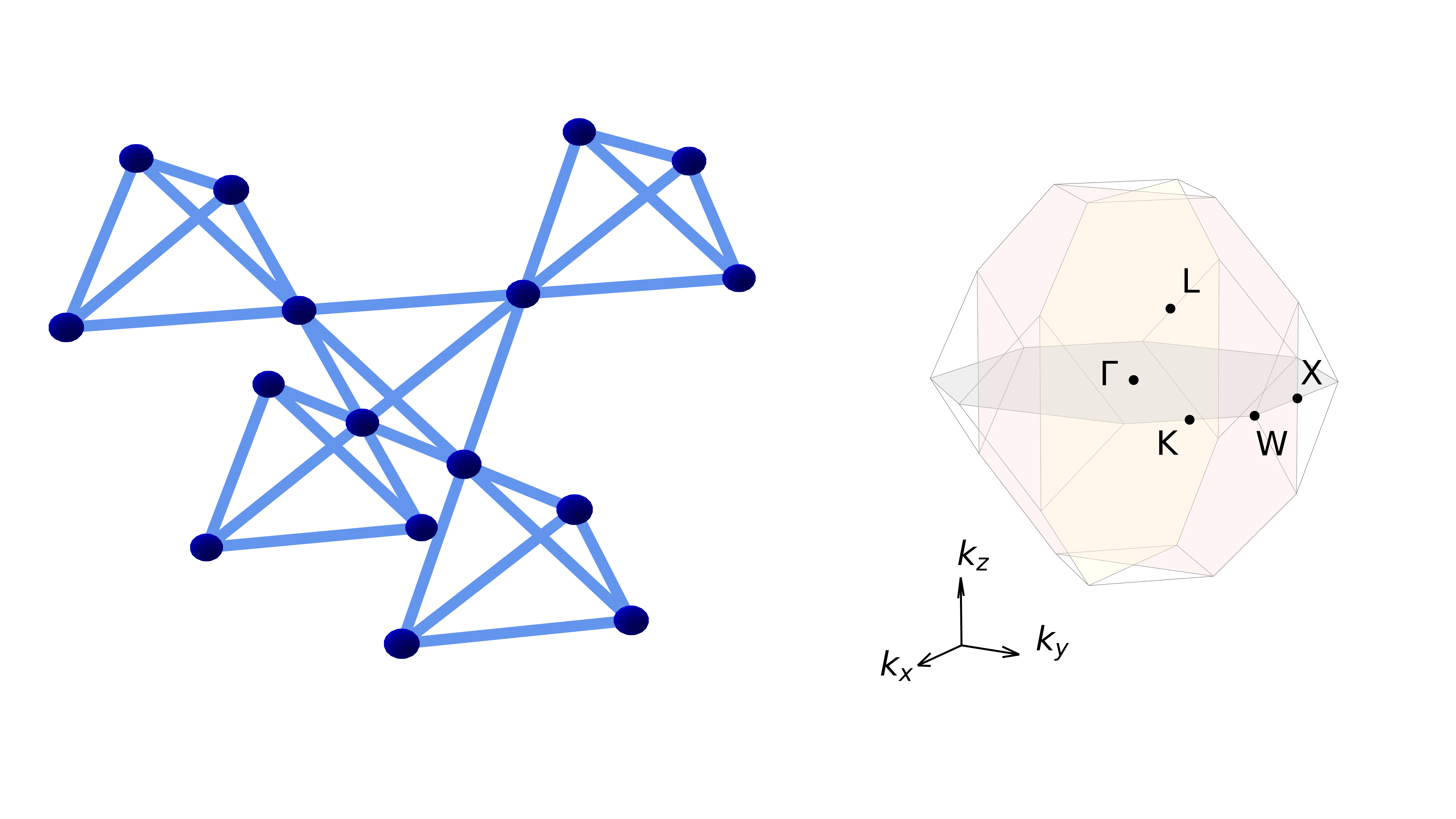}
\hss}
\caption{
The pyrochlore lattice and its first Brillouin Zone (space group $Fd\bar{3}m$ $\# 227$).
In the Brillouin zone the position of the high symmetry points are: $\Gamma =$ $(0,0,0)$, $X =$ $(0,2\pi,0)$,
$L =$ $(\pi,\pi,\pi)$, $W=$ $(\pi, 2\pi,0)$, 
$K =$ $(\frac{3\pi}{2},\frac{3\pi}{2},0)$.
}
\label{Fig:Lattice227}
\end{figure}

As an example of a 3-fold degeneracy we consider the case of the pyrochlore antiferromagnet with all-in/all-out (AIAO) magnetic order. The pyrochlore lattice is a lattice of corner-sharing tetrahedra and the AIAO order has propagation vector $\bs{k}=0$ with moments pointing into or out from the tetrahedral centers. A simple model leading to this magnetic order is the nearest neighbor antiferromagnetic Heisenberg coupling and the symmetry-allowed nearest neighbor Ising exchange 
\beq
H = \sum_{\langle ia,jb \rangle} J\bs{S}_{ia}\cdot \bs{S}_{jb} + K \left( \bs{S}_{ia}\cdot \hat{\bs{z}}_a \right)\left( \bs{S}_{jb}\cdot \hat{\bs{z}}_b \right)
\eeq
where $i,j$ are primitive fcc sites and $a,b$ are tetrahedral sublattice indices, $\hat{\bs{z}}_a$ is the local $\langle 111\rangle $ direction on sublattice $a$ and $J>0$, $K<0$. 

\begin{figure}[tp]
  \centering
  \includegraphics[width=0.75\columnwidth]{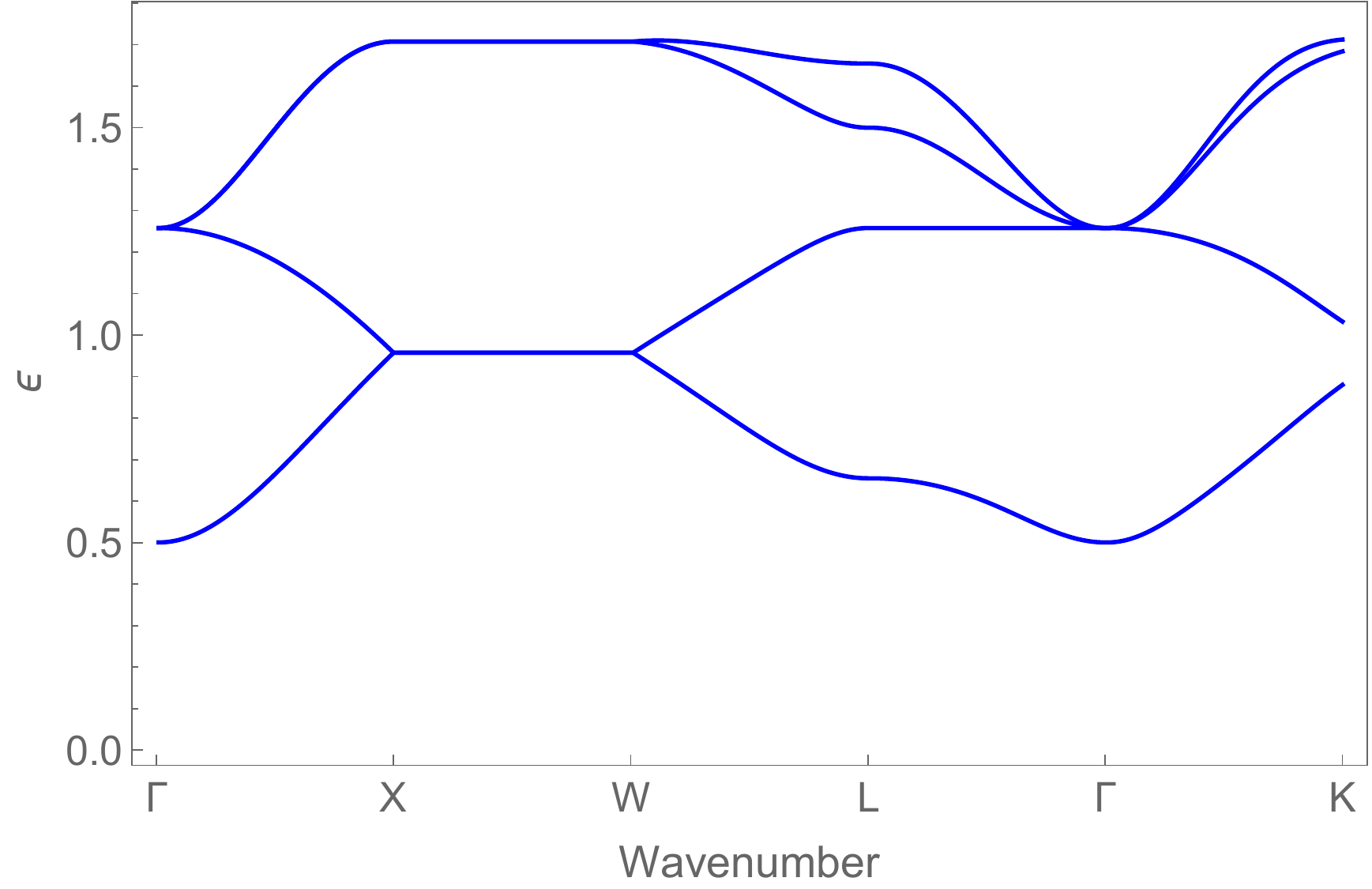}
  \caption{\label{fig:pyrochlore_aiao}
Dispersion relations of magnons in the pyrochlore lattice AIAO phase for $J=1$ and $K=-1/6$. There is a three-fold degenerate point in the upper three bands at $\Gamma$ as well as two two-fold degenerate lines along XW and one nodal line $\Gamma-X$.  
  }
\end{figure}

The magnetic order breaks down the paramagnetic symmetries to a type III magnetic space group, $227.131$ (BNS) that retains the $C_3$ and $C_2$ symmetries of the original lattice but the $2$-fold screw about $[110]$ chains and $S_4$ about $[001]$ survive only in combination with a time reversal operation. These symmetries enforce a $3$-fold degenerate point at $\Gamma$ that is shown in Fig.~\ref{fig:pyrochlore_aiao} \cite{jian2018}. 

\subsection{Nodal Topology: Garnet Structure}

The table of six-fold degenerate points reveals such a symmetry-enforce point for magnetic space group $230.148$ with magnetic structure on the $24$c Wyckoff positions. The lattice is a BCC Bravais lattice with a $12$ site basis.
(0,0,0) + (1/2,1/2,1/2) +
\begin{align*}
(1/8,0,1/4) \hspace{0.5cm} (3/8,0,3/4) \hspace{0.5cm} (1/4,1/8,0) \hspace{0.5cm} (3/4,3/8,0) \\
(0,1/4,1/8)	 \hspace{0.5cm} (0,3/4,3/8) \hspace{0.5cm} (7/8,0,3/4) \hspace{0.5cm} (5/8,0,1/4) \\
(3/4,7/8,0) \hspace{0.5cm} (1/4,5/8,0) \hspace{0.5cm} (0,3/4,7/8) \hspace{0.5cm} (0,1/4,5/8).
\end{align*}
This structure arises in the garnets with formula R$_3$M$_5$O$_12$ on the R sites. The site symmetry of the $230$ space group at $24$c consists of three $C_2$ rotations at each site. The magnetic structure compatible with $230.148$ is shown in Fig.~$2$ in the main text. This structure arises in Dy$_3$Al$_5$O$_12$. 

To compute spin waves for this structure with the requisite symmetries, we should include all symmetry-allowed exchange couplings. To nearest neighbor the $24$ sites split into two independent sublattices of corner sharing triangles $-$ the garnet hyperkagome. The fact that some $C_2$ rotations leave bonds invariant constrains the type of exchange couplings. In the absence of symmetry, there are nine possible couplings per bond. The $C_2$ rotations mix the bonds within each sublattice. This means that the bonds are equivalent and by fixing the interactions on a single bond we may generate the interactions on an entire sublattice. One may show that there are six allowed couplings to nearest neighbor. These are:
\begin{align*} 
(1) \hspace{0.5cm} S_i^y S_j^y \\
(2) \hspace{0.5cm}S_i^x S_j^x + S_i^z S_j^z  \\
(3) \hspace{0.5cm}S_i^x S_j^z \\
(4) \hspace{0.5cm}S_i^z S_j^x  \\
(5) \hspace{0.5cm}S_i^z S_j^y + S_i^y S_j^x \\  
(6) \hspace{0.5cm}S_i^x S_j^y + S_i^y S_j^z
\end{align*}
on the bond joining sites $(1/8,0,1/4)$ and $(0,1/4,1/8)$. The third of these couplings is an effective Ising coupling that fixes the required magnetic structure up to time reversal on the decoupled sublattice. The structure is stable to small additional couplings and we further include a Heisenberg exchange between the two sublattices. 

For a model with the $J_3$ nearest neighbor coupling, antiferromagnetic Heisenberg exchange between nearest neighbors and small additional nearest neighbor couplings, we compute the linear spin wave spectrum. The spectrum along various high symmetry paths is shown in Fig.~$2$ in the main text. Symmetry predicts the following pattern of degeneracies 
\begin{align*}
& \Gamma  = 3,3,3,3 \\
& H(1,1,1)  = 2,2,2,6 \\
& N(1/2,1/2,0)  = 2,2,2,2,2,2 \\
& P(1/2,1/2,1/2)  = 2,2,2,2,2,2 \\
& \Delta(0,v,0)  = 1,1,1,1,2,2,2,2 \\
& G(1+u,1-u,1)  = 2,2,2,2,2,2
\end{align*}
and these are confirmed by direct calculation for the aforementioned couplings. Note the presence of a six-fold degenerate point at $H$ on the zone boundary which is protected by non-symmorphic symmetries and which is clearly visible in the figure. This point is linearly dispersing and can be thought of as a doubled spin-one Weyl point. The $3-$fold degenerate points at $\Gamma$ by contrast are forced to have quadratic dispersion by the inversion symmetry of the lattice.

\bibliography{ebrmagnonrefs}

\clearpage

{\makeatletter
\renewcommand\table@hook{\small}
\renewcommand\arraystretch{0.60}
\makeatother
\begin{table*}[!htb]
\begin{tabular}{|c|c|c|c|c|c|c|c|c|c|}
 \hline
 \textbf{MSGs} & \textbf{MWP} &
\textbf{MSGs} & \textbf{MWP} &
\textbf{MSGs} & \textbf{MWP} &
\textbf{MSGs} & \textbf{MWP}&
\textbf{MSGs} & \textbf{MWP} \\
 \hline
 195.3 &   &  218.82 &  II &  221.97 &   &  223.104 &   &  224.115 &   \\
 200.17 &   &  218.83 &   &  222.98 &  12d, 8c &  223.105 &  II &  230.145 &  24d, 16a \\
 201.21 &  8c &  218.84 &  12d &  222.99 &  II &  223.107 &  8e &  230.146 &  II \\
 207.43 &  12d, 8c &  220.90 &  II &  222.101 &  8c &  223.108 &   &  230.148 &  24c, 16b, 16a \\
 208.47 &   &  220.91 &  16c &  222.102 &   &  223.109 &   &  230.149 &   \\
 215.73 &   &   &   &  222.103 &  12d, 8c &   &   &   &   \\
 \hline
\end{tabular}
\caption{
List of all single-valued magnetic space group (type I,II,III,IV) with $6$-fold high symmetry points, therefore all possible cases in magnons. The first column (MSGs) uses the BNS convention for the magnetic space group. The second (MWP) indicates the maximal Wyckoff position, if the magnon degenerate EBR is directly deducible from it (compatibility with magnetic order).
If blank space then the degenerate EBR can only be induced in a composite magnon BR from a lower symmetric Wyckoff position.
The type $II$ group with pure time-reversal are also listed for completeness, signed with $II$, but they are not relevant for ordered magnonic systems with strong spin-orbit as considered in this paper.
}
\label{tab:NodalPoint6fold}
\end{table*}
}

{\makeatletter
\renewcommand\table@hook{\small}
\renewcommand\arraystretch{0.60}
\makeatother
\begin{table*}[!htb]
\begin{tabular}{|c|c|c|c|c|c|c|c|c|c|}
 \hline
 \textbf{MSGs} & \textbf{MWP} &
\textbf{MSGs} & \textbf{MWP} &
\textbf{MSGs} & \textbf{MWP} &
\textbf{MSGs} & \textbf{MWP}&
\textbf{MSGs} & \textbf{MWP} \\
 \hline
 49.274 &  4c, 4d, 4g &  53.331 &  4b, 4a, 8g &  57.390 &  8c, 8e &  60.429 &  8c, 8e &  125.374 &  8e \\
 49.276 &  4c, 4d &  53.335 &  4b, 4a, 8e &  58.401 &  4c, 4a &  61.434 &  II &  128.408 &  4d, 4a \\
 50.286 &  8d, 8h &  53.336 &  4b, 4a, 4e &  58.402 &  4c, 4b, 4a &  61.435 &   &  129.420 &  8d \\
 50.288 &  8e &  54.349 &  8c, 8e &  59.413 &  8c &  61.438 &  8c, 8d, 8b &  132.458 &   \\
 51.300 &  4b, 4a &  54.352 &  8c, 8d, 8a, 8e &  59.414 &  8d &  62.442 &  II &  133.470 &  8e \\
 51.301 &  4b, 4a &  57.378 &  II &  60.418 &  II &  62.445 &  4c &  136.504 &   \\
 52.316 &  8c, 8d, 8a, 8e &  57.379 &  4c, 4d &  60.420 &  4c &  62.452 &  8c, 8d, 8b &  137.516 &  8d \\
 52.318 &  4c, 8d &  57.386 &  8c, 8d, 8a &  60.427 &  8c, 8d, 8a &  124.362 &  4c, 4d &  205.34 &  II \\
 53.330 &  4c, 4a, 4f, 4e &   &   &   &   &   &   &   &   \\
 \hline
\end{tabular}
\caption{
List of all single-valued magnetic space group (type I,II,III,IV) with $4$-fold high symmetry lines, therefore all possible cases in magnons. The first column (MSGs) uses the BNS convention for the magnetic space group. The second (MWP) indicates the maximal Wyckoff position, if the magnon degenerate EBR is directly deducible from it (compatibility with magnetic order).
If blank space then the degenerate EBR can only be induced in a composite magnon BR from a lower symmetric Wyckoff position.
The type $II$ group with pure time-reversal are also listed for completeness, signed with $II$, but they are not relevant for ordered magnonic systems with strong spin-orbit as considered in this paper.
}
\label{tab:NodalLine4fold}
\end{table*}
}

\clearpage

{\makeatletter
\renewcommand\table@hook{\small}
\renewcommand\arraystretch{0.60}
\makeatother
\begin{table*}[!htb]
\begin{tabular}{|c|c|c|c|c|c|c|c|}
 \hline
 \textbf{MSGs} & \textbf{MWP} &
\textbf{MSGs} & \textbf{MWP} &
\textbf{MSGs} & \textbf{MWP} &
\textbf{MSGs} & \textbf{MWP} \\
 \hline
 195.1 &   &  206.39 &  24d &  216.76 &   &  224.113 &  4c, 4b, 12f \\
 195.2 &  II &  207.40 &   &  216.77 &   &  224.114 &   \\
 195.3 &   &  207.41 &  II &  217.78 &  12d &  224.115 &   \\
 196.4 &   &  207.42 &   &  217.79 &  II &  225.116 &   \\
 196.5 &  II &  207.43 &  8c, 12d &  217.80 &   &  225.117 &  II \\
 196.6 &  24c, 24d &  208.44 &   &  218.81 &  6c, 6d &  225.118 &   \\
 197.7 &   &  208.45 &  II &  218.82 &  II &  225.119 &  24d \\
 197.8 &  II &  208.46 &  4c, 4b, 6e, 6f &  218.83 &   &  225.120 &   \\
 198.9 &  4a &  208.47 &   &  218.84 &  12d &  225.121 &   \\
 198.10 &  II &  209.48 &   &  219.85 &  24c, 24d &  226.122 &  24d \\
 198.11 &  12b, 8a &  209.49 &  II &  219.86 &  II &  226.123 &  II \\
 199.12 &  12b, 8a &  209.50 &  24d &  219.87 &   &  226.124 &  24c \\
 199.13 &  II &  209.51 &  24d &  219.88 &  24c, 24d &  226.125 &   \\
 200.14 &   &  210.52 &   &  220.89 &  12a, 12b, 16c &  226.126 &   \\
 200.15 &  II &  210.53 &  II &  220.90 &  II &  226.127 &  24c \\
 200.16 &   &  210.54 &  16d, 16c &  220.91 &  16c &  227.128 &   \\
 200.17 &   &  210.55 &  48d, 32c, 48f, 48e &  221.92 &   &  227.129 &  II \\
 201.18 &  4c, 4b &  211.56 &   &  221.93 &  II &  227.130 &   \\
 201.19 &  II &  211.57 &  II &  221.94 &   &  227.131 &  16d, 16c \\
 201.20 &   &  211.58 &  8c, 12d &  221.95 &   &  227.132 &   \\
 201.21 &  8c &  212.59 &   &  221.96 &   &  227.133 &  96f \\
 202.22 &  24d &  212.60 &  II &  221.97 &   &  228.134 &  48d, 32c \\
 202.23 &  II &  212.61 &  4b, 4a &  222.98 &  8c, 12d &  228.135 &  II \\
 202.24 &   &  212.62 &  8b, 12d, 12c &  222.99 &  II &  228.136 &  32b, 48d \\
 202.25 &  24c &  213.63 &   &  222.100 &  12d &  228.137 &  32b, 32c \\
 203.26 &  16d, 16c &  213.64 &  II &  222.101 &  8c &  228.138 &   \\
 203.27 &  II &  213.65 &  4b, 4a &  222.102 &   &  228.139 &  32b, 48d, 96f \\
 203.28 &   &  213.66 &  12d, 12c, 8a &  222.103 &  8c, 12d &  229.140 &   \\
 203.29 &  32b, 48d &  214.67 &   &  223.104 &   &  229.141 &  II \\
 204.30 &  8c &  214.68 &  II &  223.105 &  II &  229.142 &  12d \\
 204.31 &  II &  214.69 &  12d, 8b, 12c, 8a &  223.106 &  8e, 6c, 6d &  229.143 &  8c \\
 204.32 &   &  215.70 &   &  223.107 &  8e &  229.144 &   \\
 205.33 &  4b, 4a &  215.71 &  II &  223.108 &   &  230.145 &  24d, 16a \\
 205.34 &  II &  215.72 &   &  223.109 &   &  230.146 &  II \\
 205.35 &   &  215.73 &   &  224.110 &   &  230.147 &  24c, 24d, 16b \\
 205.36 &  24d, 8b &  216.74 &   &  224.111 &  II &  230.148 &  24c, 16b, 16a \\
 206.37 &  24d, 8b, 8a &  216.75 &  II &  224.112 &  12f &  230.149 &   \\
 206.38 &  II &   &   &   &   &   &   \\
 \hline
\end{tabular}
\caption{
List of all single-valued magnetic space group (type I,II,III,IV) with $3$-fold high symmetry points, therefore all possible cases in magnons. The first column (MSGs) uses the BNS convention for the magnetic space group. The second (MWP) indicates the maximal Wyckoff position, if the magnon degenerate EBR is directly deducible from it (compatibility with magnetic order).
If blank space then the degenerate EBR can only be induced in a composite magnon BR from a lower symmetric Wyckoff position.
The type $II$ group with pure time-reversal are also listed for completeness, signed with $II$, but they are not relevant for ordered magnonic systems with strong spin-orbit as considered in this paper.
}
\label{tab:NodalPoint3fold}
\end{table*}
}

\clearpage

{\makeatletter
\renewcommand\table@hook{\small}
\renewcommand\arraystretch{0.60}
\makeatother
\begin{longtable}{|c|c|c|c|c|c|c|c|}
\caption{List of all single-valued magnetic space group (type I,II,III,IV) with $4$-fold high symmetry points, therefore all possible cases in strong SO magnons. The first column (MSGs) uses the BNS convention for the magnetic space group. The second (MWP) indicates the maximal Wyckoff position, if the magnon degenerate EBR is directly deducible from it (compatibility with magnetic order).
If blank space then the degenerate EBR can only be induced in a composite magnon BR from a lower symmetric Wyckoff position.
The type $II$ group with pure time-reversal are also listed for completeness, signed with $II$, but they are not relevant for ordered magnonic systems with strong spin-orbit as considered in this paper.}\\
\hline
 \textbf{MSGs} & \textbf{MWP} &
\textbf{MSGs} & \textbf{MWP} &
\textbf{MSGs} & \textbf{MWP} &
\textbf{MSGs} & \textbf{MWP}\\
\hline
\endfirsthead
\multicolumn{8}{c}%
{\tablename\ \thetable\ -- \textit{Continued from previous page}} \\
\hline
\textbf{MSGs} & \textbf{MWP} &
\textbf{MSGs} & \textbf{MWP} &
\textbf{MSGs} & \textbf{MWP} &
\textbf{MSGs} & \textbf{MWP} \\
\hline
\endhead
\hline \multicolumn{8}{r}{
} \\
\endfoot

 11.55 &  4c, 4a, 4e &  56.374 &  4b, 4a, 8e &  110.246 &  II &  141.560 &  16d, 16e, 16c \\
 13.71 &  4d, 4a, 4f, 4e &  57.378 &  II &  110.247 &  8a &  142.562 &  II \\
 13.74 &  4b, 4a, 4e &  57.379 &  4c, 4d &  111.256 &  4f &  142.564 &  16e, 16c \\
 14.80 &  4c, 4a &  57.382 &  4c, 4b, 4a, 4d &  111.258 &   &  142.566 &  8b, 16e, 8a \\
 17.11 &  4c, 4b, 4a, 4d &  57.383 &  4c, 4b, 4a, 4d &  112.265 &  8i &  142.569 &  16e \\
 17.14 &  4b, 4a &  57.386 &  8c, 8d, 8a &  112.266 &   &  156.52 &   \\
 17.15 &  4c, 4b, 4a &  57.387 &  4c, 4d, 4f, 4e &  113.272 &  4a &  157.56 &  4b \\
 18.21 &  4b, 4a &  57.390 &  8c, 8e &  113.274 &  4c, 4d &  158.58 &  II \\
 18.22 &  4b, 4a &  57.391 &  4c, 8d &  114.276 &  II &  159.62 &  II \\
 19.26 &  II &  57.392 &  4b, 4a, 8e &  114.280 &  4c, 4b &  160.68 &   \\
 19.28 &  4b, 4a &  58.400 &  4c, 4d, 8g &  115.288 &   &  161.70 &  II \\
 28.93 &  4c, 4b, 4a &  58.401 &  4c, 4a &  115.290 &   &  162.78 &  4d, 6f \\
 28.94 &  4c, 4b, 4a &  58.402 &  4c, 4b, 4a &  116.298 &  4d &  163.80 &  II \\
 28.95 &  4b, 4a &  58.403 &  4c, 4b, 4a, 4d &  117.304 &  4d, 4a &  163.82 &   \\
 28.96 &  4c, 4b, 4a &  59.412 &  4d, 4b, 4f &  117.306 &  4b, 4a &  164.90 &  6e \\
 28.98 &  4b, 4a &  59.413 &  8c &  118.312 &  4d, 4b &  165.92 &  II \\
 29.100 &  II &  59.414 &  8d &  118.313 &  4c, 4b, 4d, 4f &  165.94 &  4d \\
 29.104 &  4b, 4a &  60.418 &  II &  123.348 &   &  166.102 &  18e \\
 29.105 &  8a &  60.420 &  4c &  123.350 &   &  167.104 &  II \\
 29.107 &  4a &  60.422 &  4c, 4b, 4a &  124.352 &  II &  167.106 &   \\
 29.109 &  4a &  60.423 &  4c, 4b, 4a &  124.354 &  4f, 4e &  176.144 &  II \\
 29.110 &  4b, 4a &  60.424 &  4c, 4b, 4a &  124.355 &  4e &  176.148 &  4c, 6g \\
 30.116 &  4b, 4a &  60.425 &  4c &  124.359 &   &  183.190 &   \\
 30.117 &  4c, 4b, 4a, 4d &  60.426 &  8c, 8d, 8b, 8e &  124.361 &  4b, 8f &  184.192 &  II \\
 30.118 &  4c, 4b, 4a &  60.427 &  8c, 8d, 8a &  124.362 &  4c, 4d &  184.193 &  4b, 6c \\
 30.120 &  4a &  60.428 &  4c, 4d, 8g &  125.372 &  4b, 8e &  184.194 &  4b, 6c \\
 30.121 &  4c, 4b, 4a &  60.429 &  8c, 8e &  125.374 &  8e &  185.198 &  II \\
 30.122 &  4b, 4a &  60.430 &  4b, 4a, 8h, 8c &  126.376 &  II &  185.199 &  4b \\
 31.128 &  4b, 4a &  60.431 &  4c, 4b, 4a &  126.378 &  4c, 8f &  185.202 &  4b, 6c \\
 31.129 &  4a &  60.432 &  4c, 4b, 4a, 4d &  126.379 &  8f &  186.204 &  II \\
 31.130 &  4c, 4b, 4a &  61.434 &  II &  126.383 &   &  186.206 &   \\
 31.131 &  4b, 4a &  61.435 &   &  126.384 &  4c, 8e, 4a &  186.208 &  6c \\
 31.133 &  4a &  61.436 &  4b, 4a &  126.385 &  8e, 8f &  187.214 &   \\
 32.139 &  4b, 4a &  61.437 &   &  126.386 &  8f &  188.216 &  II \\
 32.142 &  4a &  61.438 &  8c, 8d, 8b &  127.396 &  4a &  188.217 &   \\
 32.143 &  4b, 4a &  61.439 &  4b, 4a, 8e &  127.398 &  4c, 4a &  189.226 &  4c \\
 33.145 &  II &  61.440 &  8c, 8d, 8b, 8e &  128.400 &  II &  190.228 &  II \\
 33.149 &  4a &  62.442 &  II &  128.402 &  4c &  190.230 &   \\
 33.150 &  8a &  62.445 &  4c &  128.403 &  4c, 4d &  191.242 &   \\
 33.151 &  4b, 4a &  62.446 &  4c, 4b, 4a &  128.406 &   &  192.244 &  II \\
 33.153 &  4a &  62.447 &  4c, 4b, 4a &  128.407 &   &  192.246 &  4c, 4d, 6f \\
 33.154 &  4a &  62.448 &  4c, 4b, 4a &  128.408 &  4d, 4a &  192.247 &  4d, 6f \\
 34.160 &  4b, 4a &  62.450 &  4b, 4a, 8d &  128.409 &  4c, 4d, 8f &  192.248 &  6g, 6f \\
 34.161 &  4b, 4a &  62.451 &  4d, 4b &  129.420 &  8d &  192.249 &  4c, 6g, 6f \\
 34.162 &  4b, 4a &  62.452 &  8c, 8d, 8b &  129.422 &  8f &  192.251 &   \\
 34.163 &  4c, 4b, 4a &  62.454 &  4c, 4b, 4a &  130.424 &  II &  193.253 &  4c, 6f \\
 48.262 &  4b, 4a, 8e &  62.455 &  8c, 8e &  130.426 &  8d &  193.254 &  II \\
 48.263 &  4f, 4e &  63.467 &  8d, 8a, 8f &  130.427 &  4a, 8d &  193.256 &  4c, 4d \\
 49.272 &  4c, 4a, 4f, 4h &  67.508 &  8c, 8e, 8b &  130.428 &  4b, 4a &  193.258 &  6f \\
 49.274 &  4c, 4d, 4g &  68.518 &  8c, 8e, 8g, 8a &  130.429 &  4c, 4b, 4a, 8d &  193.261 &   \\
 49.275 &  4c, 4d &  73.549 &  II &  130.430 &   &  193.262 &  4c, 6g \\
 49.276 &  4c, 4d &  73.552 &  8c, 8d, 8e &  130.431 &  4c &  194.263 &  6g \\
 50.284 &  4d, 4b, 4f, 4h &  74.561 &  8e, 8d, 8b, 8a &  130.432 &  4c, 4a, 8d &  194.264 &  II \\
 50.285 &  4c, 8e, 4d &  84.56 &  4e &  130.433 &  8e, 4a, 8f &  194.267 &   \\
 50.286 &  8d, 8h &  84.58 &  4c, 4d &  130.434 &  8e, 4a &  194.269 &  6g \\
 50.288 &  8e &  85.64 &  4c, 4a, 8d &  131.444 &   &  194.271 &   \\
 51.299 &  4c, 4a, 4f &  85.66 &  8f &  131.445 &  8e &  194.272 &  6g \\
 51.300 &  4b, 4a &  86.72 &  4b, 8d &  131.446 &   &  198.10 &  II \\
 51.301 &  4b, 4a &  86.74 &  8f &  132.456 &  4f &  198.11 &  12b, 8a \\
 51.303 &  4b, 4a, 4f, 4e &  89.94 &   &  132.457 &  4b &  205.34 &  II \\
 51.304 &  4b, 4a &  90.100 &  4c, 4b &  132.458 &   &  205.35 &   \\
 52.306 &  II &  91.109 &  8c, 8b, 8a &  133.460 &  II &  205.36 &  24d, 8b \\
 52.310 &  4c, 4b, 4a, 4d &  91.110 &  4b, 4a, 8f &  133.462 &  4b, 8e &  206.38 &  II \\
 52.311 &  4c, 4b, 4a, 4d &  92.112 &  II &  133.463 &  8e &  206.39 &  24d \\
 52.313 &  4c, 4d &  92.115 &  4a &  133.467 &   &  212.59 &   \\
 52.314 &  4b, 4a, 8g &  92.116 &  4b, 4a, 8d &  133.468 &  4d, 8e &  212.60 &  II \\
 52.315 &  4c, 4d, 8f, 4b, 4a &  93.126 &   &  133.469 &  8e, 8f &  212.62 &  8b, 12d, 12c \\
 52.316 &  8c, 8d, 8a, 8e &  94.132 &  4b &  133.470 &  8e &  213.63 &   \\
 52.317 &  4c, 4b, 4a, 4d &  95.141 &  8c, 8b, 8a &  134.480 &  8e &  213.64 &  II \\
 52.318 &  4c, 8d &  95.142 &  4b, 4a, 8f &  134.481 &  4c, 4d &  213.66 &  12d, 12c, 8a \\
 52.319 &  4b, 4a, 8d, 8h &  96.144 &  II &  134.482 &  8f &  215.73 &   \\
 52.320 &  4b, 4a, 4e &  96.147 &  4a &  135.484 &  II &  216.77 &   \\
 53.330 &  4c, 4a, 4f, 4e &  96.148 &  4b, 4a, 8d &  135.486 &  4c, 4a &  218.82 &  II \\
 53.331 &  4b, 4a, 8g &  99.168 &   &  135.489 &  4c, 4b, 4a, 4d &  219.86 &  II \\
 53.332 &  4c, 4f, 4g, 4h, 4a, 4e &  99.170 &   &  135.490 &   &  220.89 &  12a, 12b, 16c \\
 53.333 &  4b, 4a, 4f, 4e &  100.176 &  4a &  135.491 &   &  220.90 &  II \\
 53.335 &  4b, 4a, 8e &  100.178 &  4a &  135.492 &  4d &  221.97 &   \\
 53.336 &  4b, 4a, 4e &  101.184 &  4c &  135.494 &  4d, 4b &  222.99 &  II \\
 54.338 &  II &  101.186 &   &  136.504 &   &  222.101 &  8c \\
 54.342 &  4c, 4d, 4b, 4a, 4e &  102.192 &   &  136.505 &  4a, 4e &  222.102 &   \\
 54.344 &  4c, 4d, 4b, 4a, 4e &  102.193 &  4c, 4b &  136.506 &  4c, 4d &  223.105 &  II \\
 54.345 &  4c, 4d, 4e &  102.194 &  4b &  137.508 &  II &  223.108 &   \\
 54.347 &  8c, 8d, 8a, 8e &  103.196 &  II &  137.510 &  8e &  223.109 &   \\
 54.348 &  4b, 4a, 4f, 4e &  103.201 &  8c, 4a &  137.511 &  8e &  224.115 &   \\
 54.349 &  8c, 8e &  103.202 &  4a &  137.514 &   &  225.121 &   \\
 54.351 &  4b, 4a, 8h, 8c &  104.204 &  II &  137.516 &  8d &  226.123 &  II \\
 54.352 &  8c, 8d, 8a, 8e &  104.208 &  4b, 4a &  137.517 &  8e, 8f &  226.126 &   \\
 55.361 &  4c, 4a &  104.209 &  4b, 8c &  137.518 &  8f &  227.133 &  96f \\
 55.362 &  4b, 4a, 8e &  105.216 &   &  138.520 &  II &  228.135 &  II \\
 55.364 &  4c, 4b, 4a, 4d &  105.217 &  8c &  138.522 &  4c, 4d &  228.137 &  32b, 32c \\
 56.366 &  II &  105.218 &   &  138.524 &  4b, 4a &  228.138 &   \\
 56.369 &  4c, 4b, 4a, 4d &  106.220 &  II &  138.525 &  4c, 4d, 4b, 4a, 4e &  230.145 &  24d, 16a \\
 56.370 &  4c, 4b, 4a, 4d &  106.224 &  4b &  138.527 &  4e &  230.146 &  II \\
 56.371 &  4c, 4d &  106.226 &  4b &  138.528 &  4b, 8d &  230.147 &  24c, 24d, 16b \\
 56.372 &  8c, 8d, 8b, 8e &  109.244 &  16b &  138.530 &  8e &  230.149 &   \\
 56.373 &  4b, 4a, 8c &   &   &   &   &   &   \\
\label{tab:NodalPoint4fold}
\end{longtable}
} 

\clearpage

{\makeatletter
\renewcommand\table@hook{\small}
\renewcommand\arraystretch{0.60}
\makeatother
\begin{longtable}{|c|c|c|c|c|c|c|c|}
\caption{List of all single-valued magnetic space group (type I,II,III,IV) with $2$-fold high symmetry planes, therefore all possible cases in magnons. The first column (MSGs) uses the BNS convention for the magnetic space group. The second (MWP) indicates the maximal Wyckoff position, if the magnon degenerate EBR is directly deducible from it (compatibility with magnetic order).
If blank space then the degenerate EBR can only be induced in a composite magnon BR from a lower symmetric Wyckoff position.
The type $II$ group with pure time-reversal are also listed for completeness, signed with $II$, but they are not relevant for ordered magnonic systems with strong spin-orbit as considered in this paper.}\\
\hline
 \textbf{MSGs} & \textbf{MWP} &
\textbf{MSGs} & \textbf{MWP} &
\textbf{MSGs} & \textbf{MWP} &
\textbf{MSGs} & \textbf{MWP}\\
\hline
\endfirsthead
\multicolumn{8}{c}%
{\tablename\ \thetable\ -- \textit{Continued from previous page}} \\
\hline
\textbf{MSGs} & \textbf{MWP} &
\textbf{MSGs} & \textbf{MWP} &
\textbf{MSGs} & \textbf{MWP} &
\textbf{MSGs} & \textbf{MWP} \\
\hline
\endhead
\hline 
\endfoot

 3.5 &  2c, 2a, 2d, 2b &  51.304 &  4b, 4a &  64.471 &  8e &  128.400 &  II \\
 3.6 &  2a, 2b &  52.306 &  II &  64.472 &  8e &  128.401 &  4d \\
 4.8 &  II &  52.307 &  4c, 4d &  64.475 &  4b, 4a, 8e, 8c &  128.402 &  4c \\
 4.9 &  2a &  52.309 &  4c, 4d &  64.476 &  4b, 4a, 8e, 8c &  128.405 &  4c, 4d, 2a, 2b \\
 4.10 &  4a &  52.310 &  4c, 4b, 4a, 4d &  64.479 &  8c, 8e, 8b, 8f &  128.406 &   \\
 10.48 &  2c, 2g, 2a, 2d &  52.311 &  4c, 4b, 4a, 4d &  65.488 &  8e &  128.408 &  4d, 4a \\
 10.49 &  2c, 2a, 2d, 2b &  52.314 &  4b, 4a, 8g &  65.490 &  8e, 8f &  128.410 &  4c, 2a, 2b \\
 11.51 &  II &  52.316 &  8c, 8d, 8a, 8e &  66.498 &  4b, 4a, 8e &  129.412 &  II \\
 11.52 &  2e &  52.317 &  4c, 4b, 4a, 4d &  66.500 &  4b, 4a, 8e &  129.413 &   \\
 11.54 &  2d, 2a, 2b, 2c, 2e &  52.318 &  4c, 8d &  67.508 &  8c, 8e, 8b &  129.414 &  4d, 4e \\
 11.55 &  4c, 4a, 4e &  52.319 &  4b, 4a, 8d, 8h &  67.510 &  8c, 8d &  129.417 &  4d, 2a, 2b, 2c, 4e \\
 13.71 &  4d, 4a, 4f, 4e &  52.320 &  4b, 4a, 4e &  68.518 &  8c, 8e, 8g, 8a &  129.418 &   \\
 13.73 &  4f, 4e &  53.322 &  II &  68.520 &  8c, 8d, 8f &  129.420 &  8d \\
 13.74 &  4b, 4a, 4e &  53.323 &  4g &  75.4 &  4c, 2a, 2b &  129.422 &  8f \\
 14.76 &  II &  53.324 &  4g &  75.6 &  4b, 2a &  130.424 &  II \\
 14.77 &   &  53.327 &  2d, 4g, 2a, 2b, 2c &  76.8 &  II &  130.425 &  4c, 4a \\
 14.79 &  2c, 2a, 2d, 2b &  53.328 &  2d, 4g, 2a, 2b, 2c &  76.11 &  8a &  130.426 &  8d \\
 14.80 &  4c, 4a &  53.330 &  4c, 4a, 4f, 4e &  77.16 &  4c &  130.429 &  4c, 4b, 4a, 8d \\
 14.82 &  4c, 4d, 4e &  53.331 &  4b, 4a, 8g &  77.18 &  4b &  130.430 &  4b \\
 16.4 &  2f, 2e, 2h, 2b &  53.333 &  4b, 4a, 4f, 4e &  78.20 &  II &  130.432 &  4c, 4a, 8d \\
 16.5 &  4k &  53.334 &  2c, 2a, 2d, 2b &  78.23 &  8a &  130.434 &  4b, 4a, 8e \\
 16.6 &   &  53.335 &  4b, 4a, 8e &  81.36 &  2c, 4g, 2a &  131.444 &   \\
 17.8 &  II &  53.336 &  4b, 4a, 4e &  81.38 &  2a, 2b &  131.445 &  8e \\
 17.10 &  2c, 2a, 2d, 2b &  54.338 &  II &  83.48 &  2c, 2a, 4e &  131.446 &   \\
 17.11 &  4c, 4b, 4a, 4d &  54.340 &  4c, 4d, 4e &  83.50 &  4c, 2a, 2b &  132.456 &  4f \\
 17.13 &  2d, 4k, 2a, 2b, 2c &  54.341 &  4c, 4d, 4e &  84.56 &  4e &  132.457 &  4b \\
 17.14 &  4b, 4a &  54.342 &  4c, 4d, 4b, 4a, 4e &  84.58 &  4c, 4d &  132.458 &   \\
 17.15 &  4c, 4b, 4a &  54.344 &  4c, 4d, 4b, 4a, 4e &  85.64 &  4c, 4a, 8d &  133.468 &  4d, 8e \\
 18.17 &  II &  54.347 &  8c, 8d, 8a, 8e &  85.66 &  4d, 8f &  133.469 &  4b, 8e, 8f \\
 18.18 &  2a, 2b &  54.348 &  4b, 4a, 4f, 4e &  86.72 &  4b, 8d &  133.470 &  8e \\
 18.19 &  2a, 2b &  54.349 &  8c, 8e &  86.74 &  8f &  134.480 &  8e \\
 18.20 &  4c, 4b, 4a, 4d &  54.350 &  4g, 4f, 4b, 4a, 4e &  89.92 &  2d, 4f, 2b &  134.481 &  4c, 4d \\
 18.21 &  4b, 4a &  54.351 &  4b, 4a, 8h, 8c &  89.93 &  4f, 4e &  134.482 &  8f \\
 18.22 &  4b, 4a &  54.352 &  8c, 8d, 8a, 8e &  89.94 &  4d &  135.484 &  II \\
 18.24 &  2c, 2a, 2d, 2b &  55.354 &  II &  90.96 &  II &  135.485 &  4d \\
 19.26 &  II &  55.355 &   &  90.98 &  2c, 2a, 2b &  135.486 &  4c, 4a \\
 19.27 &  4a &  55.356 &   &  90.99 &   &  135.489 &  4c, 4b, 4a, 4d \\
 19.28 &  4b, 4a &  55.357 &  2c, 2a, 2d, 2b &  90.100 &  4c, 4b &  135.490 &  4b \\
 19.29 &  4b, 4a &  55.358 &  2c, 2a, 2d, 2b &  90.102 &  4c, 2a, 2b &  135.492 &  4d \\
 20.32 &  II &  55.360 &  4c, 4d, 4f, 4e &  91.104 &  II &  135.494 &  4d, 4b \\
 20.34 &  4b, 4a &  55.361 &  4c, 4a &  91.109 &  8c, 8b, 8a &  136.496 &  II \\
 20.36 &  8c, 8d, 8b, 8a &  55.362 &  4b, 4a, 8e &  91.110 &  4b, 4a, 8f &  136.497 &   \\
 21.42 &  4c, 4d, 8k &  55.364 &  4c, 4b, 4a, 4d &  92.112 &  II &  136.498 &  4c \\
 21.44 &  4c, 4d &  56.366 &  II &  92.114 &  4a &  136.501 &  4c, 4d, 2a, 2b \\
 25.61 &   &  56.367 &  4c, 4d &  92.115 &  4a &  136.502 &  4d \\
 25.64 &   &  56.368 &  4c, 4d &  92.116 &  4b, 4a, 8d &  136.504 &   \\
 25.65 &   &  56.369 &  4c, 4b, 4a, 4d &  92.117 &  8c, 8b, 8a &  136.506 &  4c, 4d \\
 26.67 &  II &  56.370 &  4c, 4b, 4a, 4d &  93.124 &  4f &  137.508 &  II \\
 26.68 &  2a, 2b &  56.372 &  8c, 8d, 8b, 8e &  93.125 &  4c, 4b, 4d &  137.509 &   \\
 26.69 &  2a, 2b &  56.373 &  4b, 4a, 8c &  93.126 &   &  137.510 &  4d, 8e \\
 26.71 &  4b, 4a &  56.374 &  4b, 4a, 8e &  94.128 &  II &  137.513 &  4d, 8e, 2a, 2b \\
 26.72 &  4b, 4a &  56.376 &  4b, 4a, 8e &  94.130 &  4d, 2a, 2b &  137.514 &   \\
 26.76 &  4a &  57.378 &  II &  94.131 &  4d &  137.516 &  8d \\
 27.82 &  2c, 2a, 2d, 2b &  57.379 &  4c, 4d &  94.132 &  4b &  137.518 &  8f \\
 27.85 &  4c, 4b, 4a &  57.380 &  4c, 4d &  94.134 &  4c, 4d &  138.520 &  II \\
 27.86 &  4b, 4a &  57.381 &  4c, 4d &  95.136 &  II &  138.521 &  4a \\
 28.94 &  4c, 4b, 4a &  57.382 &  4c, 4b, 4a, 4d &  95.141 &  8c, 8b, 8a &  138.522 &  4c, 4d \\
 28.95 &  4b, 4a &  57.383 &  4c, 4b, 4a, 4d &  95.142 &  4b, 4a, 8f &  138.525 &  4c, 4d, 4b, 4a, 4e \\
 28.96 &  4c, 4b, 4a &  57.384 &  4c, 4b, 4a, 4d &  96.144 &  II &  138.526 &  4b, 4e \\
 28.98 &  4b, 4a &  57.386 &  8c, 8d, 8a &  96.146 &  4a &  138.528 &  4b, 8d \\
 29.100 &  II &  57.387 &  4c, 4d, 4f, 4e &  96.147 &  4a &  138.530 &  8e \\
 29.101 &  4a &  57.388 &  4d, 4b, 4f, 4e &  96.148 &  4b, 4a, 8d &  168.112 &  4b, 2a, 6c \\
 29.102 &  4a &  57.390 &  8c, 8e &  96.149 &  8c, 8b, 8a &  169.114 &  II \\
 29.104 &  4b, 4a &  57.391 &  4c, 8d &  99.168 &   &  169.115 &  6a \\
 29.105 &  8a &  57.392 &  4b, 4a, 8e &  99.170 &   &  170.118 &  II \\
 29.109 &  4a &  58.394 &  II &  100.176 &  4a &  170.119 &  6a \\
 30.118 &  4c, 4b, 4a &  58.395 &   &  100.178 &  4a &  171.124 &  6b, 6a \\
 30.119 &  2a, 2b &  58.396 &   &  101.184 &  4c &  172.128 &  6b, 6a \\
 30.120 &  4a &  58.397 &  2c, 2a, 2d, 2b &  101.186 &   &  173.130 &  II \\
 30.122 &  4b, 4a &  58.398 &  2c, 2a, 2d, 2b &  102.192 &   &  173.131 &  2a, 2b \\
 31.124 &  II &  58.400 &  4c, 4d, 8g &  102.194 &  4b &  175.142 &  4c, 2a, 6f \\
 31.125 &  2a &  58.401 &  4c, 4a &  103.200 &  4c, 2a, 2b &  176.144 &  II \\
 31.126 &  2a &  58.402 &  4c, 4b, 4a &  103.202 &  4b, 4a &  176.145 &  2c, 2a, 2d \\
 31.128 &  4b, 4a &  58.404 &  2c, 2a, 2d, 2b &  104.208 &  4b, 4a &  176.147 &  6g, 2b \\
 31.129 &  4a &  59.406 &  II &  104.210 &  4b, 2a &  177.154 &  4d, 6g, 2b \\
 31.133 &  4a &  59.407 &  2a, 2b &  105.216 &   &  178.156 &  II \\
 32.139 &  4b, 4a &  59.408 &   &  105.218 &   &  178.157 &  6b, 6a \\
 32.142 &  4a &  59.409 &  4c, 4d, 2a, 2b &  106.224 &  4b &  178.158 &  6b, 6a \\
 32.143 &  4b, 4a &  59.410 &  4c, 4d, 2a, 2b &  106.226 &  4b &  179.162 &  II \\
 33.145 &  II &  59.412 &  4d, 4b, 4f &  111.256 &  4f &  179.163 &  6b, 6a \\
 33.146 &  4a &  59.413 &  8c &  111.257 &  4g &  179.164 &  6b, 6a \\
 33.147 &  4a &  59.414 &  8d &  111.258 &   &  180.172 &  6b, 6d \\
 33.149 &  4a &  59.416 &  8k &  112.264 &  2d, 2a, 4e &  181.178 &  6b, 6d \\
 33.150 &  8a &  60.418 &  II &  112.265 &  4c, 8i &  182.180 &  II \\
 33.154 &  4a &  60.419 &  4c &  112.266 &  4d &  182.181 &  2a \\
 34.161 &  4b, 4a &  60.420 &  4c &  113.268 &  II &  182.182 &  2c, 2d, 2b \\
 34.162 &  4b, 4a &  60.421 &  4c &  113.269 &   &  183.190 &   \\
 34.164 &  2a, 2b &  60.422 &  4c, 4b, 4a &  113.271 &  2c, 2a, 2b &  184.196 &  4b, 2a, 6c \\
 35.169 &  8c &  60.423 &  4c, 4b, 4a &  113.272 &  4a &  185.198 &  II \\
 35.171 &  8b &  60.424 &  4c, 4b, 4a &  113.274 &  4c, 4d &  185.199 &  4b \\
 36.173 &  II &  60.426 &  8c, 8d, 8b, 8e &  114.276 &  II &  185.200 &  4b, 2a \\
 36.174 &  4a &  60.427 &  8c, 8d, 8a &  114.277 &  4d &  186.204 &  II \\
 36.175 &  4a &  60.428 &  4c, 4d, 8g &  114.279 &  4d, 2a, 2b &  186.205 &  2a, 2b \\
 36.178 &  8b, 8a &  60.429 &  8c, 8e &  114.280 &  4c, 4b &  186.206 &   \\
 37.184 &  4b, 4a, 8c &  60.431 &  4c, 4b, 4a &  114.282 &  4c, 2a, 2b &  191.242 &   \\
 37.186 &  4a, 8b &  60.432 &  4c, 4b, 4a, 4d &  115.288 &   &  192.252 &  4c, 2a, 6f \\
 47.254 &   &  61.434 &  II &  115.290 &   &  193.254 &  II \\
 47.255 &   &  61.435 &   &  116.296 &  2c, 4g, 2d &  193.256 &  4c, 4d \\
 47.256 &   &  61.436 &  4b, 4a &  116.298 &  4d, 4b &  193.257 &  4c, 2a \\
 48.262 &  4b, 4a, 8e &  61.438 &  8c, 8d, 8b &  117.304 &  4d, 4a &  193.258 &  6f \\
 48.263 &  4f, 4e &  61.439 &  4b, 4a, 8e &  117.306 &  4b, 4a &  193.259 &  4d, 6f, 2b \\
 48.264 &  8k &  62.442 &  II &  118.312 &  4d, 4b &  194.264 &  II \\
 49.272 &  4c, 4a, 4f, 4h &  62.443 &  4c &  118.314 &  2c, 2d &  194.266 &  2c, 2d, 2b \\
 49.273 &  2f, 2e, 2a, 2b &  62.444 &  4c &  123.348 &   &  194.267 &   \\
 49.274 &  4c, 4d, 4g &  62.445 &  4c &  123.349 &   &  194.268 &  6g, 2a \\
 49.275 &  4c, 4d &  62.446 &  4c, 4b, 4a &  123.350 &   &  194.269 &  6g \\
 49.276 &  4c, 4d &  62.447 &  4c, 4b, 4a &  124.360 &  2c, 4f, 2a &  195.3 &   \\
 50.284 &  4d, 4b, 4f, 4h &  62.448 &  4c, 4b, 4a &  124.361 &  4b, 8f &  198.10 &  II \\
 50.285 &  4c, 8e, 4d &  62.450 &  4b, 4a, 8d &  124.362 &  4c, 4d &  200.17 &   \\
 50.286 &  8d, 8h &  62.451 &  4d, 4b &  125.372 &  4b, 8e &  201.21 &  8c \\
 50.287 &  4f, 4e &  62.452 &  8c, 8d, 8b &  125.373 &  4f, 4e &  205.34 &  II \\
 50.288 &  8e &  62.453 &  4c, 8d &  125.374 &  8e &  207.43 &  8c, 12d \\
 51.290 &  II &  62.454 &  4c, 4b, 4a &  126.384 &  4c, 8e, 4a &  208.47 &   \\
 51.292 &  2f, 2e &  62.455 &  8c, 8e &  126.385 &  8e, 8f &  212.60 &  II \\
 51.293 &   &  63.458 &  II &  126.386 &  4d, 8f &  213.64 &  II \\
 51.294 &  2d, 2a, 2b, 2f, 2c, 2e &  63.459 &  4c &  127.388 &  II &  215.73 &   \\
 51.296 &  2d, 2a, 2b, 2f, 2c, 2e &  63.460 &   &  127.389 &   &  218.84 &  12d \\
 51.299 &  4c, 4a, 4f &  63.463 &  4c, 4b, 4a, 8d &  127.390 &   &  221.97 &   \\
 51.300 &  4b, 4a &  63.464 &  4c, 4b, 4a, 8d &  127.393 &  2c, 2a, 2d, 2b &  222.103 &  8c, 12d \\
 51.301 &  4b, 4a &  63.467 &  8d, 8a, 8f &  127.394 &   &  223.109 &   \\
 51.302 &  4f, 4e &  64.470 &  II &  127.396 &  4a &  224.115 &   \\
 51.303 &  4b, 4a, 4f, 4e &   &   &  127.398 &  4c, 4a &   &   \\
\label{tab:NodalPlane2fold}
\end{longtable}
}

\clearpage

{\makeatletter
\renewcommand\table@hook{\small}
\renewcommand\arraystretch{0.60}
\makeatother
\begin{longtable}{|cccc||cccc||cccc|}
\caption{List of magnetic space groups and relative Wyckoff positions which host a single decomposable EBR for spin waves. The column "MSGs" indicate the number of the space group in BNS setting, "MWP" the maximal Wyckoff position, "MPG" the magnetic point group isomorphic to the site symmetry group and finally "Bands" the number of bands composing the EBR.
Total of 445 EBRs out of the 1907 single-valued ones.}\\
\hline
\textbf{MSGs} & \textbf{MWP} & \textbf{MPG} & \textbf{Bands} &
\textbf{MSGs} & \textbf{MWP} & \textbf{MPG} & \textbf{Bands} &
\textbf{MSGs} & \textbf{MWP} & \textbf{MPG} & \textbf{Bands}\\
\hline
\endfirsthead
\multicolumn{12}{c}%
{\tablename\ \thetable\ -- \textit{Continued from previous page}} \\
\hline
\textbf{MSGs} & \textbf{MWP} & \textbf{MPG} & \textbf{Bands} &
\textbf{MSGs} & \textbf{MWP} & \textbf{MPG} & \textbf{Bands} &
\textbf{MSGs} & \textbf{MWP} & \textbf{MPG} & \textbf{Bands}\\
\hline
\endhead
\hline \multicolumn{12}{r}{
} \\
\endfoot
 9.41 &  8a &  $1$ &  4 &  89.92 &  4f &  $2 '2 '2$ &  4 &  147.16 &  6e &  $-1$ &  6 \\
 12.62 &  4e &  $-1$ &  2 &  90.101 &  4e &  $2 '2 '2$ &  4 &  148.17 &  9d &  $-1$ &  3 \\
 12.62 &  4f &  $-1$ &  2 &  90.101 &  4f &  $2 '2 '2$ &  4 &  148.17 &  9e &  $-1$ &  3 \\
 15.90 &  8f &  $-1$ &  4 &  90.102 &  4c &  $2 '2 '2$ &  4 &  148.20 &  18e &  $-1$ &  6 \\
 15.91 &  8a &  $-1$ &  4 &  91.109 &  8a &  $2$ &  8 &  150.27 &  2d &  $3$ &  2 \\
 15.91 &  8b &  $-1$ &  4 &  91.109 &  8b &  $2 '$ &  8 &  150.28 &  4d &  $3$ &  4 \\
 15.91 &  8e &  $2$ &  4 &  92.116 &  8d &  $2 '$ &  8 &  157.55 &  2b &  $3$ &  2 \\
 15.91 &  8f &  $2 '$ &  4 &  92.117 &  8a &  $2$ &  8 &  159.64 &  4b &  $3$ &  4 \\
 18.23 &  4k &  $2 '$ &  4 &  92.117 &  8b &  $2 '$ &  8 &  162.73 &  3f &  $2/m$ &  3 \\
 20.35 &  8k &  $2 '$ &  4 &  92.117 &  8c &  $2 '$ &  8 &  162.73 &  3g &  $2/m$ &  3 \\
 21.40 &  4k &  $2$ &  2 &  92.118 &  8f &  $2 '$ &  8 &  162.77 &  2c &  $32 '$ &  2 \\
 21.41 &  4k &  $2 '$ &  2 &  94.129 &  4d &  $2$ &  4 &  162.77 &  2d &  $32 '$ &  2 \\
 21.42 &  8k &  $2$ &  4 &  94.130 &  4d &  $2$ &  4 &  162.77 &  3f &  $2 '/m '$ &  3 \\
 23.52 &  8k &  $2 '$ &  4 &  94.133 &  4c &  $2 '2 '2$ &  4 &  162.77 &  3g &  $2 '/m '$ &  3 \\
 24.56 &  8k &  $2$ &  4 &  94.133 &  4d &  $2 '2 '2$ &  4 &  162.78 &  6f &  $2/m$ &  6 \\
 29.105 &  8a &  $1$ &  8 &  94.133 &  4e &  $2 '2 '2$ &  4 &  163.79 &  6g &  $-1$ &  6 \\
 32.141 &  4c &  $2 '$ &  4 &  95.141 &  8a &  $2$ &  8 &  163.83 &  6g &  $-1$ &  6 \\
 35.167 &  4c &  $2 '$ &  2 &  95.141 &  8b &  $2 '$ &  8 &  163.84 &  4c &  $32 '$ &  4 \\
 35.168 &  4c &  $2$ &  2 &  96.148 &  8d &  $2 '$ &  8 &  163.84 &  6f &  $2 '/m '$ &  6 \\
 36.177 &  8c &  $2 '$ &  4 &  96.149 &  8a &  $2$ &  8 &  164.85 &  3e &  $2/m$ &  3 \\
 36.179 &  8b &  $2 '$ &  4 &  96.149 &  8b &  $2 '$ &  8 &  164.85 &  3f &  $2/m$ &  3 \\
 37.184 &  8c &  $2$ &  4 &  96.149 &  8c &  $2 '$ &  8 &  164.89 &  2d &  $3m '$ &  2 \\
 37.185 &  8a &  $2$ &  4 &  96.150 &  8f &  $2 '$ &  8 &  164.89 &  3e &  $2 '/m '$ &  3 \\
 37.185 &  8b &  $2 '$ &  4 &  97.154 &  4c &  $2 '2 '2$ &  2 &  164.89 &  3f &  $2 '/m '$ &  3 \\
 37.185 &  8c &  $2 '$ &  4 &  97.154 &  4d &  $2 '2 '2$ &  2 &  164.90 &  6e &  $2/m$ &  6 \\
 37.185 &  8d &  $2$ &  4 &  97.156 &  8e &  $2 '2 '2$ &  4 &  165.91 &  4d &  $3$ &  4 \\
 37.186 &  8b &  $2$ &  4 &  97.156 &  8f &  $2 '2 '2$ &  4 &  165.91 &  6e &  $-1$ &  6 \\
 41.217 &  8a &  $2$ &  4 &  98.160 &  8f &  $2 '$ &  4 &  165.93 &  4d &  $3$ &  4 \\
 41.217 &  8b &  $2 '$ &  4 &  98.162 &  8a &  $2 '2 '2$ &  4 &  165.95 &  4d &  $3$ &  4 \\
 41.217 &  8c &  $m '$ &  4 &  98.162 &  8b &  $2 '2 '2$ &  4 &  165.95 &  6e &  $-1$ &  6 \\
 41.218 &  8b &  $2 '$ &  4 &  98.162 &  8c &  $2 '2 '2$ &  4 &  165.96 &  4d &  $3m '$ &  4 \\
 42.221 &  8b &  $2 '$ &  2 &  99.167 &  2c &  $m 'm '2$ &  2 &  165.96 &  6e &  $2 '/m '$ &  6 \\
 42.222 &  8b &  $2$ &  2 &  100.177 &  4c &  $m 'm2 '$ &  4 &  166.97 &  9d &  $2/m$ &  3 \\
 43.228 &  16a &  $2$ &  4 &  101.182 &  4c &  $2$ &  4 &  166.97 &  9e &  $2/m$ &  3 \\
 43.228 &  16b &  $2 '$ &  4 &  101.183 &  4c &  $2$ &  4 &  166.101 &  9d &  $2 '/m '$ &  3 \\
 45.239 &  8c &  $2 '$ &  4 &  102.190 &  4b &  $2$ &  4 &  166.101 &  9e &  $2 '/m '$ &  3 \\
 45.240 &  8a &  $2$ &  4 &  102.191 &  4b &  $2$ &  4 &  166.102 &  18e &  $2/m$ &  6 \\
 45.240 &  8b &  $2 '$ &  4 &  103.195 &  4c &  $2$ &  4 &  167.103 &  18d &  $-1$ &  6 \\
 45.240 &  8c &  $m '$ &  4 &  103.199 &  4c &  $2$ &  4 &  167.107 &  18d &  $-1$ &  6 \\
 48.262 &  8e &  $-1$ &  8 &  103.200 &  4c &  $m 'm '2$ &  4 &  167.108 &  18e &  $2 '/m '$ &  6 \\
 48.264 &  8k &  $-1$ &  8 &  103.201 &  8c &  $2 '$ &  8 &  168.109 &  2b &  $3$ &  2 \\
 50.281 &  4e &  $-1$ &  4 &  104.203 &  4b &  $2$ &  4 &  168.109 &  3c &  $2$ &  3 \\
 50.281 &  4f &  $-1$ &  4 &  104.207 &  4b &  $2$ &  4 &  168.111 &  3c &  $2 '$ &  3 \\
 50.286 &  8h &  $2 '$ &  8 &  104.209 &  8c &  $2 '$ &  8 &  168.112 &  4b &  $3$ &  4 \\
 52.314 &  8g &  $2$ &  8 &  104.210 &  4b &  $m 'm '2$ &  4 &  168.112 &  6c &  $2$ &  6 \\
 52.315 &  8f &  $-1$ &  8 &  105.217 &  8c &  $2 '$ &  8 &  173.132 &  6c &  $2 '$ &  6 \\
 52.318 &  8d &  $-1$ &  8 &  106.219 &  4b &  $2$ &  4 &  175.137 &  2c &  $-6$ &  2 \\
 52.319 &  8d &  $-1$ &  8 &  106.222 &  4a &  $2$ &  4 &  175.137 &  2d &  $-6$ &  2 \\
 52.319 &  8h &  $2$ &  8 &  106.225 &  4a &  $m 'm '2$ &  4 &  175.137 &  3f &  $2/m$ &  3 \\
 53.323 &  4g &  $2 '$ &  4 &  106.225 &  8c &  $2 '$ &  8 &  175.137 &  3g &  $2/m$ &  3 \\
 53.325 &  4g &  $2 '$ &  4 &  107.231 &  4b &  $m 'm '2$ &  2 &  175.141 &  3f &  $2 '/m '$ &  3 \\
 53.326 &  4g &  $2 '$ &  4 &  108.238 &  8c &  $m 'm2 '$ &  4 &  175.141 &  3g &  $2 '/m '$ &  3 \\
 53.328 &  4g &  $2$ &  4 &  109.244 &  16b &  $2 '$ &  8 &  175.142 &  4c &  $-6$ &  4 \\
 54.347 &  8c &  $2$ &  8 &  110.245 &  8a &  $2$ &  4 &  175.142 &  6f &  $2/m$ &  6 \\
 54.347 &  8e &  $2 '$ &  8 &  110.248 &  8a &  $2$ &  4 &  176.143 &  6g &  $-1$ &  6 \\
 56.373 &  8c &  $-1$ &  8 &  110.250 &  8a &  $m 'm '2$ &  4 &  176.147 &  6g &  $-1$ &  6 \\
 56.374 &  8e &  $2$ &  8 &  110.250 &  16b &  $2 '$ &  8 &  176.148 &  6g &  $2 '/m '$ &  6 \\
 56.376 &  8e &  $-1$ &  8 &  111.255 &  2e &  $2 '2 '2$ &  2 &  177.151 &  3f &  $2 '2 '2$ &  3 \\
 58.400 &  8g &  $2 '$ &  8 &  111.255 &  2f &  $2 '2 '2$ &  2 &  177.151 &  3g &  $2 '2 '2$ &  3 \\
 59.409 &  4c &  $-1$ &  4 &  112.264 &  4e &  $2 '2 '2$ &  4 &  177.152 &  3f &  $2 '2 '2$ &  3 \\
 59.409 &  4d &  $-1$ &  4 &  112.265 &  8i &  $2 '$ &  8 &  177.152 &  3g &  $2 '2 '2$ &  3 \\
 60.426 &  8b &  $-1$ &  8 &  113.273 &  4g &  $m 'm2 '$ &  4 &  177.153 &  2c &  $32 '$ &  2 \\
 60.426 &  8c &  $2 '$ &  8 &  114.275 &  4d &  $2$ &  4 &  177.153 &  2d &  $32 '$ &  2 \\
 60.426 &  8d &  $2 '$ &  8 &  114.277 &  4d &  $2$ &  4 &  177.153 &  3f &  $2 '2 '2$ &  3 \\
 60.426 &  8e &  $2$ &  8 &  114.279 &  4d &  $2$ &  4 &  177.153 &  3g &  $2 '2 '2$ &  3 \\
 60.428 &  8g &  $2 '$ &  8 &  114.281 &  8i &  $2 '$ &  8 &  177.154 &  4d &  $32 '$ &  4 \\
 60.429 &  8c &  $-1$ &  8 &  114.282 &  4c &  $2 '2 '2$ &  4 &  177.154 &  6g &  $2 '2 '2$ &  6 \\
 60.429 &  8e &  $2 '$ &  8 &  115.287 &  2g &  $m 'm '2$ &  2 &  182.184 &  6f &  $2 '2 '2$ &  6 \\
 60.430 &  8c &  $-1$ &  8 &  116.291 &  4i &  $2$ &  4 &  182.184 &  6g &  $2 '2 '2$ &  6 \\
 60.430 &  8h &  $2 '$ &  8 &  116.294 &  4i &  $2$ &  4 &  183.187 &  3c &  $m 'm2 '$ &  3 \\
 61.438 &  8b &  $-1$ &  8 &  116.295 &  4i &  $2$ &  4 &  183.188 &  3c &  $m 'm2 '$ &  3 \\
 61.438 &  8c &  $2 '$ &  8 &  116.296 &  4g &  $m 'm '2$ &  4 &  183.189 &  2b &  $3m '$ &  2 \\
 61.438 &  8d &  $m '$ &  8 &  117.305 &  4e &  $2 '2 '2$ &  4 &  183.189 &  3c &  $m 'm '2$ &  3 \\
 61.439 &  8e &  $2 '$ &  8 &  117.305 &  4f &  $2 '2 '2$ &  4 &  184.191 &  4b &  $3$ &  4 \\
 61.440 &  8b &  $-1$ &  8 &  120.326 &  8e &  $2 '2 '2$ &  4 &  184.191 &  6c &  $2$ &  6 \\
 61.440 &  8c &  $2 '$ &  8 &  120.326 &  8f &  $2 '2 '2$ &  4 &  184.193 &  6c &  $2 '$ &  6 \\
 61.440 &  8d &  $2 '$ &  8 &  121.331 &  4c &  $2 '2 '2$ &  2 &  184.194 &  6c &  $2 '$ &  6 \\
 61.440 &  8e &  $2 '$ &  8 &  121.331 &  4d &  $-4$ &  2 &  184.195 &  4b &  $3$ &  4 \\
 62.455 &  8e &  $2 '$ &  8 &  121.332 &  8g &  $m 'm2 '$ &  4 &  184.195 &  6c &  $2$ &  6 \\
 63.462 &  8d &  $-1$ &  4 &  122.335 &  8d &  $2 '$ &  4 &  184.196 &  4b &  $3m '$ &  4 \\
 63.464 &  8d &  $-1$ &  4 &  122.337 &  8d &  $2 '$ &  4 &  184.196 &  6c &  $m 'm '2$ &  6 \\
 64.471 &  8e &  $2 '$ &  4 &  123.345 &  2e &  $m 'm 'm$ &  2 &  185.200 &  4b &  $3$ &  4 \\
 64.473 &  8e &  $2 '$ &  4 &  123.345 &  2f &  $m 'm 'm$ &  2 &  185.201 &  4b &  $3$ &  4 \\
 64.474 &  8c &  $-1$ &  4 &  124.351 &  4e &  $2/m$ &  4 &  185.202 &  6c &  $m 'm2 '$ &  6 \\
 64.474 &  8e &  $2 '$ &  4 &  124.353 &  4f &  $2 '2 '2$ &  4 &  186.208 &  6c &  $m 'm2 '$ &  6 \\
 64.475 &  8e &  $2 '$ &  4 &  124.357 &  4e &  $2/m$ &  4 &  189.225 &  2c &  $-6$ &  2 \\
 64.476 &  8c &  $-1$ &  4 &  124.357 &  4f &  $2 '2 '2$ &  4 &  189.225 &  2d &  $-6$ &  2 \\
 64.476 &  8e &  $2$ &  4 &  124.360 &  4f &  $m 'm 'm$ &  4 &  190.232 &  4d &  $-6$ &  4 \\
 64.477 &  8e &  $2$ &  4 &  125.369 &  4e &  $2 '/m '$ &  4 &  191.238 &  3f &  $m 'm 'm$ &  3 \\
 65.485 &  4e &  $2/m$ &  2 &  125.369 &  4f &  $2 '/m '$ &  4 &  191.238 &  3g &  $m 'm 'm$ &  3 \\
 65.485 &  4f &  $2/m$ &  2 &  126.375 &  8f &  $-1$ &  8 &  191.239 &  3f &  $m 'm 'm$ &  3 \\
 66.498 &  8e &  $2/m$ &  4 &  126.377 &  4c &  $2 '2 '2$ &  4 &  191.239 &  3g &  $m 'm 'm$ &  3 \\
 66.500 &  8e &  $2/m$ &  4 &  126.381 &  4c &  $2 '2 '2$ &  4 &  191.240 &  2c &  $-6m '2 '$ &  2 \\
 68.513 &  8h &  $2 '$ &  4 &  126.381 &  8f &  $-1$ &  8 &  191.240 &  2d &  $-6m '2 '$ &  2 \\
 68.514 &  8h &  $2$ &  4 &  126.384 &  8e &  $2 '/m '$ &  8 &  191.240 &  3f &  $m 'm 'm$ &  3 \\
 68.515 &  8h &  $2$ &  4 &  126.385 &  8f &  $2 '2 '2$ &  8 &  191.240 &  3g &  $m 'm 'm$ &  3 \\
 68.516 &  8c &  $-1$ &  4 &  126.386 &  8f &  $2 '/m '$ &  8 &  192.243 &  4d &  $-6$ &  4 \\
 68.516 &  8d &  $-1$ &  4 &  128.399 &  4c &  $2/m$ &  4 &  192.243 &  6g &  $2/m$ &  6 \\
 68.516 &  8h &  $2 '$ &  4 &  128.401 &  4d &  $2 '2 '2$ &  4 &  192.245 &  4c &  $32 '$ &  4 \\
 68.517 &  8h &  $2$ &  4 &  128.405 &  4c &  $2/m$ &  4 &  192.245 &  6f &  $2 '2 '2$ &  6 \\
 68.519 &  8f &  $2 '2 '2$ &  4 &  128.409 &  8f &  $2 '2 '2$ &  8 &  192.246 &  6f &  $2 '2 '2$ &  6 \\
 68.519 &  8g &  $2 '2 '2$ &  4 &  128.410 &  4c &  $m 'm 'm$ &  4 &  192.247 &  6f &  $2 '2 '2$ &  6 \\
 69.524 &  8e &  $2/m$ &  2 &  129.417 &  4d &  $2 '/m '$ &  4 &  192.248 &  6f &  $2 '2 '2$ &  6 \\
 69.524 &  8f &  $2 '2 '2$ &  2 &  129.417 &  4e &  $2 '/m '$ &  4 &  192.248 &  6g &  $2 '/m '$ &  6 \\
 70.532 &  16b &  $2 '2 '2$ &  4 &  130.423 &  8d &  $-1$ &  8 &  192.249 &  6f &  $2 '2 '2$ &  6 \\
 70.532 &  16c &  $2 '2 '2$ &  4 &  130.429 &  8d &  $-1$ &  8 &  192.249 &  6g &  $2 '/m '$ &  6 \\
 70.532 &  16d &  $2 '2 '2$ &  4 &  130.432 &  8d &  $2 '/m '$ &  8 &  192.250 &  4c &  $32 '$ &  4 \\
 70.532 &  32e &  $-1$ &  8 &  130.433 &  8f &  $2 '2 '2$ &  8 &  192.250 &  4d &  $-6$ &  4 \\
 71.536 &  8k &  $-1$ &  4 &  130.434 &  8e &  $2 '/m '$ &  8 &  192.250 &  6f &  $2 '2 '2$ &  6 \\
 72.543 &  8e &  $-1$ &  4 &  132.451 &  4f &  $2/m$ &  4 &  192.250 &  6g &  $2/m$ &  6 \\
 72.544 &  8e &  $-1$ &  4 &  132.453 &  4e &  $2 '2 '2$ &  4 &  192.252 &  4c &  $-6m '2 '$ &  4 \\
 73.550 &  8c &  $2$ &  4 &  132.453 &  4f &  $2/m$ &  4 &  192.252 &  6f &  $m 'm 'm$ &  6 \\
 73.550 &  8d &  $2 '$ &  4 &  132.454 &  4e &  $2 '2 '2$ &  4 &  193.253 &  6f &  $2/m$ &  6 \\
 73.550 &  8e &  $2 '$ &  4 &  133.461 &  4b &  $2 '2 '2$ &  4 &  193.257 &  4c &  $-6$ &  4 \\
 75.1 &  2c &  $2$ &  2 &  133.462 &  8e &  $-1$ &  8 &  193.258 &  6f &  $2/m$ &  6 \\
 75.4 &  4c &  $2$ &  4 &  133.465 &  8e &  $-1$ &  8 &  193.259 &  4d &  $32 '$ &  4 \\
 75.5 &  4c &  $2 '$ &  4 &  133.466 &  4a &  $2 '2 '2$ &  4 &  193.259 &  6f &  $2 '/m '$ &  6 \\
 75.6 &  4b &  $2$ &  4 &  133.469 &  8e &  $2 '2 '2$ &  8 &  193.260 &  4c &  $-6$ &  4 \\
 76.11 &  8a &  $1$ &  8 &  134.477 &  4c &  $2 '2 '2$ &  4 &  193.260 &  4d &  $32 '$ &  4 \\
 77.17 &  4a &  $2$ &  4 &  134.477 &  4d &  $2 '2 '2$ &  4 &  193.260 &  6f &  $2 '/m '$ &  6 \\
 77.17 &  4b &  $2$ &  4 &  134.478 &  4c &  $2 '2 '2$ &  4 &  193.262 &  6g &  $m 'm 'm$ &  6 \\
 77.17 &  4c &  $2 '$ &  4 &  135.485 &  4d &  $2 '2 '2$ &  4 &  194.263 &  6g &  $2/m$ &  6 \\
 78.23 &  8a &  $1$ &  8 &  135.493 &  8e &  $2 '2 '2$ &  8 &  194.268 &  6g &  $2 '/m '$ &  6 \\
 79.25 &  4b &  $2$ &  2 &  136.499 &  4c &  $2/m$ &  4 &  194.269 &  6g &  $2/m$ &  6 \\
 79.28 &  8c &  $2 '$ &  4 &  136.501 &  4c &  $2/m$ &  4 &  194.270 &  6g &  $2 '/m '$ &  6 \\
 80.32 &  8a &  $2 '$ &  4 &  137.510 &  8e &  $-1$ &  8 &  194.272 &  6g &  $m 'm 'm$ &  6 \\
 80.32 &  8b &  $2 '$ &  4 &  137.512 &  4d &  $m 'm '2$ &  4 &  198.11 &  12b &  $2 '$ &  12 \\
 80.32 &  8c &  $2$ &  4 &  137.513 &  4d &  $m 'm '2$ &  4 &  203.29 &  48d &  $2 '2 '2$ &  12 \\
 81.33 &  2g &  $2$ &  2 &  137.513 &  8e &  $-1$ &  8 &  205.36 &  24d &  $2 '$ &  24 \\
 81.36 &  4g &  $2$ &  4 &  137.517 &  8e &  $2 '2 '2$ &  8 &  206.37 &  24d &  $2$ &  12 \\
 81.37 &  4g &  $2 '$ &  4 &  139.537 &  4c &  $m 'm 'm$ &  2 &  207.43 &  12d &  $2 '2 '2$ &  12 \\
 82.42 &  8g &  $2 '$ &  4 &  139.537 &  8f &  $2 '/m '$ &  4 &  210.55 &  32c &  $32 '$ &  8 \\
 83.43 &  2e &  $2/m$ &  2 &  140.547 &  8e &  $2 '/m '$ &  4 &  210.55 &  48d &  $2 '2 '2$ &  12 \\
 83.43 &  2f &  $2/m$ &  2 &  141.560 &  16c &  $2 '2 '2$ &  8 &  210.55 &  48e &  $2 '2 '2$ &  12 \\
 83.48 &  4e &  $2/m$ &  4 &  142.563 &  8b &  $2 '2 '2$ &  4 &  210.55 &  48f &  $2 '2 '2$ &  12 \\
 83.50 &  4c &  $2/m$ &  4 &  142.563 &  16e &  $2 '$ &  8 &  211.58 &  8c &  $32 '$ &  4 \\
 85.59 &  4d &  $-1$ &  4 &  142.564 &  16c &  $-1$ &  8 &  211.58 &  12d &  $2 '2 '2$ &  6 \\
 85.59 &  4e &  $-1$ &  4 &  142.564 &  16e &  $2 '$ &  8 &  212.62 &  12c &  $2 '2 '2$ &  12 \\
 85.64 &  8d &  $-1$ &  8 &  142.566 &  16e &  $2$ &  8 &  212.62 &  12d &  $2 '2 '2$ &  12 \\
 85.66 &  8f &  $-1$ &  8 &  142.567 &  16c &  $-1$ &  8 &  213.66 &  12c &  $2 '2 '2$ &  12 \\
 86.67 &  4e &  $2$ &  4 &  142.567 &  16e &  $2 '$ &  8 &  213.66 &  12d &  $2 '2 '2$ &  12 \\
 86.70 &  4e &  $2$ &  4 &  142.568 &  16e &  $2 '$ &  8 &  218.84 &  12d &  $-4$ &  12 \\
 86.71 &  4e &  $2$ &  4 &  142.569 &  16e &  $2$ &  8 &  222.98 &  12d &  $-4$ &  12 \\
 87.75 &  4c &  $2/m$ &  2 &  142.570 &  16c &  $2 '2 '2$ &  8 &  222.103 &  12d &  $-42 'm '$ &  12 \\
 87.75 &  8f &  $-1$ &  4 &  147.13 &  2d &  $3$ &  2 &  223.106 &  8e &  $32 '$ &  8 \\
 88.86 &  16c &  $-1$ &  8 &  147.13 &  3e &  $-1$ &  3 &  224.113 &  12f &  $2 '2 '2$ &  12 \\
 88.86 &  16e &  $2 '$ &  8 &  147.13 &  3f &  $-1$ &  3 &  227.133 &  96f &  $2 '2 '2$ &  24 \\
 89.90 &  2e &  $2 '2 '2$ &  2 &  147.16 &  4d &  $3$ &  4 &  228.139 &  96f &  $2 '2 '2$ &  24 \\
 89.90 &  2f &  $2 '2 '2$ &  2 &   &   &  $$ &   &   &   &  $$ &   \\
\label{tab:EBRsplitted}
\end{longtable}
}